\documentclass[twoside,12pt]{article}
\usepackage{epsfig}

\newcommand{\be}{\begin{equation}}
\newcommand{\ee}{\end{equation}}
\newcommand{\bea}{\begin{eqnarray}}
\newcommand{\eea}{\end{eqnarray}}

\begin{document}

\title{ \vspace{1cm} Progress in high-energy cosmic ray physics}
\author{S.\ Mollerach and E.\ Roulet\\
  \\
  Centro At\'omico Bariloche, CONICET, Argentina}
\maketitle
\begin{abstract}
  We review some of the recent progress in our knowledge about high-energy cosmic rays, with an emphasis on the interpretation of the different observational results. We discuss the effects that are relevant to shape the cosmic ray spectrum and the explanations proposed to account for its features and for the observed changes in composition. The physics of air-showers is summarized and we also present the results obtained on the proton-air cross section and on the muon content of the showers. We discuss the cosmic ray propagation through magnetic fields, the effects of diffusion and of magnetic lensing, the cosmic ray interactions with background radiation fields and the production of secondary neutrinos and photons. We also consider the cosmic ray anisotropies, both at large and small angular scales, presenting the results obtained from the TeV up to the highest energies and discuss the models proposed to explain their origin.
\end{abstract}
\section{Introduction}

Cosmic rays were discovered more than a century ago as particles arriving to the Earth from outer space. They consist mostly of ionized atomic nuclei, although electrons, positrons, antiprotons, gamma rays and neutrinos also arrive. We will focus here on the nuclear component, and the main observables that can be studied to learn about it are the spectrum (i.e. the distribution in energy), the composition (i.e. the distribution of nuclear masses at any given energy) and the anisotropies (i.e. the features in the distribution of arrival directions at different energies). 

The spectrum of the primary cosmic rays (CRs) extends from somewhat below one GeV up to beyond 100~EeV, i.e. spanning more than 11 orders of magnitude\footnote{Energy units are, in increasing powers of 1000, (k,M,G,T,P,E)eV, where in particular ${\rm EeV}\equiv 10^{18}$~eV. We use in this review natural units with $\hbar=c=k_B=1$.}.  In order to observe them, different techniques are adopted, depending on the CR energies under study.  Below a few hundred~TeV, CRs have very large fluxes, where for instance  $\Phi(>{\rm TeV})\simeq 5\times 10^6~{\rm m}^{-2}$sr$^{-1}$yr$^{-1}$, and can be directly detected with instruments on balloons or satellites before they interact in the atmosphere  (see \cite{maestro} for a recent overview and references). The precise mass or charge (sometimes even the isotope) can  be measured combining detectors such as magnetic spectrometers, calorimeters, transition radiation detectors, etc. At CR energies above few hundred TeV the fluxes become so low, with for instance $\Phi(>{\rm PeV})\simeq 50~{\rm m}^{-2}$sr$^{-1}$yr$^{-1}$, that the direct detection is no longer practical. However, when energetic CRs interact in the atmosphere they lead to the generation of air showers, also named atmospheric cascades,  consisting of a very large number of secondaries, some of which may eventually reach ground level. These showers include a hadronic component consisting of  protons, neutrons, pions, kaons and heavier mesons and baryons, an electromagnetic component consisting of electrons, positrons and  photons as well as a muonic component and neutrinos.  Hence, detectors at high elevation mountains, such as those  in the Tibet YBJ laboratory, are used to study CRs with TeV--PeV energies, while lower elevations are preferred for the study of the more penetrating showers from CRs with higher energies  (see \cite{verzi} for a recent overview and references). 

Besides the sampling of the particles at ground level with arrays of detectors, also the light emitted by the passage through the atmosphere of the electrons and positrons  in the shower can be detected using telescopes on the ground. This light is produced  as forward Cherenkov radiation when these particles travel faster than the speed of light in air or as isotropic fluorescence due to the emission from molecular nitrogen that got excited by the passage of the charged particles. Actually, imaging  the Cherenkov emission with telescopes, in observatories such as H.E.S.S. or in the future with CTA (see \cite{gamma} for a recent overview and references), it is also possible to do astronomy in the energy range between tens of GeV and up to tens of TeV by discriminating the showers initiated by photons from the much larger background from hadronic showers. Using arrays of non-imaging detectors of atmospheric  Cherenkov light, such as those in TUNKA or Yakutsk, it is possible to study CRs from PeV up to EeV energies, while the detection of fluorescence light becomes competitive above 100~PeV.  Also the radio emission at frequencies from 30 to 80~MHz, produced by the charge separation of the electrons and positrons of the shower by the Earth magnetic field, as well as due to the negative charge excess from the atomic electrons entrained by the electromagnetic component of the shower (Askaryan effect), is becoming a complementary way to detect CRs with energies in excess of 10~PeV \cite{schroeder}.

Regarding the direct detection experiments, some of the recent ones are those on balloons, such as ATIC, TRACER or CREAM, satellite-based ones such as PAMELA  or those in the International Space Station, such as AMS or the ISS-CREAM  experiment. We note that underground detectors, such as the IceCube experiment in the South Pole, can detect the muonic component of air showers. Although they are mainly focused on the detection of neutrinos, they also study CRs with energies above the TeV. 
Among the surface detector arrays, the IceTop experiment above the IceCube detector, consisting of frozen water-Cherenkov detectors (WCD) spread over an area of about 1~km$^2$, and the KASCADE and KASCADE-Grande detectors in Karlsruhe, that used scintillator detectors deployed over an area of about 0.04 and 0.5~km$^2$ respectively,  focused in the  range from  few PeV up to EeV energies. The AGASA experiment in Japan covered an area of 100~km$^2$ with a sparser array of scintillators, exploring energies above 100~PeV. The Haverah-Park experiment in England pioneered the technique of using WCD to observe the Cherenkov emission from the charged particles traversing the water in the detector while the Fly's Eye and Hi-Res experiments in Utah pioneered that of observing the air fluorescence UV light. 

Since the fluxes of CRs with energies above 1~EeV are extremely small, with e.g.  $\Phi(>{\rm EeV})\simeq 10~{\rm km}^{-2}$sr$^{-1}$yr$^{-1}$, huge detectors are required to observe a significant number of these events.  At present the largest detector array in operation is the Pierre Auger Observatory in Argentina, sampling an area of 3000~km$^2$ with WCD on the surface and several telescopes that observe the atmospheric fluorescence. The Telescope Array (TA) is the largest one in the northern hemisphere, currently sampling an area of 700~km$^2$ with an array of scintillators  and having also fluorescence telescopes. These hybrid detectors have the great advantage that they can study the lateral distribution of the showers at ground level with the surface detector (SD), as well as the longitudinal development of the shower  in the atmosphere with the fluorescence detector (FD). Acting the atmosphere almost as a calorimeter for the air showers, the FD leads to a good determination of the CR energies and hence enables a robust calibration of the energy assignment performed with the SD. Note that  the duty cycle of FD is only about 15\%, corresponding to the moonless nights with no clouds, while the SD is operational in principle all the time, and hence provides most of the events. The study of the shower development with FD allows,  in addition, to obtain important information on the CR composition, since at a given energy proton showers are  more penetrating (and also fluctuate more) than those from heavier nuclei. Hybrid operation also allows for many cross-checks and studies of systematics, which significantly affected the results from previous observatories using just one single technique.

For the near future the Pierre Auger Observatory is  carrying out an upgrade, adding scintillators on top of the WCD and updating the electronics, so as to be able to get composition information on an event by event basis with the SD and this should also allow it to improve the knowledge of the physics of the air showers. The Telescope Array is going to be enlarged so as to cover an area comparable to that of the Pierre Auger Observatory. This will allow to have a more uniform coverage of the whole CR sky. Several developments are also being performed to study the feasibility of new detection techniques to achieve  even larger exposures, such as the observation of atmospheric fluorescence from orbiting telescopes or the use of radio detection.

The main issues one would like to understand with these studies are:
\begin{itemize}
\item To identify the sites where CRs are produced, something which is quite difficult due to the fact that CRs are charged particles and are hence significantly deflected by intervening magnetic fields. Since these fields are not well known and the CR charges are uncertain, these deflections are poorly constrained. However, since the deflections are expected to decrease with increasing energies, at the highest energies observed, close to 100~EeV, they could become sufficiently small  that the CRs may point back not far from their sources. Also a multi-messenger approach, searching for gamma rays or neutrinos produced in association with the CR acceleration process, may be useful to pin-down the sources.

\item To identify the acceleration mechanism, i.e. whether it is due to electrostatic acceleration in pulsars, magnetars or supermassive black holes or  alternatively acceleration associated to shock waves, such as those produced in supernova (SN) explosions, gamma ray bursts (GRB), in accretion disks and jets close to the central supermassive black holes in active galactic nuclei (AGN) or in the hotspots at the termination of the jets or in the radio lobes,  in shocks inside galaxy clusters, in collisions of galaxies or in galaxy winds.

\item To explain the origin of the spectral features so as to understand the CR confinement in the Galaxy, the eventual transition from Galactic to extragalactic CRs, the maximum energies achievable at the sources as well as the possible attenuation of the fluxes during the CR propagation.

\item To determine the composition of the CRs and its evolution with energy, which can give indications about the properties of the medium in the acceleration region and the effects of the propagation, such as the confinement near the sources or the impact of the interactions that produce secondary nuclei.

\item The detailed study of anisotropies can also help to learn about the intervening magnetic fields, both in the Galaxy and in the intergalactic space.

\item Since the CRs can reach the highest energies observed in nature, of order 100~EeV, which are $10^7$ times larger than those reached at the large hadron collider (LHC), they  can probe physics in the ultra-relativistic regime, with Lorentz factors up to $10^{10}$ (or even $10^{11}$ if protons were present up to the highest CR energies). The center of mass energies involved in the CR collisions with the air nuclei are up to two orders of magnitude larger than those achieved in the LHC $pp$ collisions, and hence CRs allow one to test models of hadronic interactions well beyond the energy regime tested at colliders.
  
  \end{itemize}

In this review we will mainly consider the CRs with energies above 1~PeV, with particular emphasis on those above  1~EeV, the so-called ultrahigh-energy cosmic rays (UHECRs). Several useful books  exist about the physics and astrophysics of CRs, among which we can mention \cite{berezinsky,gaisser}.

\section{The physics of air showers}

We now consider the main aspects of the development of air showers in the atmosphere in order to understand how the different observables are related to the properties of the primary CRs. 
We will eventually be interested in the showers produced by hadronic CRs, but it is convenient to discuss first the case of a purely electromagnetic (em) shower, i.e. one initiated by a photon or an electron, which is simpler and still has many similar features.

\begin{figure}
\begin{center}
\begin{minipage}[t]{8 cm}
\centerline{\epsfig{file=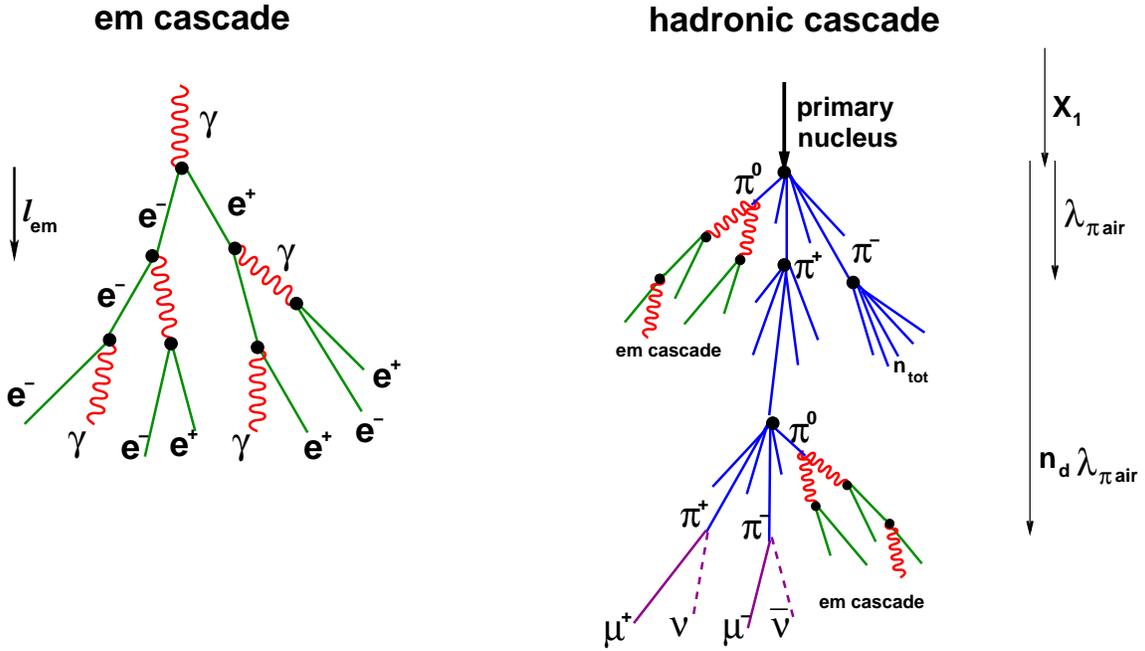,scale=0.5}}
\end{minipage}
\begin{minipage}[t]{16.5 cm}
\caption{Schematic view of the development of electromagnetic and hadronic air showers. The black dots correspond to the air nuclei with which the shower particles interact.\label{airsh.fig}}
\end{minipage}
\end{center}
\end{figure}

When a high-energy photon interacts in the air, which consists of about 78\% N and 21\% O, it will do so mostly with the electric charge of the atomic nuclei to produce $e^+e^-$ pairs, with the Compton scattering with electrons having only a small effect. In turn, the electrons and positrons will interact with nuclei
to produce photons by bremsstrahlung, with the interactions with electrons that produce ionization and excitation becoming relevant only for energies smaller than the critical value $E_{\rm c}\simeq 84$~MeV. In the simplified picture of the model of Heitler \cite{heitler}, one assumes that after a given interaction step $\ell_{\rm em}$ the two particles produced in the processes $\gamma\to e^+e^-$ or $e^\pm \to e^\pm \gamma$ share equally the energy. In this way the overall number of particles in the shower is doubled at each interaction stage and their average energy is halved, as schematically depicted in the left side of Fig.~\ref{airsh.fig}. Hence, after $n$ generations one has that the number of particles $N$ (including electrons, positrons and photons) is $N=2^n$ and their energies are $E=E_0/N$, with $E_0$ being the energy of the primary. The exponential growth in the number of secondaries continues until the typical energy per particle gets smaller than $E_{\rm c}$, so that the interactions with atomic electrons becomes relevant and the overall energy of the shower starts  to be dissipated.

Usually the distance along the shower is measured as a column density $X$, in g~cm$^{-2}$, measuring the amount of air traversed from the top of the atmosphere and in the direction of the propagation. For reference, the total vertical column density at sea level is about $10^3$\,g~cm$^{-2}$, and the total column density for a shower at zenith angle $\theta=60^\circ$ is twice as large. Similarly, the interaction lengths are specified in g~cm$^{-2}$ as $\lambda_{\rm int}\equiv \rho_{\rm air}/n_{\rm air}\sigma_{\rm int}={\bar A} m_p/\sigma_{\rm int}$, with the air density being $\rho_{\rm air}={\bar A} m_p n_{\rm air}$ where ${\bar A}\simeq 14.5$ is the average mass number of the air. In this way, the main characteristics of the electromagnetic showers can be studied quite independently from the CR arrival direction (and are similar to those observed in solids once the corresponding changes in $\lambda_{\rm int}$ and $E_{\rm c}$ are taken into account). 

The energy losses of charged particles in a medium are usually parametrized as d$E$/d$X=-\alpha-E/\xi$, where $\alpha\simeq 2$~MeV/(g~cm$^{-2})$ accounts for the ionization losses and $\xi$ parametrizes the radiative (mostly bremsstrahlung) losses. For electrons propagating in air  the radiation length is  $\xi\simeq 37$\,g~cm$^{-2}$ and is usually denoted as $X_0$.  Given that for $E\gg E_{\rm c}$  the energy of an electron decreases as $E\propto \exp(-X/X_0)$, one finds\footnote{Note that since it is easy to emit soft photons, the cross section for bremsstrahlung emission is actually infrared divergent and needs to be handled carefully in MC simulation programs,  but the radiation length $X_0$ defining the average energy loss length is well defined, and so is $\ell_{\rm em}$. The pair production interaction length is in turn $\lambda_{ee}=(9/7)X_0$. A consequence of the copious production of low energy photons is that their number is usually much larger than that of electrons and positrons, which is at variance with the simplified model of Heitler.} that the length over which the energy is halved is $\ell_{\rm em}\simeq {\rm ln}2\,X_0\simeq 25$~g~cm$^{-2}$. Note that the critical energy, corresponding to the one for which the ionization losses become equal to the radiation ones, is $E_{\rm c}\simeq\alpha X_0$.

The maximum in the development of the electromagnetic showers is reached in this model when all the particles have an energy $E_{\rm c}$, with their total number being 
\be N_{\rm max}\simeq \frac{E_0}{E_{\rm c}}\simeq 10^{10}\frac{E_0}{\rm EeV}.
\ee
This happens after a number of generations $n_{\rm max}\simeq \log N_{\rm max}/\log 2$, which is quite large, typically 30--40. The depth of the shower maximum is obtained as
\be
X_{\rm max}\simeq n_{\rm max}\ell_{\rm em}=X_0\ln \left(\frac{E_0}{E_{\rm c}}\right)\simeq 850\frac{\rm g}{\rm cm^2}+D_{10}^{\rm em}\log\left(\frac{E_0}{\rm EeV}\right).
\ee
Here we introduced the elongation rate $D_{10}\equiv {\rm d}\langle X_{\rm max}\rangle /{\rm d} \log E$, which represents the change in average depth of shower maximum per decade of energy, and for electromagnetic showers in this model is $D_{10}^{\rm em}=X_0\ln 10\simeq 85$\,g~cm$^{-2}$. 

There are two additional features that become relevant for electromagnetic showers at ultrahigh-energies. One is  the so-called Landau-Pomeranchuk-Migdal (LPM) effect \cite{lpm}, which arises from the suppression of the bremsstrahlung emission and pair production due to a destructive interference between the  scattering amplitudes from  different air nuclei, leading to an  enhanced growth of $X_{\rm max}$ at energies beyond 10~EeV. The other is the fact that photons with energies beyond 50~EeV may interact with the geomagnetic field to produce electron-positron pairs, and hence, at these energies, the showers could start developing even before entering the atmosphere \cite{preshower}.  

Turning now to the case of showers initiated by CR protons (we will later consider the case of heavier nuclei), one of the main differences that appears is that in the hadronic interactions a large number of secondaries are produced, with the total multiplicity of particles per collision, $n_{\rm tot}$, reaching values of several tens, or even a few hundred at the highest energies. This implies that although the typical hadronic interaction lengths are larger than the electromagnetic radiation length, with e.g. the one associated to proton-air interactions at PeV energies being $\lambda_{p\,{\rm air}}\simeq 80$~g~cm$^{-2}$, the showers develop faster than in the electromagnetic case. Most of the secondaries in a proton nucleus inelastic interaction are pions, with similar amounts of $\pi^0$,  $\pi^+$ and  $\pi^-$, although heavier scalar mesons such as kaons, vector mesons such as $\eta$ and $\rho$ and heavy baryonic resonances also get produced.  

One may then consider as a simplified model of hadronic showers \cite{ma05} one in which at each generation a number $n_{\rm tot}$ of pions gets produced, having all equal energies and with $n_{\rm tot}/3$ being neutral pions. The  $\pi^0$ almost immediately decay into two photons, which produce electromagnetic sub-showers similar to those discussed previously. In turn, the charged pions can decay only through weak interactions via e.g. $\pi^-\to\mu^-\bar{\nu}_\mu$ and hence, as long as their decay length remains larger than their interaction length, they will reinteract rather than decay.  This happens for $\gamma c\tau_{\pi}>\lambda_{\pi\,{\rm air}}/\rho_{\rm air}$, with the Lorentz factor being $\gamma=E/m_\pi$, the charged pion lifetime being $\tau_\pi\simeq 26$~ns and $\lambda_{\pi\,{\rm air}}\simeq 1.5\lambda_{p\,{\rm air}}\simeq 120$~g~cm$^{-2}$ (since the $\pi p$ cross section is about 2/3 the $pp$ cross section). This implies that pions will reinteract as long as their energy satisfies $E>E_{\rm d}\simeq 100$~GeV(10$^{-4}$~g~cm$^{-3}/\rho_{\rm air})$. Hence, at the  heights above 10~km where the initial development of the shower takes place, charged pions will reinteract for energies in excess of $\sim 30$~GeV (for showers with large zenith angles, which develop higher in the atmosphere, pions may be able to decay at even higher energies).  The number of generations of hadronic interactions taking place before the charged pions can finally decay is $n_{\rm d}\simeq \log(E_0/E_{\rm d})/\log(n_{\rm tot})$, so that if one adopts as a typical average multiplicity $n_{\rm tot}\simeq 20$, one gets that  $n_{\rm d}\simeq  5$--6 for EeV primaries\footnote{A more accurate estimate of the pion decay energy can be performed \cite{ka12} using that the density of the atmosphere is approximately exponential, with $\rho(z)=\rho_0 \exp(-z/h_0)$, with $h_0\simeq 10$~km. One has then that $\rho(z)=X\cos\theta/h_0$ for a shower with zenith $\theta$ after traversing a slant depth $X$. Setting $X=n_{\rm d}\lambda_{\pi\,{\rm air}}$ at the time of decay, one gets $E_{\rm d}\simeq m_\pi h_0/(c\tau_\pi n_{\rm d}\cos\theta)\simeq 150$~GeV/$n_{\rm d}\cos\theta\simeq 30$~GeV/cos$\theta$, which is actually independent from the interaction length.}. A pictorial view of the main characteristics of  hadronic showers is given in the right side of Fig.~\ref{airsh.fig}.

Since at every step about 1/3 of the energy is channeled into the electromagnetic component, after $n$ generations one has that
\be 
E_{\rm em}\simeq E_0\left[1-\left(\frac{2}{3}\right)^n\right],
\ee 
and hence we see that after  six generations about 90\% of the initial energy gets transferred to the em component of the shower, with the remaining 10\% being essentially the muons and neutrinos from the charged pion decays. Note that low energy muons may also decay before reaching the ground, via $\mu\to e\bar{\nu}_e\nu_\mu$. This happens for $\gamma c \tau_\mu<h$, i.e. for muon  energies smaller than $\sim 3$~GeV (adopting  a typical height of muon production of $h\simeq 20$~km and using that $\tau_\mu\simeq 2\,\mu$s).  
One should keep in mind that the energy losses of the muons due to ionization in the atmosphere after traversing a column density $\Delta X$ are $\alpha\Delta X\simeq 1$~GeV$(\Delta X/500$\,g~cm$^{-2}$), which may not be negligible at low energies\footnote{The ionization losses of the muons are similar to those of the electrons, with $\alpha\simeq 2$~MeV/(g~cm$^{-2}$), while the radiation losses due to bremsstrahlung are suppressed by the factor $(m_e/m_\mu)^2$, so that the critical energy for muons is much larger, of order $E_{\rm c}(\mu)\simeq 3.6$~TeV.}. Actually, we note that  most of the charged cosmic radiation reaching the ground consists of  secondary muons with  few GeV energies  resulting from the decays of pions produced in showers from relatively low energy ($\sim 100$~GeV per nucleon) hadronic CRs.

The longitudinal development of hadronic showers is more complex than that of the electromagnetic ones since the hadronic core continuously feeds the em component, the  multiplicities of secondary particles depend on energy, there is the competition between decays and reinteractions, etc. A simple estimate of the depth of maximum  of the em component of the  shower can be obtained as
\be 
X_{\rm max}\simeq X_1+X_0\,\ln \left(\frac{E_0}{2 n_{\rm tot}E_{\rm c}}\right),
\label{xmaxp}
\ee 
where $X_1$ is the depth of the first interaction and the second term corresponds to the depth of maximum of the em sub-showers resulting from the photons, with average energies $E_\gamma\simeq E_0/2n_{\rm tot}$, produced in the decays of the first generation of $\pi^0$. This approximate expression reveals how the depth of shower maximum depends on some of the main features of the hadronic interactions at high energies, such as the inelastic proton-air cross section through $X_1={\bar A}m_p/\sigma^{inel}_{p\,{\rm air}}$ or on the multiplicity of the interactions $n_{\rm tot}$. More detailed treatments should actually also account for the inelasticity of the interactions, since a significant amount of the energy of the incoming particle is usually carried away by a leading particle, as well as other aspects such as the detailed distribution of the energies of the secondaries, etc. \cite{ul11}. Specific predictions are then performed using different air showers simulation codes, such as AIRES \cite{aires} or CORSIKA \cite{corsika}, which incorporate different models to describe the hadronic interactions, such as QGSJET \cite{qgsjet}, EPOS \cite{epos} or  SIBYLL \cite{sibyll}, which were developed with emphasis in air shower physics. These models predict the main features of the interactions and are regularly updated \cite{de11,pi13} in order that they better account for the measurements performed at high energy hadron colliders, such as the Tevatron or the LHC. One has to keep in mind however that the phase space explored at colliders is mostly the one transverse to the beam, where the large transferred energies could allow for instance for the production of the heavy particles that these colliders aim to discover. This is the so-called central region, characterized by small values, $|\eta|<3$, of the pseudorapidity variable $\eta\equiv -\ln\tan(\theta/2)$, where $\theta$ is the scattering angle in the laboratory frame. On the other hand, one has that on average most of the energy flows into the forward region of the collision, corresponding to larger pseudorapidities, and hence this region is the most relevant one for the development of the air showers. However, being close to the beam axis  this forward region is more difficult to study at colliders, although a lot of effort is now being devoted to perform measurements close to the beam, with detectors such as LHCf near ATLAS or TOTEM next to CMS, to understand in particular the properties of the interactions relevant for high-energy CRs. Measurements in the central region, such as that of the pseudorapidity density of the multiplicity of charged secondaries, d$N_{\rm ch}/{\rm d}\eta$, are also quite useful to constrain hadronic models. Another issue that has to be kept in mind is that, to describe UHECR interactions, one needs to extrapolate these models up to energies well beyond the ones that have been tested at present. In particular, the center of mass (CM)  energy in the initial collision of a CR proton with a nucleon in an air atom is $\sqrt{s}=\sqrt{2m_pE}\simeq 10$~TeV$\sqrt{E/0.1{\rm EeV}}$. The CM energies being explored at the LHC, at which $\sqrt{s_{pp}}\simeq 14$~TeV, correspond then to those reached in  collisions involving the constituent nucleons of CRs with $E\simeq A\times 10^{17}$~eV, where $A$ is the mass number of the CR.

An interesting expression which can be inferred from eq.~(\ref{xmaxp}) relates the elongation rate of proton induced showers, $D_{10}^p$, to that of electromagnetic showers,
\be 
D_{10}^p\simeq D_{10}^{\rm em}+\frac{{\rm d}X_1}{{\rm d}\log E}-\frac{{\rm d}\ln 2n_{\rm tot}}{{\rm d}\log E}.
\ee 
Since $X_1$ becomes smaller with increasing energies due to the increase of the cross section while the multiplicity $n_{\rm tot}$ increases with energy, one finds that the elongation rate of proton showers is smaller than that of electromagnetic showers. Indeed, from detailed numerical simulations it turns out that $D_{10}^p\simeq 55$\,g~cm$^{-2}$, which is significantly smaller than $D_{10}^{\rm em}\simeq 85$\,g~cm$^{-2}$.

Turning now to the case in which the primary CRs are heavier nuclei, the main properties of the development of the resulting air showers can be understood on the basis of the superposition model, i.e. considering that the shower produced by a CR nucleus with mass number $A$ and energy $E_0$ is equivalent to a superposition of $A$ showers produced by protons of energy $E_0/A$. This  implies that
\be 
\langle X_{\rm max}^A(E)\rangle \simeq \langle X_{\rm max}^p(E/A)\rangle .
\ee 
This relation is reproduced reasonably well in detailed numerical simulations. If one neglects a possible energy dependence of the proton elongation rate, this relation also implies that the elongation rate of a given mass component $A$ should be similar to that of protons, $D_{10}^A\simeq D_{10}^p$. On the other hand, the measurement of smaller  values of the elongation rate would indicate a change towards a heavier mass component with increasing energies, while higher elongation rates would be a signature of the mass composition becoming lighter.

Another property which is important is that the fluctuations in the values of $X_{\rm max}$ between showers of the same energies are quite different for different primary masses. In particular, the $X_{\rm max}$ spread in proton showers arises mostly from the fluctuations in the first interaction point, with additional contributions arising from the fluctuations in the subsequent development of the em component. In the case of showers from heavy primaries, which can  be considered as a superposition of $A$ proton showers whose fluctuations get largely averaged,  the overall variance Var$(X^A_{\rm max})$ results considerably smaller than  Var$(X^p_{\rm max})$. In particular, while Var$(X^p_{\rm max})\simeq 60$\,g\,cm$^{-2}$ one has e.g. that Var$(X^{Fe}_{\rm max})\simeq 20$\,g\,cm$^{-2}$. When the composition is a mixture of different nuclear masses,  although the average depth of shower maximum can be obtained as the weighted average of the maxima of the different nuclear species, the total spread in $X_{\rm max}$ values will be larger than the spread in each mass component alone. Hence, the interpretation of the measurements of Var$(X_{\rm max})$ depends a lot on the amount of admixture between different masses present in the CR fluxes. 

It is relevant to keep in mind that the superposition model is just an approximation, since the interactions of the different nucleons in a nucleus are not independent from each other. For instance, if one nucleon in the `front' part of the nucleus does not interact, it is likely that those `behind' it will not interact either and, on the other hand, when an interaction does take place the number of `wounded' nucleons is usually larger than one. As a result, the average depth of first interaction of a nucleus with energy $E$  is usually larger than the smallest among the depths of interaction of $A$ independent proton showers with energy $E/A$. Due to these facts it is not simple to obtain an estimate of Var$(X_{\rm max}^A)$ on the basis of the superposition model \cite{semisup}.

\begin{figure}[tb]
\begin{center}
\begin{minipage}[t]{8 cm}
\centerline{\epsfig{file=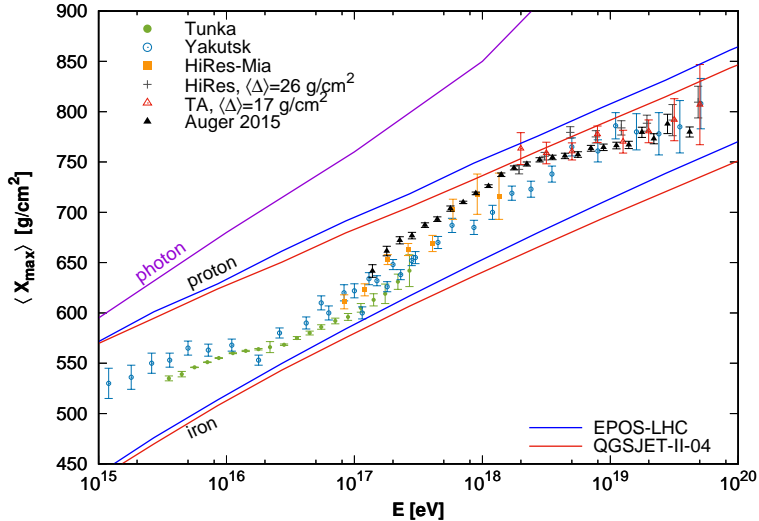,scale=1}}
\end{minipage}
\caption{Measurements of $\langle X_{\rm max}\rangle$ as a function of energy (data taken from \cite{ka12} and \cite{augerxmax}). Also shown are the predictions from photon initiated showers, as well as those from protons and Fe nuclei using different recent hadronic models. HiRes
and TA data have been corrected for detector effects, shifting the $X_{max}$ values by the amount $\langle \Delta\rangle$ indicated in each case (see \cite{ka12}).
 \label{xmaxvse}}
\end{center}
\end{figure}

In Fig.~\ref{xmaxvse} we show measured values of $\langle X_{\rm max}\rangle\,{\rm vs.}\, E$ from different experiments. Fluorescence telescopes have measurements above 0.1~EeV, while the measurements extending to lower energies come from non-imaging Cherenkov telescopes. These last experiments are essentially arrays, with typical separations of several tens of meters, of large phototubes facing up which detect the forward emission of Cherenkov light by the electrons in the shower, which leaves a footprint on the ground with typical radius of about 120~m. The total reconstructed light is proportional to the shower energy while the inner radial slope of its intensity is sensitive to the depth of shower maximum.
Also shown are the expectations for showers initiated by photons, protons and iron nuclei, illustrating the main features previously discussed. It is apparent that the $X_{\rm max}$ values lie between the predictions from protons and those of Fe, having some well defined changes in the trends with energy that will be further discussed in the next section.

Another important ingredient of the hadronic showers is their muonic component. In the simplified picture in which all muons are produced in the decays of the charged pions with energies $E_{\rm d}$, which in the case of proton initiated showers will correspond to those in the generation $n_{\rm d}$ such that $E_{\rm d}\simeq E_0/(n_{\rm tot})^{n_{\rm d}}$, the total number of muons will then be given by $N_\mu=(n_{\rm ch})^{n_{\rm d}}$,
with $n_{\rm ch}\simeq 2 n_{\rm tot}/3$ being the average multiplicity of charged pions per hadronic interaction. 
Note that since the number of generations $n_{\rm d}$ increases approximately logarithmically with energy, this implies that as the energy increases a larger fraction of the energy will be converted into the em component, and consequently a smaller fraction will end up into muons. As a result, the growth of $N_\mu$ with energy is 
less pronounced than linear. Indeed, writing $N_\mu=(E_0/E_{\rm d})^\alpha$ one gets that
\be 
N_\mu=(n_{\rm ch})^{n_{\rm d}}=(E_0/E_{\rm d})^\alpha=(n_{\rm tot})^{n_{\rm d}\alpha},
\ee 
so that the exponent determining the energy dependence of $N_\mu$ is 
\be
\alpha=\frac{\ln n_{\rm ch}}{\ln n_{\rm tot}}\simeq 1-\frac{\ln 1.5}{\ln n_{\rm tot}}\simeq 0.87\div 0.91.
\ee
In the case of nuclei with mass number $A$, using the superposition approximation one has that
\be 
N_\mu^A=A\left(\frac{E_0}{AE_{\rm d}}\right)^\alpha=A^{1-\alpha}N_\mu^p.
\ee 
This implies that for a given CR energy the number of muons becomes larger the heavier is the primary. This behavior can be understood from the fact that the nucleons that compose a heavy nucleus, having just part of the total energy,  will give rise to less generations before the charged pions decay. Hence, a smaller fraction of the energy will end up in the electromagnetic component, with a corresponding larger production of muons.
In particular, for the case of Fe nuclei one would expect that $N_\mu^{Fe}/N_\mu^p\simeq 56^{1-\alpha}\simeq 1.5$.
This property gives a  very useful handle to infer the composition of the primaries using surface detectors capable of determining the muon component, as was the case in KASCADE and as is the aim of the Auger upgrade.

Regarding the shape of the longitudinal development of the electromagnetic component of an hadronic shower (actually of the $e^+$ and $e^-$ component that leads to the fluorescence that can be measured with FD telescopes), a useful parametrization that is used to fit it is the Gaisser-Hillas function
\be 
N_{e}=N_{\rm max}\left(\frac{X-X_1}{X_{\rm max}-X_1}\right)^{\frac{X_{\rm max}-X_1}{\lambda}}\exp\left(\frac{X_{\rm max}-X}{\lambda}\right),
\ee 
with $\lambda\simeq 70$\,g\,cm$^{-2}$ being an effective attenuation length. 

\begin{figure}[tb]
\begin{center}
\begin{minipage}[t]{8 cm}
\centerline{\epsfig{file=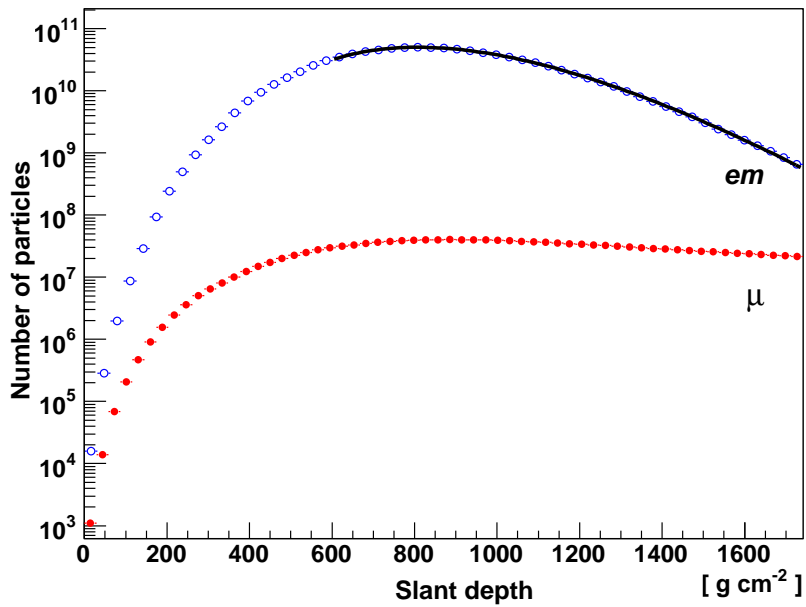,scale=.6}}
\end{minipage}
\caption{Average longitudinal profiles of 10~EeV proton initiated showers (adapted from \cite{augerweather}). A fit with a Gaisser-Hillas function to the electromagnetic component (solid line) is also shown. The average shower maximum lies at about 800~g\,cm$^{-2}$. Also note the much flatter profile of the muonic component.
 \label{longprof}}
\end{center}
\end{figure}

Regarding the longitudinal profile of the muonic component, the depth of  maximum production of muons by pion decays is about $X_{\rm max}^\mu\simeq n_{\rm d}\lambda_{\pi\,{\rm air}}$ ($\simeq 500$--600\,g\,cm$^{-2}$ at EeV energies).  At larger depths the longitudinal profile of the total number of muons remains relatively flat since their production gets suppressed and only the very low energy muons decay. The electromagnetic and muonic average longitudinal profiles for 10~EeV proton initiated showers are depicted in Fig.~\ref{longprof}.

The other important property of the showers is their lateral distribution determined at ground level. In the case of a purely electromagnetic shower, an analytical approximation to the particle density was obtained by Nishimura, Kamata and Greisen (NKG)  \cite{nk,greisen}, having the form
\be 
\frac{{\rm d}N_e}{{\rm d}r^2}\propto r_M^{-2}\left(\frac{r}{r_M}\right)^{s-2}\left(1+\frac{r}{r_M}\right)^{s-4.5},
\ee
where the radial distance from the shower core $r$ is measured in the shower plane orthogonal to the shower axis. The Moli\`ere radius $r_M\simeq (10^{-3}\,{\rm g\,cm}^{-3}/\rho_{\rm air})90$~m determines the lateral spread, which is mainly due to multiple Coulomb scattering, and the age parameter determining the slope is $s=3X/(X+2X_{\rm max})$, with $s=1$ at the shower maximum and increasing with depth. 
On the other hand, the em component of an hadronic shower, which gets constantly regenerated by the hadronic core interactions and by muon decays, has a  different shape and is more spread. The muonic component of hadronic showers, which is produced high in the atmosphere, has a relatively flat lateral profile, being actually flatter for heavy nuclei than for protons since the first develop higher. In an observatory such as Auger, which is sensitive to both em and muonic components, one then just uses an  NKG-like parameterization where the Moli\`ere radius is replaced by an effective core distance $r_{\rm c}\simeq 700$~m. The lateral distribution function (LDF) giving the measured signals $S$ as a function of radius  is parameterized as
\be
S(r)\simeq S(r_{\rm opt})\left(\frac{r}{r_{\rm opt}}\right)^\beta \left(\frac{r+r_{\rm c}}{r_{\rm opt}+r_{\rm c}}\right)^{\beta+\gamma} .
\ee
Distributions of this type, or similar alternative expressions, are used to fit the measured signals and reconstruct the showers (core position, slope of the LDF, signal amplitude). Note that at 10~EeV energies the signals can be measured up to several km from the shower core. The inferred signal at a certain optimal distance $r_{\rm opt}$ is used as an estimator of the primary CR energy. The optimal distance depends on the  separation between detectors in the array  (it is 1000~m for the Auger array with 1.5~km separation, 800~m for TA with 1.2~km separation, 450~m for the Auger Infill sub-array with 750~m separation, etc.) and is chosen so that the energy determination minimizes the dependence on the  possible shower to shower fluctuations (as well as on the unknown mass composition of the primaries). 

The relation between the signal $S(r_{\rm opt})$ and the energy of the primary is not straightforward because the attenuation of the shower before it reaches ground level depends on the zenith angle, i.e. on the amount of atmosphere traversed. One option is to compare with the results of shower simulations at different zenith angles to estimate this attenuation, as is performed by TA \cite{taspect}. The other alternative, followed by Auger, is to use the so-called constant intensity cut (CIC) method to evaluate, for a given primary energy, the dependence of the measured signal with zenith angle \cite{augernim}. The CIC relies on the fact that the CR flux is to a very good approximation isotropic, and hence the measured flux above a given energy (corresponding to a zenith dependent $S(r_{\rm opt})$) should be the same for different  zenith angles. 

After accounting for the attenuation effects, one can use hybrid events to calibrate the signals with the actual energies obtained by the FD telescope measurements. These energies are determined by fitting the longitudinal profile of the showers, whose integral gives essentially the total electromagnetic energy of the shower. One has also to account for the `missing energy' carried away by neutrinos and muons \cite{ba04}, that is typically about 5--15\% depending on the composition and energy of the primary. Previous surface detector arrays that had no FD counterpart, such as AGASA, Haverah Park or KASCADE, had to rely instead on shower simulations to estimate the energies. 
By determining the CR fluxes at different energies, the CR energy spectrum can then be finally obtained.

\section{The energy spectrum and cosmic ray composition}
\subsection{Main features of the all-particle spectrum}
The measured CR energy spectrum is quite steep, being approximately a power law d$\Phi/{\rm d}E\propto E^{-\gamma}$, with the spectral index having values close to $\gamma\simeq 3$. It  extends from somewhat below 1~GeV up to beyond 100~EeV, with the
differential  flux dropping by more than 30 orders of magnitude in this range. It is determined by direct measurements for energies below few hundred TeV while from studies of air showers  for higher energies. 
 The vast majority of the CRs come from outside the solar system,  although some particles produced in the Sun can be observed up to GeV energies in association with solar flares.
Below energies of $\sim 10$~GeV per nucleon (GeV/$n$)  the CR flux gets suppressed by the effect of the solar wind, which drags them away from the solar system. Since this wind is modulated by the solar activity, the observed low energy CR flux turns out to be time dependent.

\begin{figure}[tb]
\begin{center}
\begin{minipage}[t]{8 cm}
\centerline{\epsfig{file=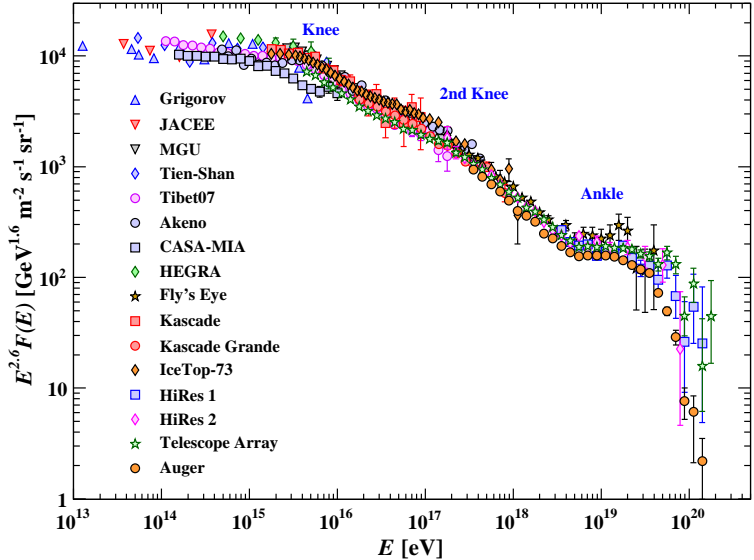,scale=0.6}}
\end{minipage}
\caption{Differential CR spectrum, multiplied by $E^{2.6}$ so as to make it flatter, as a function of energy (taken from \cite{pdg16}).\label{spectrum.fig}}
\end{center}
\end{figure}

When looking in detail to the spectrum at higher energies, see Fig.~\ref{spectrum.fig}, some spectral features become noticeable:
\begin{itemize}
\item Above few hundred GeV and up to few PeV the overall spectrum has a slope $\gamma\simeq 2.7$. 
\item At the so-called {\em knee}, at an energy of about 4~PeV, the total spectrum steepens to  $\gamma\simeq 3$.
\item At the so-called {\em second-knee} at $\sim 0.1$~EeV there is a further steepening to   $\gamma\simeq 3.3$.
\item At the so-called {\em ankle}, at about 5~EeV, the spectrum becomes harder again, with   $\gamma\simeq 2.6$.
\item At energies beyond 40~EeV a strong suppression becomes apparent, with a steepening to values of  $\gamma> 4$ taking place. At the highest energies, an exponential suppression may be present, or even a final cutoff may exist somewhere above 200~EeV, although the CR statistics becomes so small at these energies  that the picture is still unclear.
\end{itemize}
Besides these main features, there is also a hardening  at about 20~PeV, the so-called {\it low energy ankle} (\cite{leankle}), with the spectrum being actually a bit steeper, $\gamma\simeq 3.1$, from the knee up to this  feature and a bit harder,  $\gamma\simeq 2.9$, from this energy up to the second knee.

Note that the determination of the fluxes by different experiments sometimes differ significantly. This is often due to differences in the energy calibrations performed by experiments using different detection techniques or due to differences in the energy resolutions, which have a large impact because the spectrum falls very steeply.  

To understand the origin of the spectral features it is very important to be able to separate the CR fluxes according to their masses, since different nuclear species do show features at different energies.

 \subsection{Spectrum and  composition from low to high energies}
\begin{figure}[tb]
\begin{center}
\begin{minipage}[t]{8 cm}
\centerline{\epsfig{file=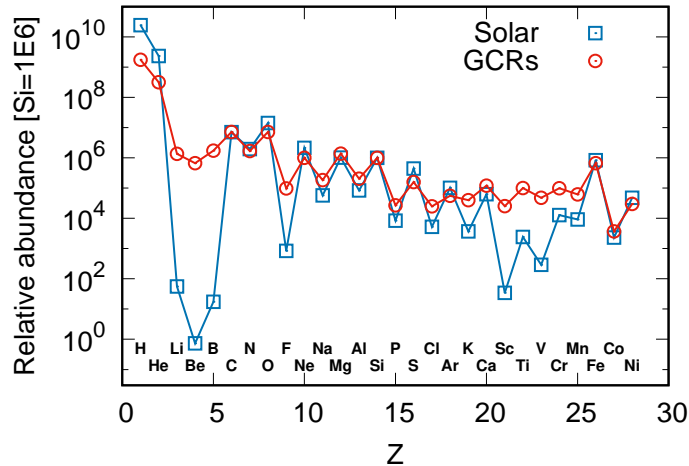,scale=1.5}}
\end{minipage}
\caption{Relative abundances of the elements from H to Ni in the solar system (from \cite{lo03}) and in low energy Galactic CRs (from  \cite{criscomp,smart}).  The abundances were normalized such that they are $10^6$ for Si.\label{compsolgcr}}
\end{center}
\end{figure}

The composition of the low energy Galactic CRs (GCRs) has many similarities with the element abundances measured in the solar system, as shown in Fig.~\ref{compsolgcr}. One may notice that nuclei with even $Z$ are generally more abundant than the neighboring odd $Z$ nuclei, which is due to the importance of He burning reactions in stellar nucleosynthesis. 
Some interesting differences appear however. The most common elements, H and He, are less abundant in GCRs than in the solar system, probably reflecting the composition of the medium in the acceleration region and also the relatively large ionization potential of these elements that may be an obstacle for their injection in the acceleration process. Another noticeable difference is that several elements that are relatively under-abundant in the solar system have much larger abundances in the GCRs. This is the consequence of the production of these elements as secondaries in the interactions of the primary CRs with the gas in the interstellar medium (ISM). For instance, Li, Be and B are produced in the spallation of the much more abundant C, N and O, in turn F is produced in the spallation of Ne and the elements from Sc to Mn in the spallation of Fe and Ni\footnote{Note that since the energy per nucleon is minimized for Fe nuclei, stellar nucleosynthesis does not produce significant amounts of nuclei heavier than Fe. Those elements are believed to be produced through rapid neutron capture processes in supernova explosions or neutron star mergers, and hence their relative abundances are quite suppressed.}. Since the spallation cross sections are known, one can estimate from the observed secondary abundances the total average amount of matter traversed by the primary CRs,   which turns out to be $\Sigma\simeq 5$\,g\,cm$^{-2}$. If one were to estimate the average density of the matter as that of the ISM, $n_H\simeq 1$\,cm$^{-3}$, this would imply that the typical residence time of the low energy CRs in the Galaxy should be $\tau\simeq \Sigma/(\beta c\, m_p n_H)$, with $\beta=v/c$ the particle velocity. In particular, for relativistic particles  this would lead to $\tau\simeq  3$~Myr. Note that the times inferred are much larger than the one that would be required for straight propagation from a Galactic source, since for instance the Galactic center lies just 0.03~Mlyr away. This shows that the propagation is indeed diffusive, with the charged CRs wandering around for very long times in the regular and turbulent Galactic magnetic fields.

There is an independent way to estimate the CR residence time by determining the relative abundances of some radioactive nuclei. Of particular interest is the abundance of $^{10}$Be, which has a half-life of 1.5~Myr. This isotope is produced in spallation of CNO nuclei in a proportion of about 20\% (the rest being $^9$Be and $^7$Be), but its observed isotopic abundance in CRs is only about 3\%, from which one can infer that the typical residence time is about 15~Myr  \cite{be10}. Since these measurements are performed for energies of about 0.1~GeV/$n$, i.e. for $\beta\simeq 0.4$, this would imply that the average density traversed is about $n_H\simeq 0.3$~cm$^{-3}$. This small value suggest that the CRs spend a considerable fraction of their time in low density regions  outside the disk of the Galaxy, in the so-called halo. 

Another very interesting result is the determination of the energy dependence of the secondary to primary CR abundances, due to the fact that for increasing energies one expects the confinement times to be smaller and hence the column densities traversed should also be correspondingly reduced. A particularly sensitive determination can be obtained from the B/C ratio, measured recently with great precision by AMS-02 \cite{amsb2c} for rigidities $R\equiv p/eZ$ in the GV--TV range\footnote{ For magnetized detectors it becomes actually convenient to refer to the particle rigidities instead of the energies. The rigidity is measured in Volts. In the relativistic case it is just proportional to the ratio $E/Z$, which is measured in eV. It is also convenient to use rigidity when processes of electromagnetic nature are relevant, such as for CR confinement or acceleration, while for studies of spallation it is convenient to use  energy per nucleon.}.  
The change in the B/C ratio from about 0.3 down to 0.05 in this rigidity range indicates that the energy dependence of the confinement time scales as $\tau_{\rm esc}\propto R^{-\delta}$, with $\delta=0.333\pm 0.014$ for rigidities larger than 20~GV. This  dependence is expected to arise from the rigidity dependence of the diffusion coefficient $D$ determining the confinement times in the Galaxy ($\tau_{\rm esc}\propto D^{-1}$). The result is consistent with the expectation that $\delta=1/3$ for the case in which the magnetic field turbulence has a Kolmogorov spectrum.

For CR energies larger than $\sim 200$~GeV/$n$ and up to few PeV the spectra of the main CR elements are well described by power laws  d$\Phi_i/{\rm d}E\propto f_i E^{-\gamma}$, with $\gamma\simeq 2.5$--2.7 and $f_i$ being the fractions  of the different elements $i$ at a given energy. 
Note also that
as the energy increases the production of secondary nuclei should get reduced, as happened with B/C, due to the  suppression of the spallation processes resulting from the reduced residence times. Hence, the relative abundance of these elements with respect to their primaries is expected to approach at PeV energies values closer to those observed in the solar system.  As a consequence,  different elements may have slightly different spectral indices, and in particular the spectrum of Fe is a bit harder than those of H and He since the large Fe spallation cross section suppresses its abundance at low energies.

The fact that the CR spectrum is approximately a power law clearly indicates that they get accelerated in non-thermal processes. It was Fermi \cite{fe49} who originally proposed that the acceleration could be the result of the encounter of the CRs with moving magnetized regions, such as molecular clouds. In the rest frame of the cloud, where there are no electric fields, the CR would enter from one direction, get isotropized by the scattering with the magnetic field and exit in another direction without changing its energy. However, when viewed from  the original frame where the magnetic cloud is moving, the CR will scatter off the cloud and  gain energy when it exits in the direction of the cloud motion while it will lose energy when exiting in an opposite direction. After many encounters with different magnetic clouds, with some energy gains and some losses, the CRs will eventually  acquire a non-thermal spectrum. However,  this original mechanism turned out to be quite inefficient, with the net energy gain per collision depending quadratically in the small cloud velocity (and hence was dubbed second-order Fermi acceleration).

It was later realized \cite{ax77,kr77,be78,bl78} that a much more advantageous situation could result in the neighborhood of shock waves through magnetized media. In this case, since the gas in the two regions (downstream and upstream of the shock) are in relative motion towards each other, when the CR crosses the shock it will have a face-on collision with the medium, it will get eventually isotropized and likely cross back the shock with a net gain of energy. This energy gain turns out to be linear in the shock velocity (which moreover in environments like supernova explosions can be of order $10^3$--$10^4$~km\,s$^{-1}$,  much larger than the typical velocities of Galactic molecular clouds). This mechanism has then been dubbed first-order Fermi acceleration, or diffusive shock acceleration (DSA). Since the shock  crossing can take place many times, the energy gain can be quite significant, and  one can also show that the resulting energy spectrum is a power law with index $\gamma_{\rm s}\simeq 2$--2.3.
If the CR sources have then a spectrum $\sim R^{-\gamma_{\rm s}}$ and the diffusion in the Galaxy, via a rigidity dependent escape time $\tau_{\rm esc}\propto R^{-\delta}$, further steepens the CR spectrum, it is expected that the final spectrum observed at Earth should scale as $\sim R^{-\gamma}$ with $\gamma=\gamma_{\rm s}+\delta$. Given that the observed spectral index below the knee is $\gamma\simeq 2.6$--2.7 and that $\delta\simeq 1/3$, one infers then that $\gamma_{\rm s}\simeq 2.3$, compatible with the expectations from the DSA.
One could also mention that if  different sources accelerate CRs up to different maximum rigidities, their superposition can  lead to an effective steepening in the overall source spectral index (see e.g. \cite{pt10}). It is only the overall combined  spectrum from all sources that should have, after propagation through the Galaxy, an effective spectral index of $\sim 2.6$--2.7 at the Earth location.  

Regarding the local CR density, it is related to the CR flux by $n_{\rm CR}=(4\pi/v)\Phi$. A dominant fraction of the GCR energy density is at energies between 100~MeV--10~GeV, where the effects of the solar wind are relevant, making the measured fluxes significantly smaller than the ones outside the solar system. One can correct for these effects and evaluate the CR kinetic  energy density as 
\be
\rho_E=\frac{4\pi}{c}\int {\rm d}E\, \frac{E}{\beta}\frac{{\rm d}\Phi}{{\rm d}E}.
\ee
Adding up the contributions from all elements one obtains $\rho_E\simeq 0.5\,{\rm eV}\,{\rm cm^{-3}}$ \cite{gaisser}\footnote{A rough estimate of the GCR energy density can be obtained integrating the energy flux using an  extrapolation of the power-law flux determined above 20~GeV/$n$ down to about 1~GeV/$n$, where the spectrum reaches its maximum, and the result is indeed comparable to the one obtained with the more accurate evaluation.}.
One can then make an estimate of the total energy in CRs in the Galaxy considering as its typical volume  $V_{\rm G}\simeq \pi R_{\rm G}^2 h_{\rm G}$, i.e. considering the CR confinement region as a cylinder of radius $R_{\rm G}\simeq 15$~kpc and height $h_{\rm G}\simeq 0.4$~kpc. This would lead to a total energy $E_{\rm G}\simeq \rho_E V_{\rm G}\simeq 6\times 10^{54}$~erg. If we assume that we are in a steady state and consider a typical CR residence time of 10~Myr, this would imply a CR production rate, or total source luminosity, of
\be
L_{\rm G}\simeq \frac{E_{\rm G}}{\tau_{\rm esc}}\simeq  2\times 10^{40}\,\frac{\rm erg}{\rm s}.
\ee
It was already pointed out by Ginzburg and Syrovatskii \cite{gi64} that the required energy could be provided by Galactic supernovae. For instance, each core collapse SN explosion emits about $3\times 10^{53}$~erg in a burst of neutrinos lasting a  few seconds, about  one foe ends up  as kinetic energy of the ejecta (where foe stands for ten to the fifty one erg), and only $\sim 10^{49}$~erg  end up as light emitted during the several months over which the SN appears as a very bright object in the sky, as bright as a whole galaxy\footnote{The less frequent type IIb SNe may produce about 3 foe of kinetic energy and accelerate CRs to larger maximum energies than the more frequent type IIP or Ib/c \cite{pt10}.}. Also  type~Ia SN, which do not leave a compact remnant nor emit significant amount of neutrinos, produce ejecta carrying  about one foe. The explosions give rise to powerful shocks propagating through the envelope and across the environment, which can last for $\sim 10^4$~yr and in which CRs can be efficiently accelerated\footnote{There is an initial period of free expansion of the shock, lasting about 200~yr and in which the shock velocity is $10^4$~km\,s$^{-1}$. Then takes place the Sedov phase, starting when the swept up mass becomes comparable to the mass of the ejecta and lasting for about $10^4$~yr. In this phase, during  which most CRs would be accelerated, the shock velocity decreases progressively as $t^{-1}$ down to values of  few hundred km\,s$^{-1}$.}. Since there are about three SN per century in the Galaxy, to account for the estimated CR luminosity one would need that about few percent of the kinetic energy of the ejecta be converted  into the kinetic energy of the accelerated  CRs, something which according to detailed simulations of SN explosions is quite feasible. This is one of the reasons why it is believed that the majority of the Galactic CRs could be produced in SN explosions. Although the initial estimates of the maximum CR energies achievable in  a SN environment were quite low, of order $R_{\rm max}\simeq 10^{13}$--$10^{14}$~V, once one takes into account that the shock can propagate into the wind of the progenitor star and  that the magnetic fields can be amplified to values much larger than those in the ISM, proton energies in excess of few PeV may be obtained (see for instance \cite{pt10}).  Evidence for acceleration of hadronic CRs in SN remnants up to TeV energies has been obtained in recent years from the observations of gamma rays with a spectrum consistent with that expected from $\pi^0$ decays \cite{snr}.
On the other hand,  the super-massive black hole in the Galactic center seems to be accelerating CRs up to PeV energies \cite{pevatron}. Another type of source where CRs may be accelerated  are the pulsars left as compact remnants of core collapse supernova explosions. These rapidly rotating neutron stars, with magnetic fields amplified to huge values during the core collapse, induce strong electric potentials that can accelerate the charged particles to very large energies, even beyond the EeV in some extreme cases, although the energy losses are an issue.

Taking into account that the acceleration process as well as the diffusive propagation are, due to their electromagnetic nature, rigidity dependent, and assuming that the different elements become completely ionized so that they get accelerated in proportion to their original abundances in the medium $f_i^0$, one can relate these abundances to the relative flux normalizations $f_i$ measured at a given energy (neglecting the possible effects of spallation processes). In particular, if the spectrum of the different elements $i$ are d$\Phi_i/{\rm d}E\propto f_i E^{-\gamma}$,  one should have that 
\be 
\frac{{\rm d}\Phi_i}{{\rm d\,log}E}(Z_iE)\simeq \frac{f^0_i}{f^0_H}\frac{{\rm d}\Phi_H}{{\rm d\,log}E}(E).
\ee
This would imply then that $f_i\simeq (f^0_i/f^0_H)Z_i^{\gamma-1}$. Hence, the fractions of the heavy elements present at a given CR energy are expected to be  much larger than the overall abundances of those elements in the GCRs (or in the solar system). 
In particular, while the overall fractions in the GCR of H:He:CNO:Fe amount to about $83.6:15.2:0.75:0.03$\%, the contributions of these elements to the fluxes at $\sim$TeV energies are in the approximate proportion $42:26:10:7$\% \cite{wi98}. These fractions are compatible with the relation derived above, except for a possible depletion of Fe in the low energy GCRs due to spallation. 

\begin{figure}[tb]
\begin{center}
\begin{minipage}[t]{8 cm}
\centerline{\epsfig{file=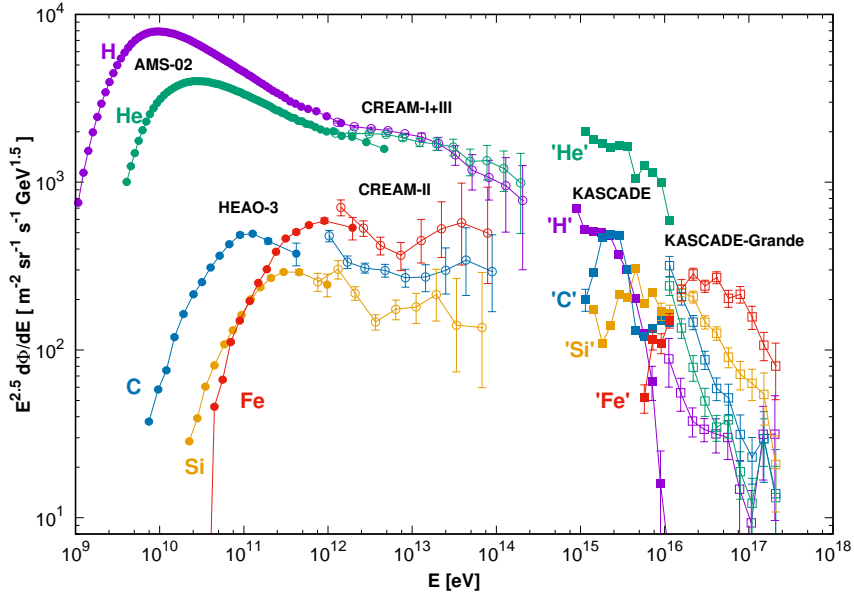,scale=1}}
\end{minipage}
\caption{Spectra of different components determined by AMS-02, HEAO-3, CREAM and the air-shower arrays KASCADE and KASCADE-Grande (see text for a detailed description and references).\label{spectrumle}}
\end{center}
\end{figure}

Fig.~\ref{spectrumle} shows some determinations of the fluxes of separate CR components as a function of the energy (the differential spectra are multiplied by $E^{2.5}$ to make them flatter). The different data correspond to the H and He fluxes determined by AMS-02 in the International Space Station \cite{amsh,amshe}, the fluxes of C, Si and Fe determined by the HEAO-3 satellite \cite{heao}, the fluxes of H and He determined by the balloon flights CREAM-I and CREAM-III \cite{cream3} and those of C, Si and Fe determined by CREAM-II \cite{cream2}. At energies beyond the PeV the measurements are from air showers, and hence do not determine directly the separate elements but rather a fit is performed to the observed electron and muon densities at the ground using a five component model (with representative elements `H', `He', `C', `Si' and `Fe'). The abundances determined in this way are then actually representative of mass groups (for instance, `C' could actually be closer to the sum of the C, N and O actual abundances). Moreover, the abundances inferred depend on the hadronic model adopted in the simulations performed to obtain the ground level electron and muon density expectations from each component. The data points for the KASCADE experiment \cite{ka08},  shown for $E<10^{16.1}$~eV,  were obtained with the models QGSJET-01 (and Fluka for air-shower particles with energies below 200~GeV), while those of KASCADE-Grande adopted QGSJET-II-02/Fluka \cite{ap13}.

From the figure one can appreciate the strong suppression of the fluxes at energies below $\sim 10$~GeV/$n$, something which is exacerbated by the effects of the solar wind. There is a hardening of the spectra above $\sim 200$~GeV/$n$, where they approach power laws $E^{-\gamma}$, with $\gamma\simeq 2.5$--2.7, dependent slightly on the element considered. For instance, at PeV energies the He component is already more abundant than the H one. The `H' spectrum becomes steeper at $\sim 4$~PeV, which coincides with the energy of the knee, and the suppression of the heavier elements is compatible with being due to a rigidity dependent effect, so that they become steeper at energies $\sim 4Z$~PeV. We would also like to point out here that the low energy ankle, at about 20~PeV, could be due to the cross-over between the steeply falling He and CNO fluxes and the harder Fe and Si fluxes. The second knee, at energies $\sim 0.1$~EeV, is interpreted as the steepening of the Fe flux,  at an energy $\sim 26\times 4$~PeV \cite{caanti}. Since the spectrum beyond the second knee falls as $E^{-3.3}$, the individual spectra should fall at least that steeply beyond their respective knees if the effect is rigidity dependent. The KASCADE-Grande experiment has observed the steepening of the Fe component at around 0.1~EeV, supporting the interpretation of the second-knee mentioned  above, and moreover they found a hardening of the light component at an energy close to this value  \cite{ap13}. As this light component, presumably of extragalactic origin, becomes increasingly important for increasing energies above 0.1~EeV, the spectrum of the heavy Galactic component may actually become steeper than  $E^{-3.3}$, eventually even having a cutoff in the EeV range associated to the maximum energies achievable at the sources. 

The actual energy at which the transition between the Galactic and the extragalactic CR components takes place is still a matter of debate. Until recently the ankle looked to be the natural place for this to happen, since the hardening of the spectrum at about 5~EeV could be associated to the emergence of the extragalactic flux above the steeply falling Galactic flux. However, the fact that at EeV energies the values of $\langle X_{\rm max}\rangle$ approach the predictions of protons (see Fig.~\ref{xmaxvse}), combined with the lack of observed anisotropies in the direction of the Galactic center or Galactic plane \cite{LS12}, suggest that this emerging light component above 0.1~EeV is not of Galactic origin, so that the transition could be taking place somewhere between the second-knee and the ankle.

The fact that the knee in the spectrum could be due to a rigidity dependent effect leading to a steepening in the overall spectrum and an associated increase in the average CR mass, was first suggested by Peters \cite{pe61}. However, the actual cause of the steepening is not yet known. It could be the consequence of a limit in the maximum energy achievable at the sources, as may be the case in acceleration in SN, or could result from a change in the energy dependence of the confinement times inside the Galaxy. In particular, at the energy of the knee, drift effects start to dominate the CR transport over the usual turbulent diffusion, changing the behavior $D(E)\propto E^\delta$ from the value $\delta\simeq 1/3$ associated to the Kolmogorov turbulence to a value $\delta\simeq 1$ associated to the antisymmetric part of the diffusion tensor\footnote{In the presence of both turbulent and regular magnetic fields, the diffusion is anisotropic, with different coefficients in the direction along ($D_\parallel$) and orthogonal ($D_\perp$) to the regular field. Moreover, drift motions lead to a macroscopic current ${\bf J}_A=D_A {\bf b}\times  \nabla n$ (with ${\bf b}={\bf B}/|{\bf B}|$), which then gives an antisymmetric contribution to the off-diagonal diffusion tensor defined through $ J_i\equiv - D_{ij}\,  \nabla_j n$. The antisymmetric diffusion component scales as the Larmor radius, so that $D_A\propto E$.} \cite{ptanti}. This mechanism would allow to account for the knee as well as for the second-knee and the spectrum slightly above it \cite{caknee,caanti}, without requiring the source spectrum to be suppressed. Since drift motions are expected to enhance the CR escape from the Galaxy, the study of the anisotropies could give a handle to discriminate among the different proposed explanations for the knee.

\subsection{Spectrum and composition at ultra-high energies}

\begin{figure}[tb]
\begin{center}
\begin{minipage}[t]{8 cm}
\centerline{\epsfig{file=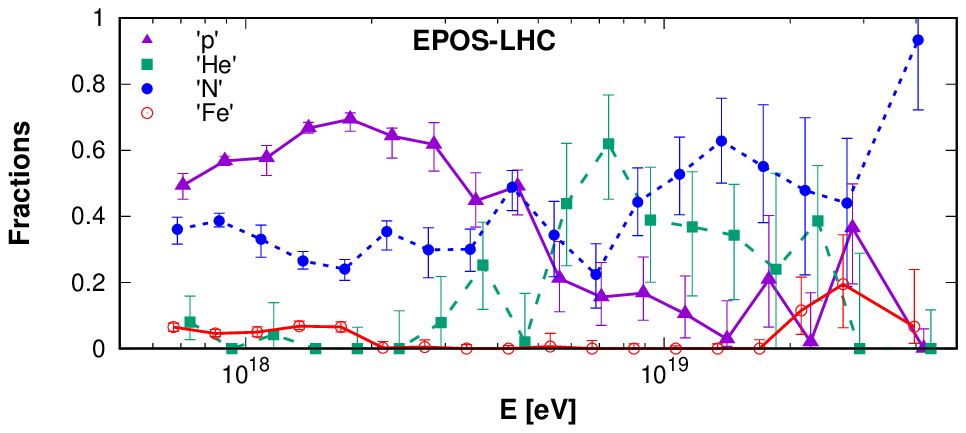,scale=0.9}\epsfig{file=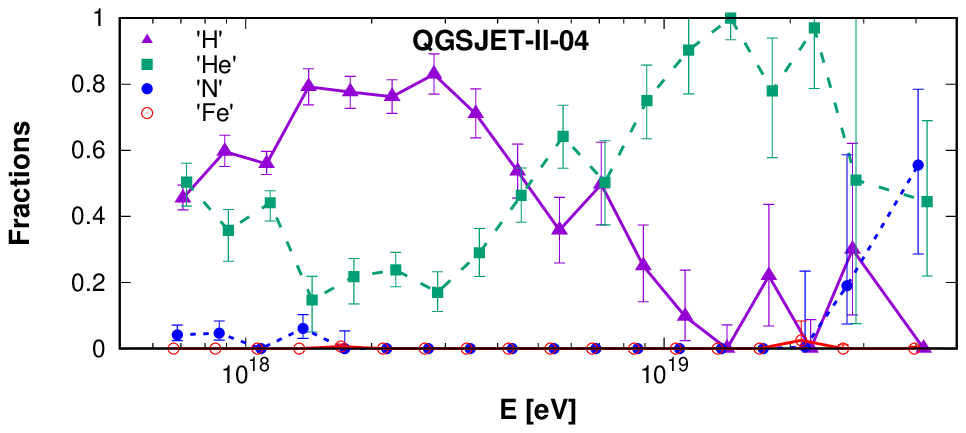,scale=0.9}}
\end{minipage}
\caption{Reconstructed fractions of different elements as a function of energy obtained by the Pierre Auger Observatory from a fit to the $X_{\rm max}$ distributions \cite{augerxmax}. Results are shown adopting the EPOS-LHC hadronic model (left panel) or the QGSJET-II-04 model (right panel). The statistical uncertainties are indicated.\label{augercomp.fig}}
\end{center}
\end{figure}

The CR spectrum at ultra-high energies presents two main features, the hardening at the ankle at about 5~EeV and the suppression starting above $\sim 40$~EeV (see Fig.~\ref{spectrum.fig}). In Fig.~\ref{augercomp.fig} we show the reconstructed fractions of four representative elements (`H', `He', `N' and `Fe') obtained by the Pierre Auger Observatory \cite{sein} by fitting the distributions  of $X_{\rm max}$ values observed in different energy bins. Although the details depend significantly  on the hadronic model adopted, as illustrated for the models EPOS-LHC (left panel) and QGSJET-II-04 (right panel),  some general features are common. In particular, the heavy component that was present at $\sim 0.1$~EeV becomes strongly suppressed and at 1--5~ EeV energies a dominant fraction is due to light elements (`H' and `He'). Above the ankle energy  a strong suppression of the proton component is observed and the composition  becomes heavier,  although the details of the mass distribution (whether it becomes dominated by `N' above 10~EeV, as for EPOS-LHC, or if large amounts of `He' survive up to higher energies, as for QGSJET),  depends on the hadronic model considered to interpret the data. 

These trends towards a heavier composition are clearly observed in the flattening of the elongation rate appearing above 2~EeV in the Auger observations depicted in Fig.~\ref{xmaxvse} ($D_{10}$ changes from values of about 80~g\,cm$^{-2}$ in the energy decade below 2~EeV to about 25~g\,cm$^{-2}$ in the energy decade above 2~EeV). Another important constraint comes from the decrease of Var$(X_{\rm max})$ observed for increasing energies \cite{augerxmax}, which also suggests a transition towards a heavier composition and requires that little mixing between different mass components be present above $\sim 10$~EeV.  Given the  small statistics achieved by FD measurements above $\sim 30$~EeV it is still not clear whether at the highest energies a small Fe component is present or even if a small contribution from a hard proton component could appear. Other results on $\langle X_{\rm max}\rangle$ in this energy range have been obtained by the Telescope Array and HiRes experiments (see Fig.~\ref{xmaxvse}), and having larger statistical uncertainties they are compatible both with the results from Auger as well as with a light-composition scenario extending up to the highest energies. Results from Yakutsk and Fly's Eye also exist in this energy range but  have less statistical significance.

Turning now to discuss the proposed explanations for the high-energy suppression, an effect like this one was originally predicted by Greisen \cite{gr66} and Zatsepin and Kuzmin \cite{za66} soon after the discovery of the cosmic microwave background  (CMB) and well before the observation of this feature in recent years \cite{hiresgzk,augergzk}.  This so-called GZK effect arises because sufficiently energetic CR protons propagating through the omnipresent 2.7\,K CMB background, with density $n_{\rm CMB}\simeq 400$~cm$^{-3}$ and average energies $\langle\varepsilon\rangle=3kT\simeq 7\times 10^{-4}$~eV, may interact with these photons to produce pions. The threshold proton energy for a process like $p+\gamma\to p+\pi^0$ (or $p+\gamma\to n+\pi^{+}$) to occur is obtained when the center of mass energy $\sqrt{s}$ is just enough to produce the pion at rest, i.e. for
\be
s_{\rm th}=(m_p+m_\pi)^2=(m_p+\varepsilon'_{\rm th})^2-{\varepsilon'}_{\rm th}^2=m_p^2+2m_p\varepsilon'_{\rm th}.
\ee
This relation also constrains the photon energy in the rest frame of the incident proton, $\varepsilon'$, to be sufficiently large so as to allow for the 
production of the massive pion. Since $\varepsilon'\simeq \gamma\varepsilon(1-\cos\theta)$, with $\theta$ the angle between the photon and the proton momenta in the laboratory frame (i.e. the CMB rest frame), one has that the most favorable situation is for a head-on collision for which $\varepsilon'\simeq 2\gamma\varepsilon$. Since $\gamma=E/m_p$ one has then
\be 
E^{\gamma\pi}_{\rm th}\simeq \frac{m_\pi m_p}{2\varepsilon}\simeq 70\,{\rm EeV}\left(\frac{10^{-3}\,{\rm eV}}{\varepsilon}\right).
\ee
Given the fact that the CMB photons have a Planckian distribution, also protons with energies down to $\sim 50$~EeV may have significant interactions with the high energy tail of the photon distribution to produce pions and in this way have their fluxes attenuated. 

\begin{figure}[tb]
\begin{center}
\begin{minipage}[t]{8 cm}
\centerline{\epsfig{file=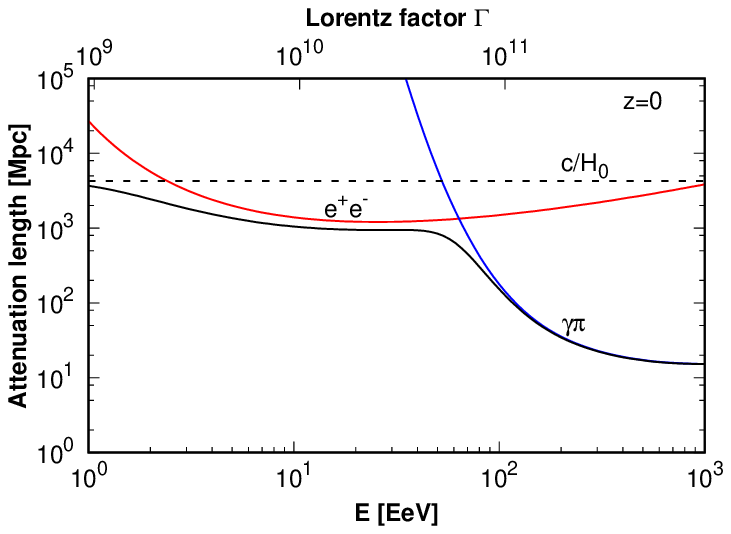,scale=1.2}\epsfig{file=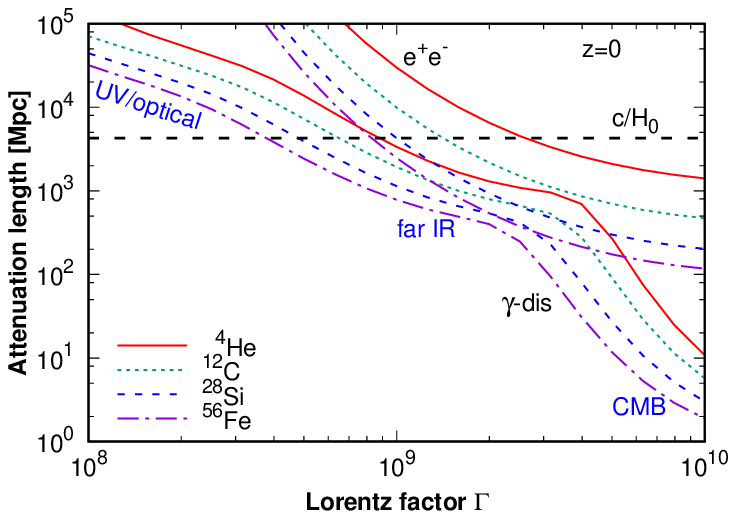,scale=1.1}}
\end{minipage}
\caption{Energy loss length for protons (left panel, as a function of energy) and heavier nuclei (right panel, as a function of the Lorentz factor), computed for redshift $z=0$ (from \cite{hmr15}). The separate contributions from pair production ($e^+e^-$) and from photo-disintegration ($\gamma$-dis), as well as the dominant photon background responsible for these lasts, are indicated. \label{eloss.fig}}
\end{center}
\end{figure}

Just above threshold the photo-pion cross section has a resonant behavior due to the production of $\Delta$ baryons, reaching values of $\sim 0.5$~mb, and then decreases down to $\sigma_{\gamma \pi}\sim 0.2$--0.3~mb. If one makes the rough estimate $n_\gamma \sigma_{\gamma \pi}\simeq 200$~cm$^{-3}\times  0.3$~mb (i.e. considering just half of the photons to be above threshold since they may be in the low energy tail or travel in a similar direction as the proton), one can estimate the interaction  mean free path as $\lambda=1/(n_\gamma\sigma_{\gamma\pi})\simeq 5$~Mpc. The most relevant quantity to study the attenuation of the CR proton flux is however the energy loss length 
\be 
\lambda_E\equiv -\frac{1}{E}\frac{{\rm d}E}{{\rm d}x},
\ee
which indicates the distance over which the energy would fall, if the losses were to be constant, to 1/$e$ of its initial value. Since the inelasticity of each photo-pion interaction, i.e. the fraction of the initial proton energy transferred to the pion, is about 20--30\%, this means that one would typically need about three interactions for this energy reduction to take place, and hence one expects $\lambda_E\simeq 15$~Mpc  when the proton is well above threshold. In Fig.~\ref{eloss.fig} (left panel) we show the results of a detailed computation of this energy loss length as a function of energy and one can see how it steeply increases with respect to this asymptotic value as the threshold energy is approached from above. 

Also shown in Fig.~\ref{eloss.fig} is the contribution to the proton energy losses due to the pair production process $p\gamma\to p\,e^{+}e^{-}$, which has a significantly smaller inelasticity, since close to the threshold only a fraction $\sim 2m_e/m_p\simeq 10^{-3}$ of the initial energy is lost in each interaction. The threshold for this process is
\be 
E^{ee}_{\rm th}\simeq \frac{m_e m_p}{\varepsilon}\simeq 0.5\,{\rm EeV}\left(\frac{10^{-3}\,{\rm eV}}{\varepsilon}\right),
\ee
and hence the effects of pair production extend down to much lower energies.
Also shown as a horizontal line is the loss length associated to the Universe expansion (redshift losses) given by $c/H_0\simeq 4$~Gpc, with $H_0\simeq 70$~km\,s$^{-1}$\,Mpc$^{-1}$ being the Hubble parameter. The redshift losses are the dominant attenuation for protons below 2~EeV while above this energy and up to about 50~EeV it is the pair production losses that dominate, and above this energy the dominant effect is that of the photo-pion production. 

\begin{figure}[tb]
\begin{center}
\begin{minipage}[t]{8 cm}
\centerline{\epsfig{file=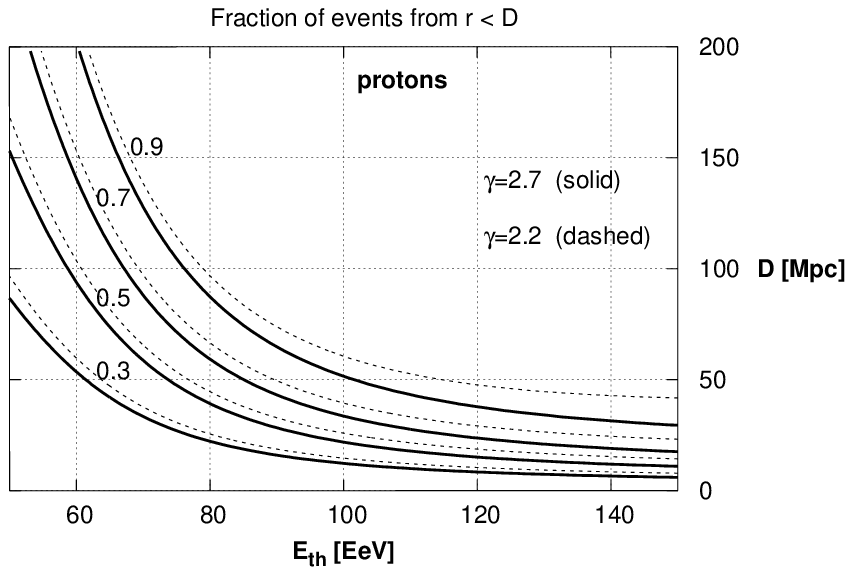,scale=1}\epsfig{file=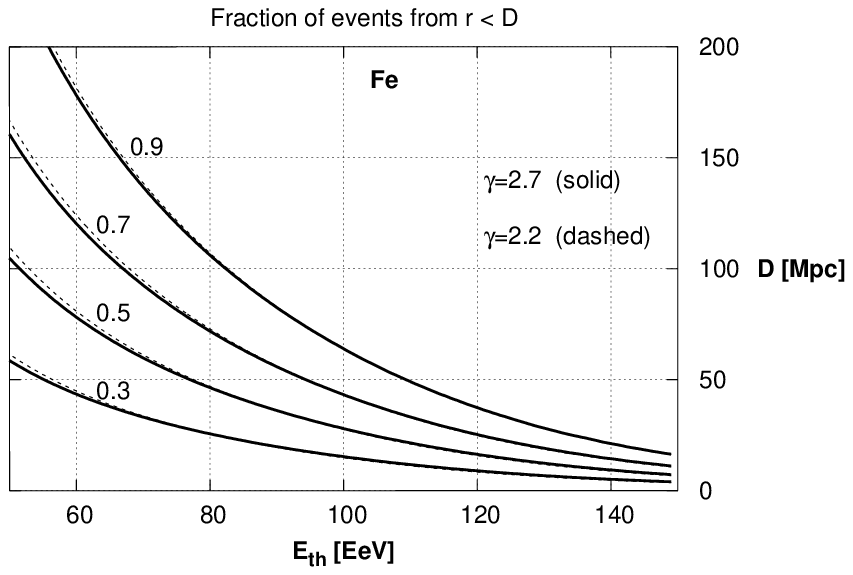,scale=1}}
\end{minipage}
\caption{Fraction of the flux observed above a threshold energy $E_{\rm th}$ originating  from sources within a distance $D$. Sources are assumed uniformly distributed and with power law spectra ($\gamma=2.2$ and 2.7  are shown). The curve labeled as 0.9 would correspond the the GZK horizon defined as the distance within which 90\% of the flux observed above a given threshold  was produced (adapted from \cite{horizon}).  Left panel is for proton sources and right panel for Fe sources.\label{gzkatt.fig}}
\end{center}
\end{figure}

The most important prediction associated to the GZK effect is that one would not expect to observe CR protons from very far away sources at energies close to or above $E_{\rm th}^{\gamma\pi}$. A GZK horizon may be defined as the distance within which 90\% of the CR flux reaching Earth above a given energy threshold was produced, assuming uniformly distributed sources of equal luminosities and having power law spectra \cite{horizon}. In Fig.~\ref{gzkatt.fig} we show (left panel) the resulting flux fractions as a function of the threshold energies, with the one labeled 0.9 corresponding to the GZK horizon defined above. The horizon distance is 200~Mpc for $E_{\rm th}=60$~EeV while it is 90~Mpc for $E_{\rm th}=80$~EeV. It is clear then that setting a sufficiently high threshold implies that only the nearby sources will contribute significantly to the fluxes, and this can be quite helpful in the searches for anisotropies. 

The other important feature in the proton energy-loss curve shown in Fig.~\ref{eloss.fig} is due to the $e^+e^-$ attenuation effects. It is seen that for energies between 5 and 50~EeV the associated attenuation length is about 1~Gpc, and increases considerably below 5~EeV, becoming larger than the redshift one below 2~EeV. Due to the pair-production process,  CR protons from sources at distances larger than $\sim 1$~Gpc would have their energies significantly degraded as they propagate, enhancing the spectrum below 5~EeV by the time they arrive to Earth. This effect has been indeed proposed as an explanation for the ankle feature \cite{dip}. For this so-called dip-model to work it would be required that the sources predominantly emit protons up to energies well beyond that of the ankle, something which does not seem to be the case. To explain in this scenario the abrupt change in slope observed in the spectrum at the ankle energy, steep spectra at the sources, with $\gamma\sim 2.3$--2.7, are required. Also a strong cosmological evolution of the sources is preferred so as to enhance the contribution arising from high redshifts and in this way get a larger enhancement of the flux below the ankle. An attractive feature of the dip-model is that it predicts a significant contribution of extragalactic protons just below the ankle energy, as is indicated by observations.

In the alternative scenarios in which the UHECRs are heavier nuclei  with mass number $A$, the effects of interactions during propagation are different since, at a given energy, the Lorentz factor scales as $\Gamma=E/Am_p$. The threshold energies for photo-pion and pair production processes are then correspondingly increased by a factor $A$. The photo-pion production with CMB photons is then only relevant, in the case of heavy nuclei, for CR energies well beyond the maximum observed ones. Regarding the pair-production process, since the associated Bethe-Heitler cross section scales as $Z^2$ the corresponding attenuation lengths become smaller, even if the inelasticities scale as $A^{-1}$. The main new ingredient is however  that the dominant attenuation effect is due to the nuclear photo-disintegration, i.e. the fact that nuclei excited by the background photons will emit nucleons (and eventually alpha particles) and in this way reduce their mass, keeping their Lorentz factor essentially constant. The associated attenuation length is then 
\be 
\lambda_E^{\gamma\hbox{-}{\rm dis}}=-\frac{1}{E}\left.\frac{{\rm d}E}{{\rm d}x}\right|_{\gamma\hbox{-}{\rm dis}}=-\frac{1}{A}\frac{{\rm d}A}{{\rm d}x}.
\ee
 The classic paper where these processes were studied in detail is that of Puget, Stecker and Bredekamp \cite{psb}. An important contribution to the photo-disintegration comes from the giant dipole resonance (GDR), which is essentially a collective nuclear oscillation mode of the protons against the neutrons. The GDR has a threshold photon energy, in the rest frame of the nucleus, of about 10~MeV (the value depends on the nucleus considered). The threshold energy is then
\be 
E^{\gamma\hbox{-}{\rm dis}}_{\rm th}\simeq  5\,{\rm EeV}\left(\frac{10^{-3}\,{\rm eV}}{\varepsilon}\right)A,
\label{ethpdis}
\ee
corresponding to Lorentz factors of about $\Gamma\simeq 5\times 10^{9}$ when the interactions are with the CMB photons.

The photo-disintegration cross section is much larger than the photo--pion one and hence the attenuation lengths of heavy nuclei become actually smaller than the proton one, being of just a few Mpc when the CMB photons can excite the GDR (Lorentz factors $\Gamma\simeq 10^{10}$). Another important aspect is that, being that these cross sections are large, besides the CMB also the extragalactic background light (EBL) becomes relevant for the attenuation of heavy nuclei at energies below the threshold for interactions with CMB photons. This EBL radiation consists mostly of two contributions, one for UV/optical wavelengths due to the emission from stars, and another at far IR wavelengths due to reemission from dust. The cosmological evolution of the EBL is not as straightforward as the CMB one, which is just due to redshift effects, and to obtain it the evolution of the different source populations has to be modeled. Given the fact that the energies of the EBL photons ($10^{-3}$--1~eV) are larger than the ones of the CMB, the associated threshold energies for the GDR are smaller in this case, see eq.~(\ref{ethpdis}). The resulting attenuation lengths are shown in the right panel of Fig.~\ref{eloss.fig}, where the main features just discussed are illustrated.  In Fig.~\ref{gzkatt.fig} (right panel) we show the corresponding GZK horizons for the case of Fe nuclei, which turn out to be  not very different from the ones of protons shown in the left panel although the underlying physical processes are quite different in the two cases. For lighter nuclei the suppression would be shifted to lower energies. 

\begin{figure}[tb]
\begin{center}
\begin{minipage}[t]{8 cm}
\centerline{\epsfig{file=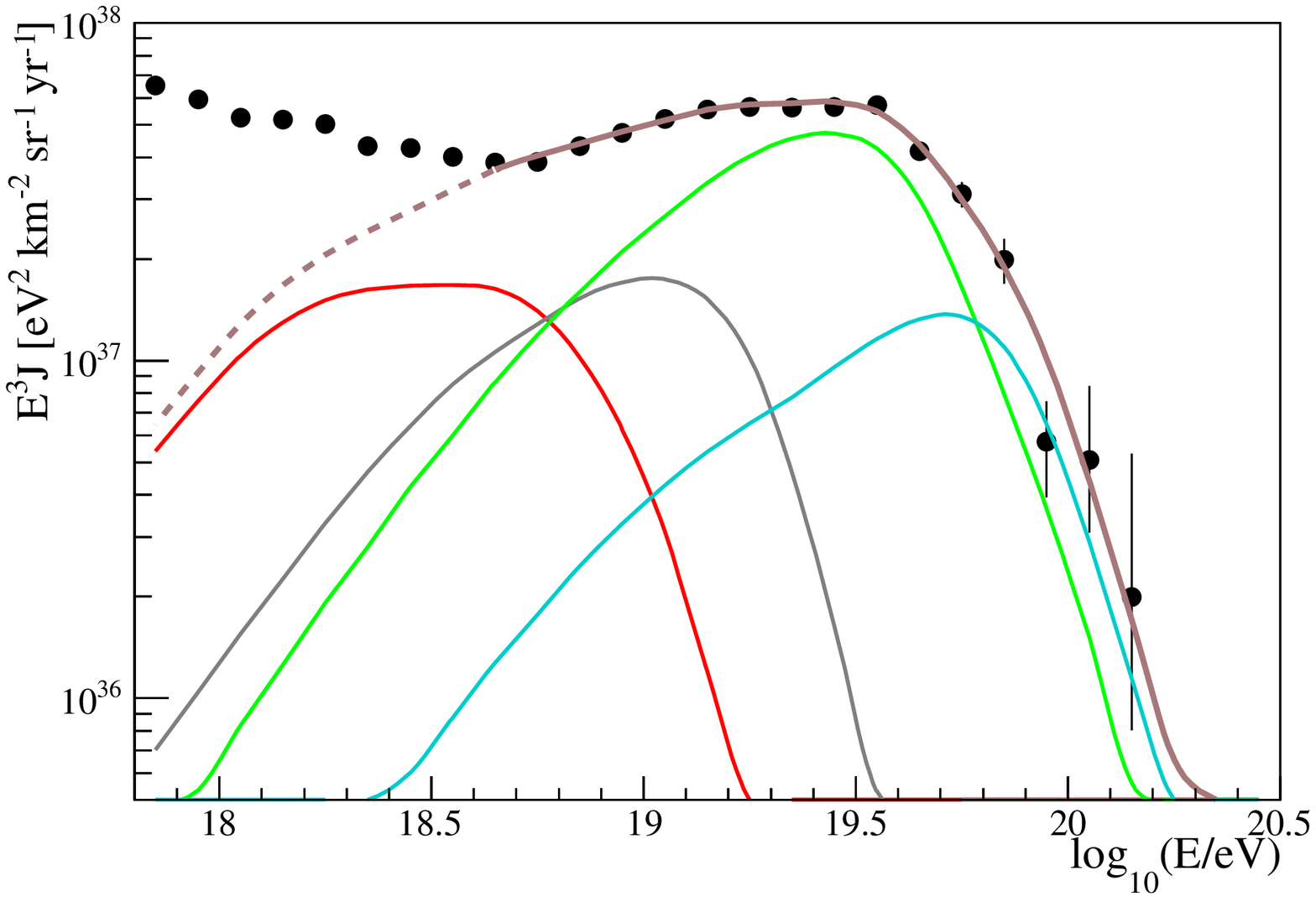,scale=.4}}\end{minipage}
\begin{minipage}[t]{16 cm}
\centerline{\epsfig{file=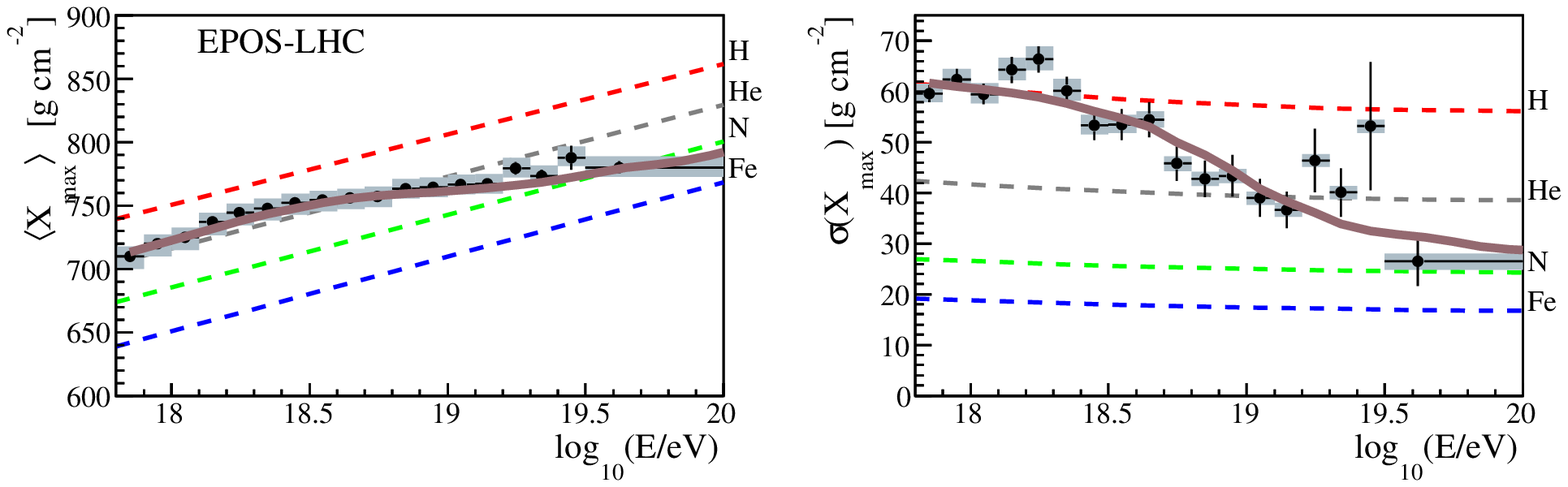,scale=.7}}
\end{minipage}
\caption{Scenario fitting the spectrum and $X_{\rm max}$ distributions measured by  Auger and interpreted according to the EPOS-LHC hadronic model (from \cite{combfit}). Top panel: the spectra of the different mass groups arriving to Earth: $A = 1$ (red), $2 \leq A \leq 4$ (grey), $5\leq A \leq 22$ (green), $23\leq A \leq 38$ (cyan). Lower panels: $X_{\rm max}$ and Var$(X_{\rm max}$) measurements and model fit. \label{combfit.fig}}
\end{center}
\end{figure}

Yet another scenario to account for the observed suppression in the spectrum would be that it be just related to the maximum rigidity attainable in the acceleration process at the sources. This is indeed a likely possibility  since it requires very extreme conditions to accelerate particles to the highest CR energies observed. This is also supported by the observation that the proton fraction appears to become suppressed above 5--10~EeV.  
A recent analysis by the Auger Collaboration \cite{combfit} has fitted simultaneously the spectrum and the $X_{\rm max}$ distributions observed above 5~EeV with a model in which the sources  emit certain fractions  of different nuclei (H, He, N, Si and Fe) with power law spectra having a rigidity cutoff $R_{\rm c}$. The nuclei are propagated from the sources to the Earth accounting for the interactions with the radiation backgrounds just discussed and the mass distribution of the arriving nuclei is obtained. The source spectral index, cutoff rigidity and elemental fractions are fitted to reproduce the observations. The results depend significantly on the hadronic model adopted to interpret the $X_{\rm max}$ results, as well as on the details of the EBL model and the cosmological source evolution considered. We show in Fig.~\ref{combfit.fig} the results obtained adopting the EPOS-LHC model, where the spectra of the different mass groups reaching Earth are depicted in the top panel. The best fitting parameters are $\gamma=0.96^{+0.08}_{-0.13}$ and log$(R_{\rm c}/{\rm V})=18.68^{+0.02}_{-0.04}$. The resulting values of $\langle X_{\rm max}\rangle$ and Var$(X_{\rm max}$)  are shown in the bottom panels, showing a quite reasonable agreement with the observed values. One can see that the inferred source spectral index is significantly smaller than the one resulting from Fermi acceleration ($\gamma\simeq 2$).  This comes mostly from the small values of Var($X_{\rm max})$ determined above 10~EeV (see bottom right panel of Fig.~\ref{combfit.fig}), which require that the simultaneous presence at a given energy of elements with significantly different masses be avoided. This can be achieved if the hard spectral shapes make the heavy components abruptly emerge as the lighter ones get suppressed above their spectral cutoff. The best fit elemental fractions obtained in this particular scenario, defined at a fixed energy below the H cutoff,  are H:He:N:Si:Fe${}\simeq 0:67.3:28.1:4.6:0$\%. One sees that a significant amount of N (or actually the CNO component) is required and that, at the highest energies, when this component gets attenuated by propagation effects as well as by its source cutoff, some heavier elements such as Si should be present. The cutoff rigidity is quite small,  about 5~EV, and hence the protons from the sources are really not much tested  because they would appear only below the energy range fitted (and hence the value $f_H\simeq 0$ has to be taken with a grain of salt). 
We note that the results for the QGSJET-II-04 model, from which one infers instead a lighter composition and a larger cutoff rigidity, requires very small values of the spectral index, $\gamma<-1$, and leads to poor fits to the observed $X_{\rm max}$ distributions \cite{combfit}.

The multicomponent fits to the spectrum and composition seem to require, besides low values of $R_{\rm c}$, hard spectral indices, $\gamma<2$, at variance with the expectations from diffusive shock acceleration. The observations may actually be explained, instead of resulting from hard spectral indices at the sources, as being due to a hardening of the spectrum as a result of the diffusion of the CRs in extragalactic magnetic fields \cite{mo13}. Indeed, since the diffusion time from a given source, $t\propto D^{-1}$, can become very large at low energies, eventually exceeding the age of the Universe even for CRs coming from the closest extragalactic sources, this could suppress the extragalactic component of the spectrum at low energies. As a result, the spectrum at the Earth would appear harder even if the spectrum at the sources was compatible with $\gamma\simeq 2$.  For this scenario to be relevant to account for the features observed at UHE the extragalactic magnetic fields should have strengths not much smaller than 1~nG. Another possibility to produce a harder spectrum is by the effect of magnetic confinement around the sources, for instance if the sources were near the cores of  galaxy clusters and the CR escape times from them were to exceed the lifetime of the Universe for $E/Z<1$~EeV \cite{hmr16b}.

Regarding the large fraction of extragalactic protons observed at few EeV energies, a possible explanation is based on the strong photo-disintegration  of heavy nuclei directly at the sources \cite{gl15,un15}. In these scenarios heavy nuclei suffer significant photo-disintegrations with radiation fields (such as UV/optical and IR) present in the acceleration region \cite{gl15} or after acceleration if they remain confined for large times by strong magnetic fields present around the sources \cite{un15}. A large number of secondary nucleons, produced by the photo-disintegration of nuclei with energy $E$ and mass number $A$, appear then at energies $E/A$, which can hence be observed as  a dominant component at energies of few EeV. For these scenarios the neutron escape mechanism \cite{nescape} is relevant, i.e. the fact that neutrons produced in interactions of charged CRs that were magnetically confined in a region can freely escape from the magnetized region. Note that the boosted neutron decay lifetime is $\gamma c\tau_n\simeq 10$\,kpc($E$/EeV), so that at EeV energies neutrons may even escape from a region as big as a galaxy before decaying. As a result of the energy dependent confinement and interactions, the nuclei exiting the source may have a harder spectrum with respect to that produced in the acceleration process while the proton spectrum can actually be steeper \cite{gl15}.

We would also like to point out here that if the maximum CR rigidity is $R_{\rm c}\simeq 5$~EV, the maximum energy for the secondary nucleons produced by photo-disintegrations is $E_{\rm max} \simeq Z e R_c/A\simeq e R_c/2\simeq 2.5$~EeV. This energy is similar to the one where the elongation rate determined by Auger decreases  \cite{augerxmax} and hence this change may be related to the suppression of the secondary nucleons above this threshold.

\subsection{UHECR source requirements}
\begin{figure}[tb]
\begin{center}
\begin{minipage}[t]{8 cm}
\centerline{\epsfig{file=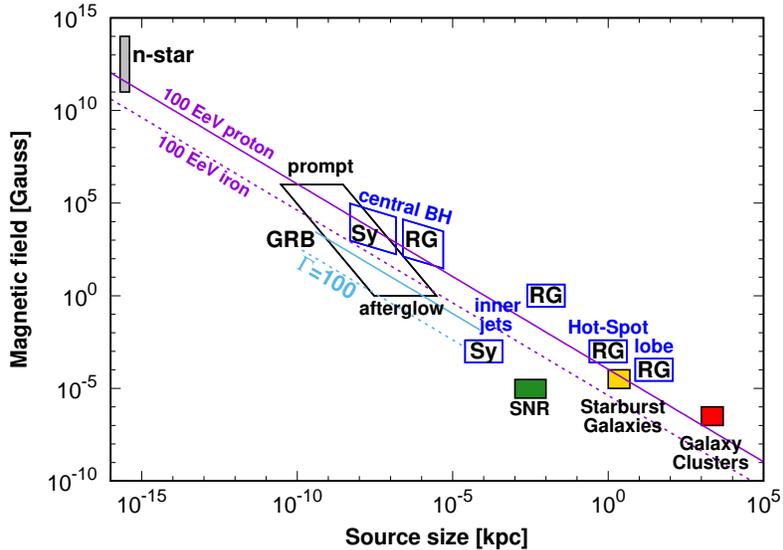,scale=1.2}}
\end{minipage}
\caption{Hillas plot showing the location of the main UHECR source candidates in the plane of their characteristic magnetic field vs. their size. To accelerate protons (iron) above 100~EeV the sources have to lie above the diagonal solid (dashed) line. Also shown are the shifted lines relevant for sources with relativistic jets, for the case of a boost factor $\Gamma=100$, relevant e.g. for prompt acceleration in  GRBs, in which case the constraint gets relaxed. The sources shown are neutron stars, GRBs accelerating CRs during the prompt emission or in the afterglow, different AGN scenarios involving radio-quiet Seyfert (Sy) or radiogalaxies (RG), for the cases of acceleration in the central black hole (BH) and the inner jets (FRI in the radio-loud case), or the hot-spots and radio lobes of FRII galaxies. The parameters corresponding to  supernova remnants (SNR), acceleration in large scale shocks in starburst galaxies or in galaxy clusters are also indicated.  \label{hillas.fig}}
\end{center}
\end{figure}

Regarding the possible sources where CRs could be accelerated up to the highest energies, there is a simple relation which states that if the acceleration region has characteristic size $L$ and magnetic field $B$ the maximum achievable energy is $E_{\rm max}\simeq \Gamma eZBL$. This is essentially the requirement that the Larmor radius be smaller than the confinement region \cite{hillas}, and the relation also allows for a possible boost by a factor $\Gamma$ in the case in which the acceleration is in a relativistic jet (for instance $\Gamma\sim 10^2$ in GRBs, or few to a few tens in relativistic AGN jets). A version of the associated `Hillas-plot' is shown in Fig.~\ref{hillas.fig}, where in the plane $B\,{\rm vs.}\,L$ different source candidates are located. Also shown are the curves above which protons or Fe nuclei can be accelerated up to 100~EeV.  It is clear that only few source types seem to be capable of accelerating protons up to the highest energies, while the acceleration of heavy nuclei is achievable for a larger variety of sources.  Some of the main candidates for acceleration up to the highest energies are GRBs and AGNs (either in the accretion disk and central black-hole or in  the jets and radio lobes), as well as reacceleration in large scale shocks in galaxy clusters or starburst galaxies. We note that a  stricter requirement results in the case of non-relativistic shocks, for which the acceleration time remains smaller than the time to diffuse out from the acceleration region  only if $E_{\rm max}< \beta_s eZBL$, with $\beta_s\equiv v_s/c$ in terms of the shock velocity $v_s$ \cite{hillas}. For $\beta_s\simeq 1$ this constraint is similar to the one from the confinement requirement plotted in Fig.~\ref{hillas.fig}.
Additional constraints on the Hillas-plot from the effects of synchrotron radiation losses have been discussed in \cite{hillas,ah02,pt10b}.

\begin{figure}[tb]
\begin{center}
\begin{minipage}[t]{8 cm}
\centerline{\epsfig{file=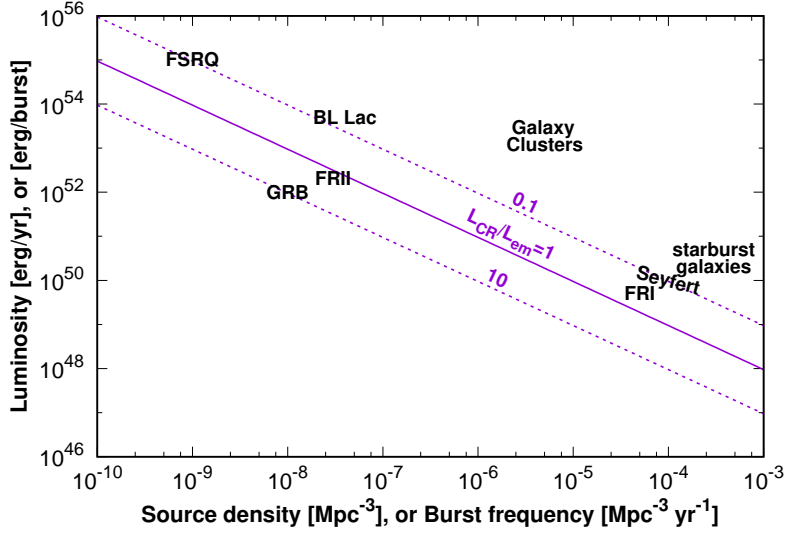,scale=1.2}}
\end{minipage}
\caption{Potential UHECR sources in a plot of their non-stellar electromagnetic  luminosities vs. source densities. The diagonal solid line indicates the condition that  the required CR luminosity, according to eq.~(\ref{lvsns}), equals the em luminosity. The dashed lines correspond to CR luminosities 10 times bigger (lower line) or smaller (upper line) than the em ones. Included are Galaxy clusters, starburst galaxies, GRBs and different types of AGNs:  radio-quiet Seyferts and radio-loud  Fanaroff Riley radiogalaxies (FRI and FRII) as well as the  blazars BL Lacertae (BL Lac) and flat spectrum radio quasars (FSRQ), in which the jets point towards the observer.  \label{sources.fig}}
\end{center}
\end{figure}

Regarding the power requirement for the sources, one may estimate the energy density associated to UHECRs from the fact that at 10~EeV the total flux is d$\Phi$/d\,ln$E\simeq 0.6$\,km$^{-2}$yr$^{-1}$sr$^{-1}$ (see Fig.~\ref{spectrum.fig}). If the extragalactic sources leading to this flux have an $E^{-2}$ spectrum, so that they contribute an equal power per logarithmic energy bin, extending  from $E_{\rm min}\sim{\rm TeV}$ up to $E_{\rm max}\sim 100\,{\rm EeV}$, the associated energy density would be 
\be 
\rho_E\simeq \frac{4\pi}{c}\left(E \frac{{\rm d}\Phi}{{\rm d\,ln}E}\right)_{10\,{\rm EeV}}\,{\rm ln}\frac{E_{\rm max}}{E_{\rm min}}\simeq 10^{55}\frac{\rm erg}{{\rm Mpc}^3}.
\ee
If the density of sources is $n_{\rm s}$ and we assume cosmic rays were emitted with luminosity $L_{\rm s}$ for a time comparable to the age of the Universe, this would imply that $\rho_E\simeq  L_{\rm s} n_{\rm s}/H_0$. Hence, the required CR source luminosity should be
\be 
L_{\rm s}\simeq 3\times 10^{42}\left(\frac{10^{-5}{\rm Mpc}^{-3}}{n_{\rm s}}\right) \frac{\rm erg}{\rm s}.
\label{lvsns}
\ee
We note that accounting for propagation energy losses in the flux normalization performed at 10~EeV could increase the required luminosity by a factor of few, while assuming steeper (harder) CR spectral shapes could increase (decrease) the overall power requirement. 
Fig.~\ref{sources.fig} shows a plot
locating different source types in the plane of luminosity (in the electromagnetic band) as a function of the source density, with the plot being inspired in a similar plot for neutrino sources presented in \cite{kowal,gaisser}.  For bursting sources the quantities in the plot would correspond to the integrated luminosity per burst as a function of the frequency of bursts per unit volume. The solid line depicted corresponds to the constraint in eq.~(\ref{lvsns}). Sources above that line could provide the required power with a CR luminosity smaller than the respective source electromagnetic luminosity.

\subsection{Secondary neutrinos and photons}
When UHECR protons interact with background radiation to produce pions, such as in $p\gamma\to n\pi^+$,  neutrinos get produced through the decay chain $\pi^+\to\mu^+ \nu_\mu\to e^+\nu_e\bar{\nu}_\mu\nu_\mu$ (and similarly for the charge conjugate channel). These neutrinos have typical energies $E_\nu\simeq E_\pi/4\simeq E_p/20$.  Hence, if the CR spectra had a sizable proton component extending beyond the photo-pion threshold for CMB interactions, a flux of neutrinos with energies peaking at few EeV would be  expected\footnote{ When the steeply falling primary spectrum leads to secondaries through a process with a threshold, the primaries with energies near the threshold give the dominant contribution to the production of secondaries.}  \cite{bz}.  Also ${\bar\nu}_e$ may be produced in the decays of the neutrons, although they will have much lower energies, $E_\nu\simeq 3\times 10^{-4}E_p$, and hence their spectra will peak at about 20~PeV \cite{engelnu}. The photo-pion production from CR protons scattering off optical and IR EBL radiation leads to the production of neutrinos in the energy range $10^{14}$--$10^{17}$~eV. Their contribution to the overall cosmogenic neutrino fluxes can be significant if the extragalactic spectrum is steep and the source evolution is strong \cite{ko10}.
For heavy nuclei interacting with the CMB photons the pion production has a threshold which is above the highest observed energies. Only if the nuclei spectra were to extend beyond $\sim 70A$~EeV would there  be an associated neutrino flux at few EeV, but it would be significantly suppressed with respect to the neutrino flux expected in scenarios in which UHECRs are pure protons. Otherwise, neutrinos from the decays of the neutrons  and other radioactive byproducts resulting from the photodisintegration interactions get produced with energies in the PeV range.

The threshold for pion production  gets smaller for photon backgrounds with energies $\varepsilon$ larger than the CMB ones, since the required Lorentz factors for pion production are $\Gamma\simeq 7\times 10^9(10^{-3}\,{\rm eV}/\varepsilon)$. The associated neutrino energies, which will be typically $E_\nu\simeq E/20A$, will be below the EeV in this case. For the photo-disintegration processes involving nuclei a copious amount of neutrons may be produced, leading to neutrinos with energies $\sim 10^{-3}E/A$. For interactions with the CMB, for which the photo-disintegration threshold corresponds to $E/A\simeq 5$~EeV, the neutrinos will appear at a few PeV. 

\begin{figure}[tb]
\begin{center}
\begin{minipage}[t]{8 cm}
\centerline{\epsfig{file=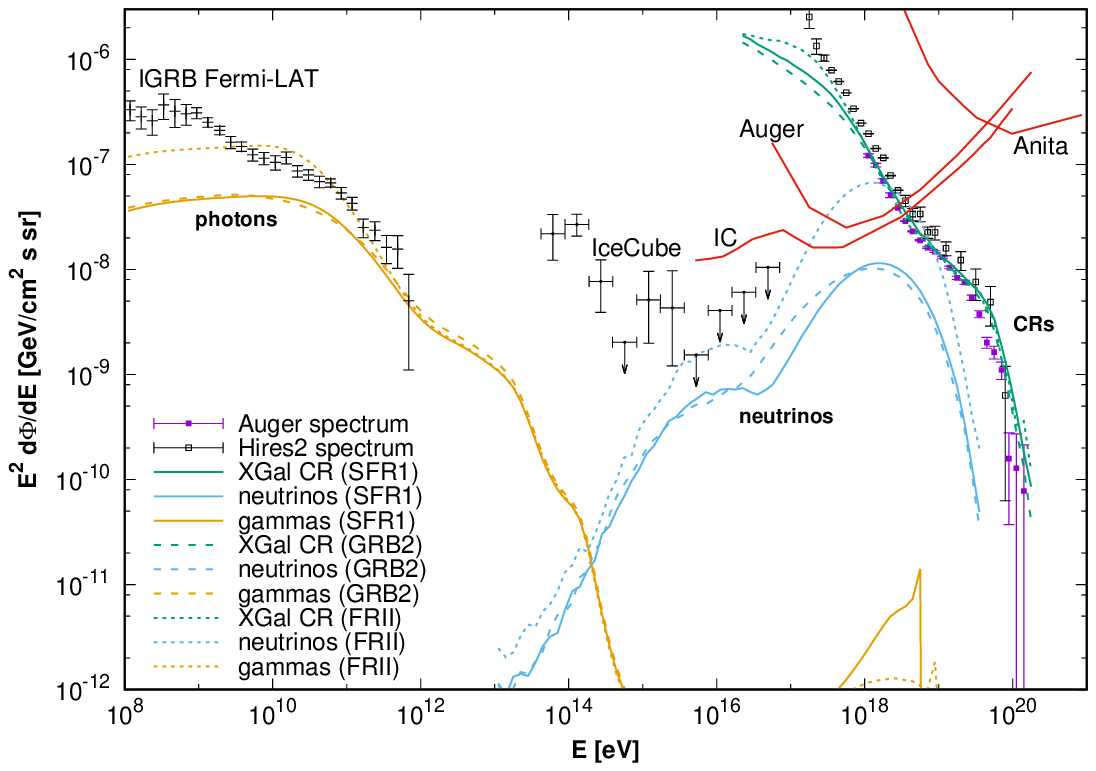,scale=.8,angle=0}\epsfig{file=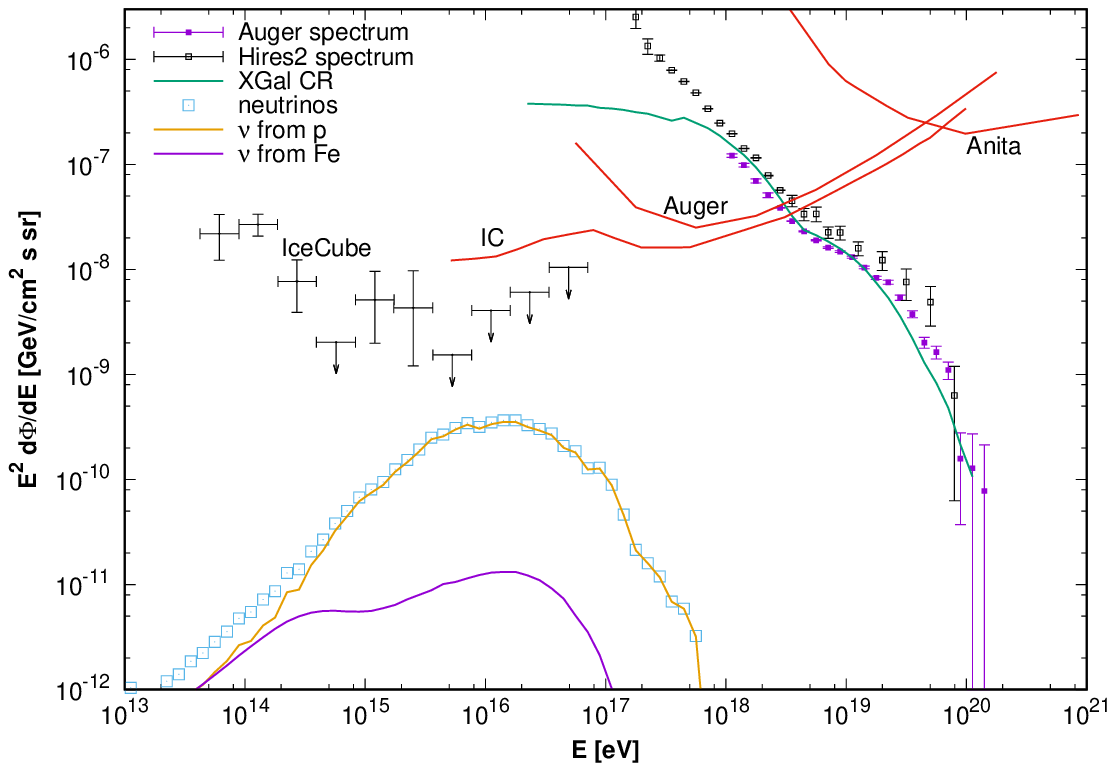,scale=.8,angle=0}}
\end{minipage}
\caption{Expected all-flavor cosmogenic neutrino fluxes as a function of energy (from \cite{cosmonu}).  Also shown are the resulting CR flux (green) and the measured fluxes by Auger and HiRes. The   astrophysical flux determined by IceCube is indicated, as well as the bounds form Auger, Anita and IceCube (90\% CL bound per energy decade).  Left panel: proton scenario with $E_{\rm max}=200$~EeV. Also the resulting photon flux for different cosmological source evolutions and the Fermi determination of the isotropic gamma ray background (IGRB) are displayed. Right panel: scenario with low rigidity cutoff ($R_{\rm max}=5$~EV), with a mixture of protons and iron primaries. \label{cosmonu.fig}}
\end{center}
\end{figure}

The study of all these channels of neutrino production is particularly interesting in view of the discovery by the IceCube Collaboration \cite{hese} of a flux of neutrinos with energies 0.1--3~PeV, in excess of the flux expected from the atmospheric neutrinos produced by CR interactions in the atmosphere. The origin of this astrophysical flux of neutrinos is still unknown. The predicted cosmogenic neutrino fluxes due to the interactions of UHECRs during propagation through the radiation backgrounds is illustrated in Fig.~\ref{cosmonu.fig} for two different CR composition scenarios, one dominated by protons up to the highest energies and one with a rigidity cutoff at $\sim 5$~EV with a transition to a heavier Fe component above the ankle energy (see \cite{cosmonu} for details). Also shown is the flux observed by IceCube. It is apparent that in both scenarios the fluxes at PeV energies are significantly below those inferred from IceCube observations, and also below the Icecube and Auger bounds at EeV energies. Note that due to neutrino oscillations, even if at the sources the neutrino flavor ratios from pion decays are in the proportion $\nu_e:\nu_\mu:\nu_\tau \simeq 1:2:0$, after propagation through cosmological distances one expects similar amounts of the three flavors. For neutrinos from neutron decays, one finds instead that the initial antineutrino flavor ratio $1:0:0$ gets converted by oscillations into $0.55:0.25:0.2$ \cite{ch09}.

The way neutrinos are searched for by Auger is by looking at air showers close to the horizon. In this case the background from hadronic CRs can be neatly rejected by looking at signals like those expected from a shower having a significant  electromagnetic component (such as large time spread of the signals in the detectors). Since neutrinos arriving horizontally can interact anywhere along their path in the atmosphere, those interacting not very far from the detector will have a significant electromagnetic component. On the other hand, the showers arising from the much more abundant hadronic CRs will start far away, since they interact when they enter  into the atmosphere. By the time the shower arrives to the detector the em component will then  be almost completely absorbed so that just narrow signals due to the surviving muonic component will result. Actually the most effective way to look for neutrinos at Auger is by the signal from tau neutrinos coming from slightly below the horizon and having suffered a charged current interaction in the crust of the Earth \cite{taushower}. The tau lepton so produced, having a typical energy of about 70\% of the neutrino energy,  has a decay length $\gamma c\tau_\tau\simeq 50\,{\rm km}(E/$EeV), and hence the tau lepton can travel few tens of km before exiting the rock, and if it then decays hadronically in the atmosphere above the detector it may be observed as a young horizontal air shower. The inclination of the incoming neutrino cannot be more than a few degrees below the horizon, since otherwise the very energetic neutrino will be strongly absorbed by the Earth.
Essentially no candidates for neutrino events have been found by Auger \cite{augernubound}, allowing to set the bounds  depicted in Fig.~\ref{cosmonu.fig}\footnote{Since these searches are essentially background free, the bounds become stronger in inverse proportion to the accumulated exposure as long as no neutrino candidate events are detected.}. Also significant bounds on 10~PeV--EeV neutrinos have been set by IceCube \cite{icnubound}.

The fact that the cosmogenic neutrinos are not sufficient to account for the astrophysical neutrino flux observed by IceCube favors the scenarios in which these neutrinos are produced directly at the sources where the CR are accelerated, or in the environments around them, rather than during their propagation through intergalactic space. Note that the relevant CR energies to produce neutrinos of few PeV via pion production is of about 0.1$A$~EeV, and if the process is due to photo-pion production (rather than $pp$ interactions) it should involve optical/UV photons.

Also note that in scenarios with low maximum rigidities, $R_{\rm c}<10$~EV, one does not expect cosmogenic neutrinos at EeV energies, and hence models where this happens are often referred to as `disappointing' \cite{disap}. Note however that even in this case the cosmogenic neutrino fluxes at energies close to 10~PeV, arising from the photo-pion interactions with the EBL, may not be much below the present bounds at these energies if the CRs at EeV energies are predominantly extragalactic protons.
On the other hand, the bounds on EeV neutrinos are at the level of the most optimistic predictions in scenarios with pure protons dominating up to 100~EeV. It is then crucial  to continue to improve the sensitivity of this kind of searches and projects such as ARA \cite{ara} or GRAND \cite{grand} will make a significant step in that direction. Note that according to Fig.~\ref{augercomp.fig} a fraction of protons at the level of 10--20\% is allowed to be present at the highest energies. It would be important to become able to test this possibility since this could have far reaching consequences for the production of neutrinos as well as for CR anisotropies.

Besides charged pions, interactions of CR protons with radiation fields also produce neutral pions  through $p\gamma\to p\pi^0$, leading then to a flux of energetic gamma rays through  $\pi^0\to\gamma\gamma$. Unlike the neutrinos, which can travel essentially unscathed through the Universe, the gamma rays interact with background photons, mostly from the CMB, producing $e^+e^-$ pairs which subsequently lead to electromagnetic cascades by inverse Compton off CMB photons and eventually also synchrotron processes by interactions with extra-galactic magnetic fields \cite{cascadegamma}. The photon mean free path for the pair production can be quite small, reaching minimum values of order 10~kpc at PeV energies, but below this energy it starts to increase due to the effects of the pair production threshold. In particular, at TeV energies the attenuation is dominated by the interactions with the more energetic EBL radiation, with the typical attenuation length being $\sim 10^2$~Mpc \cite{co97}. The electromagnetic cascades generated by the very high energy gamma rays then  typically lead to a photon diffuse background at GeV--TeV energies. Significant constraints on the photon production can be obtained by requiring that the flux from cascade photons does not exceed the diffuse gamma fluxes determined by the Fermi satellite \cite{fermibound}. The   isotropic gamma ray background (IGRB) determined by Fermi, as well as the predicted flux of cascade photons from CR proton scenarios for different assumptions on the source evolution with redshift (see \cite{cosmonu} for details), are also displayed in the left panel of Fig.~\ref{cosmonu.fig}. It has also been realized that a significant fraction of the gamma flux measured by Fermi for energies above $\sim$~GeV is due to emission from unresolved radiogalaxies, and hence the diffuse contribution from other sources may be bound to be as small as $\sim 20$\% of the total measured flux \cite{fermiresolved,ho16}, making these constraints stronger. 

If the maximum CR rigidities exceed $\sim 50$~EV, the neutral pion production off the CMB in sources not much farther than a few Mpc, for which the attenuation due to cascading is moderate, can lead to a photon flux at few EeV energies. This flux of `GZK photons' has been constrained by the Auger and TA Collaborations \cite{gammaa,gammata}, implying that  the fraction of photons in the CR flux should be smaller than few percent in the 1--10~EeV range. Although these bounds are still above the most optimistic predictions from cosmogenic production in CR interactions, they  significantly constrain exotic models for the origin of the UHECRs that  were once fashionable, such as those in which the UHECR are produced in super-heavy dark matter decays or are emitted from topological defects.

\section{Proton cross section and air shower muon content}

The detailed observation of the UHECR air showers allows to study properties of the interactions responsible for the development of the showers which are of great interest for particle physics and to better constrain the hadronic models involved. Examples of this are the determination of the  proton-air cross section, from which the proton-proton cross section can be inferred, or the study of the production of muons to determine the muon production distribution as a function of atmospheric depth, or the total number of muons reaching the ground.

\begin{figure}[tb]
\begin{center}
\begin{minipage}[t]{8 cm}
\centerline{\epsfig{file=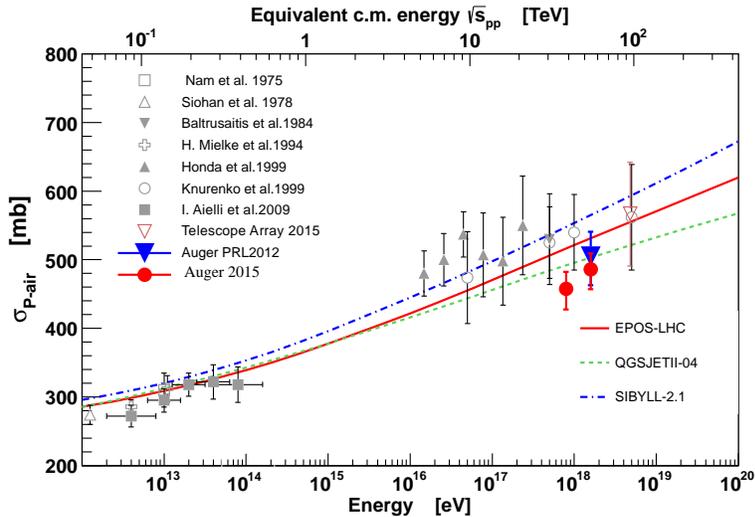,scale=.5}}
\end{minipage}
\caption{Inelastic proton-air cross section as a function of center of mass energy from various air-shower experiments (adapted from \cite{ppxsecta}). Also shown are the theoretical expectations from various hadronic models. \label{ppxsect.fig}}
\end{center}
\end{figure}

In order to determine the proton-air particle production cross section\footnote{This is the inelastic part of the cross section associated with particle production, i.e. without the small contribution in which the air nucleus just gets excited.}, FD measurements of the distribution of shower maxima are considered. Since at a given energy the lighter nuclei lead to more penetrating showers, the tail of large $X_{\rm max}$ values should be dominantly produced by light nuclei (provided that a light component contributes to the primary CR flux). Since at energies of few EeV a sizable proton component is known to exist, at these energies the tail of the $X_{\rm max}$ distribution is mostly due to proton primaries. On the other hand, at energies in excess of 10~EeV it is not clear if a proton component is present at all, and hence this kind of analysis cannot be reliably applied above this energy. The main property exploited in these studies is that the distribution of the number of showers as a function of $X_{\rm max}$ has an exponential attenuation well past the maximum of the distribution, with d$N/{\rm d}X_{\rm max}\propto \exp(-X_{\rm max}/\Lambda)$. Since the attenuation length $\Lambda$ is directly related to the particle production cross section for the first interaction of the primary, with larger cross sections leading to steeper distributions and hence smaller $\Lambda$, the determination of the slope in the tail of the distribution allows to infer the proton-air inelastic cross section. The dominant systematic effect on this measurement turns out to be the possible contribution from He primaries to the tail of the distribution. These measurements have been performed by the Auger \cite{ppxsecta} and Telescope Array \cite{ppxsectta} Collaborations, with the results shown in Fig.~\ref{ppxsect.fig} together with previous results at lower energies.  Once the proton-air cross section is determined, one may use Glauber theory to infer the proton-proton cross section. The equivalent $pp$ CM energies of these measurements reach values of 50--100~TeV, quite above the energies tested at the LHC.  The cross sections obtained  are in good agreement with the expectations from hadronic models. They do not suggest a rapid rise in the $pp$ cross section with energy, as would be required  in scenarios proposed to explain the $X_{\rm max}$ and Var$(X_{\rm max}$) Auger observations with pure proton primaries. In those hypothetical proton scenarios the small elongation rates and narrow $X_{\rm max}$ dispersions observed above 10~EeV would be the consequence of an abrupt increase in the $pp$ cross section due to new physics effects \cite{wi11}.

Regarding the studies of muon production, by looking at inclined showers ($\theta\simeq 60^\circ$) at large distances from the shower core (typically beyond $\sim 1.7$~km), it becomes possible to detect individual muons and from their timing reconstruct, essentially through triangulation, the depth along the shower axis where they were produced. Hence, by combining the results from many showers of similar energies one can obtain the distribution of muon production depths \cite{xmumaxa}. The first results obtained with this technique suggest that hadronic models tend to predict a depth of muon production significantly deeper than  is observed. In particular, the measured average depth is smaller than what is obtained in the simulations using EPOS-LHC even for Fe primaries. 

The other important measurement is that of the total number of muons at ground level \cite{nmuauger}. This can be performed for instance using hybrid events for which one has a robust determination of the electromagnetic component reaching ground. Subtracting the associated expected electromagnetic signal in the water-Cherenkov Auger detectors one infers the contribution to the signal from the muons. The result is that there is an excess of muons reaching ground even with respect to the expectations from Fe nuclei primaries. A possible relevant physical effect that could lead to an increased muon production is the enhanced production of $\rho^0$ vector mesons as leading particles (and a suppression of $\pi^0$ mesons) that was observed for instance in the pion C interactions detected by the NA61/SHINE fixed target CERN experiment \cite{rho0}. Contrary to the scalar $\pi^0$, which decays dominantly to two photons and hence feeds the em component of air showers, the spin one $\rho^0$ does not decay into two photons and moreover, having an antisymmetric wave function, it does not decay into two identical $\pi^0$, being its dominant decay channel $\rho^0\to\pi^+\pi^-$. 
This leads to a larger fraction of the energy remaining in the hadronic component in each interaction stage, 
 naturally leading to an excess in the muons with respect to what is  obtained using hadronic models that neglect the enhanced $\rho^0$ production. The most recent SIBYLL model including this effect finds indeed a significant enhancement in the number of muons produced in the shower \cite{sibyll3}. Note that since the suppression of the electromagnetic component by an enhanced $\rho^0$ production is a cumulative effect, depending on the number of generations of hadronic interactions that happen before the charged pions decay, the muon excess is expected to be reduced for lower energy primaries, and indeed no indications of a significant muon excess have been reported at energies below 0.1~EeV  (although there is always an interplay between  the expected number of muons and the inferred CR composition) \cite{msu17,ic17}. 

\section{Cosmic ray propagation in magnetic fields}

During their propagation from the sources to the Earth, cosmic rays are not only affected by the interactions with the radiation backgrounds they encounter along their trajectories, but also by the magnetic fields present. These include the magnetic fields in the environment of the sources, in the intergalactic space and in the Galaxy, and one of the  main difficulties to study these effects arises from the poor knowledge of the structure and amplitude of the relevant magnetic fields, besides the fact that the CR charges are also uncertain.

Different observations are used to measure and constrain the magnetic field in the Galaxy and in the intergalactic space: Faraday rotation, synchrotron emission, polarized dust grain emission and Zeeman splitting \cite{beck16}. Spiral galaxies, such as the Milky Way, are known to have both a regular (coherent) field as well as turbulent fields.  The regular field is usually split into a disk and a halo component, although different models are proposed to fit the observations (for a review see \cite{haverkorn15}). For our galaxy the field strength in the solar neighborhood is $\sim 6\,\mu{\rm G}$, with a regular component of $\sim 2$--$3\, \mu{\rm G}$ and a random component of $\sim 2$--$4\, \mu{\rm G}$. The amplitude of the field increases towards the Galactic center up to values of $\sim 10\,\mu{\rm G}$. The disk field has a spiral arms pattern, with some reversals between the arms. For the halo component, a toroidal and an $X$-field component are considered, although their symmetry with respect to the Galactic plane is uncertain, as is the field intensity  which takes typical values in the range of $\sim 1$--$10\,\mu{\rm G}$ (see e.g. \cite{tf17}). 
The turbulent, or random, magnetic field component can be isotropic (same dispersion in the three spatial directions) or anisotropic (different dispersions originating from the isotropic turbulent component through the compression of gas flows). It is usually characterized by its root mean square (rms) amplitude, a coherence length $l_{\rm c}$ and a spectral index that describes how the magnetic energy density $\omega$ is distributed in the different wavenumbers $k$, with $\omega(k)\propto k^{-\alpha}$. The idea that the turbulent energy is injected into the interstellar medium by stellar winds and supernova explosions on scales of 10--100~pc and then it is transferred to the smaller scales in a cascading process, naturally leads to a Kolmogorov spectrum with $\alpha=5/3$.
This is in good agreement with observations below $\sim 5$~pc, while a slight flattening of the spectrum at higher scales could appear \cite{ms96,han04}. 

The  extragalactic magnetic fields are largely unknown, both in their strength and structure. Few  observational  constraints exist, making it difficult to construct a model for them. In galaxy clusters they are probed  both through the measurement of the Faraday rotation effect on the light of background galaxies and through radio emission from diffuse synchrotron sources. The presence of large-scale magnetic fields is put in evidence by the population of relativistic electrons in the central regions, with typical size of 1~Mpc, called halos and also in some irregular extended sources in the clusters periphery, with similar size as the halos, called relics. These large-scale magnetic fields have measured amplitudes that range from a few to tens of $\mu$G in the cluster central regions \cite{fe12}, which also suggest the presence of large-scale magnetic fields of nG strength in cosmic structure filaments. The strength of the magnetic fields in voids is even more uncertain. A lower bound on the strength of intergalactic magnetic fields of  $B\ge 3\times 10^{-16}$~G  has been obtained in \cite{nv10} from the non-observation by Fermi LAT of GeV gamma-ray emission from electromagnetic cascades initiated in the intergalactic medium by TeV gamma-rays  from sources that had been observed by H.E.S.S. This comes from the request that $e^+ e^-$ produced in the cascade be deflected by the intergalactic magnetics fields at angles beyond the point spread function of the telescope. 

Turbulent magnetic fields may be produced in the Universe from the evolution of primordial seeds affected by the process of structure formation. This typically leads to magnetic fields with strength correlated with the matter density ($B\propto \rho^{2/3}$ due to the flux conservation during the collapse), and hence being enhanced in dense regions such as superclusters while suppressed in the voids. 
Different numerical simulations have obtained  predictions for the magnetic fields associated to the formation of large scale structure in the Universe, but the  amplitudes obtained for the intergalactic fields vary significantly, ranging from few nG strengths down to much smaller values \cite{si04,do04}.
The stronger constraints on the primordial magnetic fields come from the CMB anisotropy and polarization measurements. The energy momentum tensor of primordial magnetic fields is a  source for all types of cosmological perturbations (scalar, vector and tensor).  Magnetically induced   vector  perturbations do not  decay with the Universe expansion and are not suppressed by Silk damping. Their impact on the CMB temperature power spectrum is dominant on small angular scales, where the primary CMB fluctuations are  suppressed, and Planck measurements put an upper bound of $\sim 5$~nG at comoving scales of 1~Mpc \cite{pl15}.

 In the presence of a magnetic field $\vec {B}(\vec{x})$ the trajectories of cosmic rays are deflected according to the Lorentz force they experience. For a relativistic nucleus of charge $Z e$ and energy $E$, the direction of propagation $\hat n$ evolves according to
\begin{equation}
\frac{{\rm d}{\hat n}}{{\rm d}t}=\frac{Z e}{E} {\hat n} \times {\vec B} ({\vec x}).
\label{lorentz}
\end{equation}
Thus, the total deflection scales with the inverse of the energy of the particle and is proportional to its charge and to the integral along the trajectory of the orthogonal component of the magnetic field. In a regular magnetic field the deflection after traversing a distance $L$ can be written, using as reference values those  typical for UHECR propagation in the Galaxy, as
\begin{equation}
\delta \simeq 10^\circ \frac{10\, {\rm EeV}}{E/Z} \left|\int_0^L\frac{{\rm d}{\vec x}}{ {\rm kpc}} \times \frac{\vec B}{2\, \mu {\rm G}}\right|.
\label{defreg}
\end{equation}

\begin{figure}
\begin{center}
\begin{minipage}[t]{8 cm}
\centerline{\epsfig{file=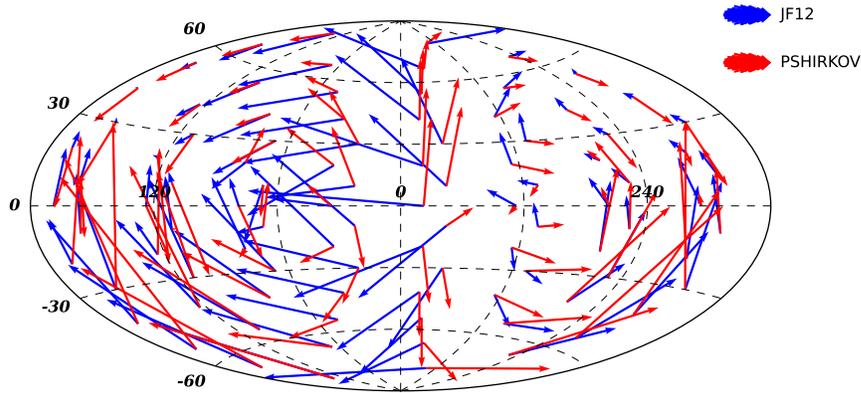,scale=0.5}}
\end{minipage}
\begin{minipage}[t]{16.5 cm}
\caption{Map in Galactic coordinates of the deflections for particles with $E/Z = 10$~EeV in the Galactic magnetic field models proposed by Jansson and Farrar (JF12) \cite{jf12} and by Pshirkov et al. \cite{ps11}. For a grid of arrival directions at Earth, the tips of the corresponding arrows indicate the arrival directions outside the Galaxy.\label{jfvsp}}
\end{minipage}
\end{center}
\end{figure}

The total deflection suffered by particles reaching the Earth from outside the Galaxy is strongly dependent on the direction of arrival and on the magnetic field model considered. We show in Fig.~\ref{jfvsp} the resulting values for the deflections of particles with $E/Z = 10$~EeV in two recently proposed models for the regular Galactic magnetic field \cite{jf12,ps11}. Although the overall size of the deflections are comparable in the two models, for some sky regions the specific results are quite different, giving a measure of the present uncertainties in these predictions.

\begin{figure}
\begin{center}
\begin{minipage}[t]{8 cm}
\centerline{\epsfig{file=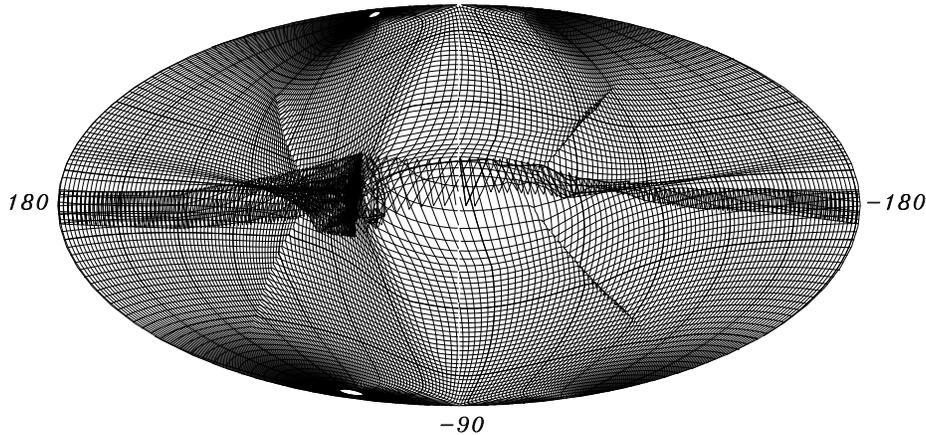,scale=0.8}}
\end{minipage}
\begin{minipage}[t]{16.5 cm}
\caption{Regular grid of arrival directions at Earth of cosmic rays with $E/Z = 20$~EeV mapped to the arrival direction outside the Galaxy after deflection in the Galactic magnetic field model proposed by Jansson and Farrar (JF12) \cite{jf12}, in Galactic coordinates.\label{skysheet}}
\end{minipage}
\end{center}
\end{figure}

Besides changing the arrival direction from which particles reach the Earth from an extragalactic source, the Galactic magnetic field can  also amplify or demagnify the observed flux and can produce multiple images of a single source \cite{hmr99}. These effects can be visualized by plotting the mapping between the CR arrival direction at the Earth and those outside the Galaxy, as shown in Fig.~\ref{skysheet}. In it a regular grid of arrival directions at the Earth is  plotted at the corresponding coordinates outside the Galaxy, computed by propagating every point of the grid along the trajectory of an antiparticle, with the indicated value of $E/Z$, up to a radius of 20~kpc from the Galactic center. The propagated grid looks stretched and folded.  Cosmic rays arriving from a direction outside the Galaxy where a fold has developed are seen at Earth from all the points of the grid that overlap in that place. The flux of sources located at directions where the grid is compressed will be amplified, while that of sources located in stretched regions will be demagnified. As the particle deflections are a function of $E/Z$, the grid will be deformed differently for different values of $E/Z$. Hence, cosmic rays from one source will arrive to the Earth from different directions and with different amplifications of their flux as a function of the energy. In particular,  for the value of $E/Z$ at which a fold, or caustic, crosses the direction towards a given individual source, a pair of highly magnified new images of the source appear at some direction in the sky. The expected signal in this case would then be a localized flux excess in a narrow energy band. Although the caustic pattern as a function of $E/Z$ is different for different Galactic magnetic field models, the typical values of $E/Z$ for which this effect is expected to become relevant is in the $E/Z \sim 10$--30~EeV range.

It is important to note that the presence of magnetic fields cannot generate anisotropies in an originally isotropic flux of cosmic rays. This is due to the Liouville theorem, that is the fact that the phase space distribution function is constant along the particle trajectories. Thus, whenever the deflections lead to a concentration of particles in a given region, the angular spread of their propagation directions will be increased in such a way that the number of particles traversing a unit area per unit solid angle is kept constant, and hence an isotropic flux remains isotropic.
However, the presence of magnetic fields do modify already existing anisotropies.

 In the energy range in which the deflection of the trajectories in the regular magnetic field is small, the inverse proportionality with the energy in eq.~(\ref{defreg}) implies that the arrival directions of cosmic rays from one source appear aligned in the sky and ordered in energy, with the most energetic ones being closer to the true source direction. The observation of this `spectrometer effect' could provide a new way to constrain the regular Galactic magnetic field and to locate the CR sources \cite{ghmr}. However, experimental searches have not led to significant signals  \cite{augermultiplets}. The measured CR composition above the ankle, showing an increased CR mass  with increasing energy, may be the reason for the non-observation of this effect up to now.
 
 The deflections of cosmic ray trajectories in turbulent magnetic fields have different characteristics due to the random orientation of the magnetic field in different domains of the size of the coherence length $l_{\rm c}$. An effective Larmor radius can be introduced as
\begin{equation}
r_L=\frac{E}{ZeB}\simeq 1.1\, \frac{E/Z}{\rm EeV}\frac{\rm nG}{B}\,{\rm Mpc},
\end{equation}
where $Ze$ is the particle charge and the field strength is characterized by its rms value $B=\sqrt{\langle B^2(x)\rangle}$. If the Larmor radius is smaller than the coherence length, the direction of the particles will be completely changed after traveling a distance $l_{\rm c}$, and the CRs will perform a random walk through the different field domains leading to a diffusive spatial propagation. In the isotropic diffusion case this can be characterized by just a scalar diffusion coefficient $D$ that describes how fast the mean square distance of the particles from the original point increases with time, $\langle r^2(t)\rangle = 6 D t$.

A relevant quantity to characterize the particle diffusion is the critical energy $E_{\rm c}$, defined such that $r_L(E_{\rm c})=l_{\rm c}$ and hence given by $E_{\rm c}=Z e B l_{\rm c}$. Specifying the typical values for Galactic (G) and extragalactic (xG) fields one then gets

\begin{equation}
E_{\rm c}^{\rm G} \simeq 80 Z\,\frac{B}{3\,\mu{\rm G}}\frac{l_{\rm c}}{30\,\rm pc}\,{\rm PeV}\ \ \ ,\ \ \ E_{\rm c}^{\rm xG} \simeq 0.9 Z\,\frac{B}{\rm nG}\frac{l_{\rm c}}{\rm Mpc}\,{\rm EeV}.
\label{ec}
\end{equation}
The critical energy separates the regimes of resonant diffusion at lower energies, in which particles have large deflections induced by their interactions with the $B$ field modes with scales comparable to the Larmor radius, and the non-resonant regime at higher energies in which the deflections after traversing a distance $l_{\rm c}$ are small, typically of order $\delta\simeq l_{\rm c}/r_L$. 
For $E<E_c$ the diffusion coefficient scales with energy as $D\propto E^{2-\alpha}$ (corresponding to $D\propto E^{1/3}$ for a Kolmogorov spectrum for which the turbulence spectral index is $\alpha=5/3$).

For CRs in the non-resonant regime, i.e. with energy $E > E_c$, the typical deflection  after propagation over a distance $L\gg l_{\rm c}$ through a random magnetic field is \cite{ac99,hmrs}
\begin{equation}
\delta_{\rm rms}\simeq \frac{BZe}{E}\sqrt{\frac{Ll_{\rm c}}{2}}\simeq 4^\circ \frac{B}{\rm nG}\frac{\rm 10\,EeV}{E/Z}\frac{\sqrt{Ll_{\rm c}}}{\rm Mpc},
\label{delta}
\end{equation}
where the last relation was specified for typical extragalactic magnetic field parameters. 
An important quantity to characterize the propagation in this regime is the diffusion length $l_{\rm D}$, defined as the distance after which the deflection of the CR trajectory becomes $\sim 1$~rad, that for isotropic diffusion is $l_{\rm D} = 3D/c$. From eq.~(\ref{delta}) we see that in this regime the diffusion length is $l_{\rm D} \simeq 2(E/BZe)^2/l_{\rm c}$ (and hence also the diffusion coefficient should grow as $D\propto E^2$). 

For $E>E_{\rm c}^{\rm xG}$ the image of a CR source lying at a distance $d_{\rm s}<l_{\rm D}$ 
will have an approximately Gaussian spread around the source direction, with an energy dependent angular width given by eq.~(\ref{delta}). The angular spread increases for farther away sources and 
for sources at distances much larger than $l_{\rm D}$ the cosmic rays may perform several turns before reaching the observer. In this last case the CR distribution from that source will be mostly isotropic, with the main feature being a dipole component in the direction towards the source \cite{hmr16}. 
Note that for a source at a distance $d_{\rm s}$ we expect that $l_{\rm D} > d_{\rm s}$ for particles with energies larger than
\begin{equation}
E_{\rm s} \simeq 3 Z  \frac{B}{\rm nG}\sqrt{\frac{l_{\rm c} d_{\rm s}}{10\, {\rm Mpc}^2}}\,{\rm EeV},
\end{equation}
and it is hence only above this energy that CRs from that source will look clustered around the  direction from where they originated.
One may estimate the typical distance to the closest sources  as the average separation between the UHECR sources, i.e. as $d_{\rm s} \simeq n_{\rm s}^{-1/3}$, with  $n_{\rm s}$ the source spatial density. One then gets  $d_{\rm s} \simeq 10$~Mpc for $n_{\rm s} = 10^{-3}$~Mpc$^{-3}$ while $d_{\rm s} \simeq 100$~Mpc for $n_{\rm s} = 10^{-6}$~Mpc$^{-3}$. Considering reasonable values for the extragalactic magnetic field around us in the Local Supercluster, $B\sqrt{l_{\rm c}}\sim {\rm nG}\sqrt{\rm Mpc}$,  protons may then have a diffusive behavior up to EeV energies  from the closest sources and up to higher energies for sources farther away. Nuclei will have a similar behavior but shifted up in energy by a factor $Z$, i.e. equal for the same rigidities.
One has to keep in mind that in addition to the deflections in the intergalactic space, also the deflections due to the Galactic magnetic field, discussed in Fig.~\ref{jfvsp}, will further affect the observed images of the CR sources.

Magnetic fields not only modify the arrival directions of cosmic rays, but they  can also modify the spectrum of the particles reaching the Earth, specially at sufficiently low  energies such that even the propagation from the closest source(s) is diffusive \cite{le04,be06,gl07,mo13}. The general result of the diffusion effects is to suppress the CR flux at low rigidities, since particles are not able to arrive to the observer from distant sources and take a much longer time than in the case of rectilinear propagation to arrive from the nearby ones. This is known as the `magnetic horizon' effect, as it limits the distance to which particles from a given source can arrive. There is however a relevant result, given by the propagation theorem \cite{propagationth}, which states that as long as the distance to the nearest sources is smaller than the other relevant length scales (diffusion length and energy loss length), the total CR flux from sources of similar strength will be the same as that expected ignoring magnetic field effects and for a continuous distribution of sources. This means that even at energies for which far away sources do not contribute anymore, as long as the nearby sources lie at a distance from the observer smaller than the diffusion length ($E>E_s$) the spectrum will not change  significantly  and only when the nearest sources get suppressed the overall spectrum will be modified.  In order for the spectrum from the closest sources to be strongly suppressed the energy loss length should be smaller than the distance traveled by the CRs. In particular, for $E/Z<1$~EeV, in which case the dominant energy losses come from the adiabatic expansion of the Universe, the spectrum will be suppressed as long as the propagation time is larger than the Hubble time, i.e. for $d_{\rm s}^2/6D>H_0^{-1}$. This happens for energies smaller than $E_H$ given by
\begin{equation}
E_{\rm H} \simeq 0.4 Z  \frac{B}{\rm nG}\sqrt{\frac{l_{\rm c} }{\rm Mpc}}\frac{d_{\rm s}}{40\, {\rm Mpc}}\,{\rm EeV}.
\end{equation}
Below this energy the magnetic horizon effect is expected to strongly suppress the flux, and as the effect is more pronounced for decreasing energies this results in a hardening of the spectrum of the CRs reaching  from a faraway  source with respect to the spectrum they had while leaving the source.
The hardening in the spectrum resulting from this effect \cite{mo13} may help to reconcile the low values ($\gamma <1.5$) of the source spectral index inferred when fitting the observed spectrum and composition above the ankle~\cite{combfit} with the standard values expected  from Fermi acceleration ($\gamma \simeq  2$).

\bigskip 
\section{Anisotropies in the arrival directions}

The distribution of the arrival directions of cosmic rays is expected to help, together with the spectrum and composition measurements, to improve our  understanding on their origin and  nature as well as on their propagation from the sources. As discussed in the last section, the trajectories are largely affected by the magnetic fields present, so that their study is strongly entangled with that of Galactic and extragalactic magnetic fields. 

At energies smaller than $E_{\rm c}$ the CRs are expected to perform a random walk in the turbulent magnetic field and after traversing a few coherence lengths they essentially lose the correlation with their initial directions. 
However, an anisotropic distribution of sources around the observer leads to a local density gradient and hence to the presence of a net CR flux given by
\begin{equation}
{\vec J} = - D_\perp {\vec \nabla_\perp} n - D_\parallel {\vec \nabla_\parallel} n + D_A {\vec b} \times {\vec \nabla} n, \label{difflux}
\end{equation}
where $D_\perp$ and $D_\parallel$ denote the diffusion coefficients perpendicular and parallel to the mean regular field ${\vec B}_{reg}$, $D_A$ is the antisymmetric
term describing Hall diffusion, $n$ is the density of cosmic rays and ${\vec b}={\vec B}_{reg}/B_{reg}$ is a unit vector along the regular magnetic field  (see e.g. \cite{ca03}). The resulting dipolar anisotropy is 
\begin{equation}
{\vec d}=-\frac{3{\vec J}}{n}
\end{equation}
For the isotropic diffusion case ($D_\perp = D_\parallel$, $D_A=0$), this will lead to a dipole anisotropy pointing towards the direction of the gradient of the density. In the anisotropic case, the directions along or perpendicular to the mean regular field are distinguished, and although in general $D_\perp$ is smaller than $D_\parallel$, their relative importance to determine the CR flux depends on the actual direction of the density gradient with respect to the regular magnetic field direction. In particular, for our galaxy the regular field follows mostly the structure of the spiral arms, and hence has a dominant azimuthal component, while the CR gradient points mostly radially or vertically, so that the impact of the perpendicular diffusion is not negligible in the study of the GCR escape from the Galaxy. The drift flux, proportional to $D_A$, is orthogonal to both the regular field and the density gradient, and it becomes non-negligible for $E/Z$ larger than few~PeV \cite{ca04}.  One can also see from eq.~(\ref{ec}) that  in the Galaxy the propagation remains diffusive up to energies $\sim 0.1Z$~EeV.

At energies in excess of $\sim 0.5Z$~EeV the Larmor radius becomes larger than the thickness of the Galactic disk, and hence one expects that Galactic sources should lead to significant anisotropies in the direction of the Galactic center and the Galactic plane (i.e. dipolar and quadrupolar excesses), the non-observation of which strongly constrains a Galactic contribution to the predominantly light CRs in the 1--5~EeV energy range \cite{LS12}. 

Above the ankle energy the extragalactic CR trajectories are expected to experience smaller deflections and to keep some `memory' of the direction where they were produced. This could  reflect into large scale anisotropies and in  clustering of CRs at small or intermediate angular scales, as well as correlation with the directions towards their sources.

\subsection{Anisotropies in the TeV to PeV energy range}

Cosmic rays are observed to arrive to the Earth almost isotropically. However, thanks to the large statistics collected in the last decades several experiments have detected small, but significant, anisotropies in the distribution of arrival directions. Many experiments can only detect variations in the flux as a function of the right ascension (RA) direction $\alpha$, but not on the declination $\delta$.\footnote{This is due to the uncertain attenuation of the showers in the atmosphere, which depends on the zenith angle and hence changes in different declination bands. Also geomagnetic effects as well as a possible tilt of the array can introduce systematic dependences on declination. On the other hand, the almost uniform integrated exposure in RA allows to become sensitive to very small intrinsic anisotropies in this coordinate.} These experiments then provide measurements of the projected anisotropy in the RA direction and eventually also give the results as a function of $\alpha$ for different declination bands. 

\begin{figure}[t]
\begin{center}
\begin{minipage}[t]{8 cm}
\centerline{\epsfig{file=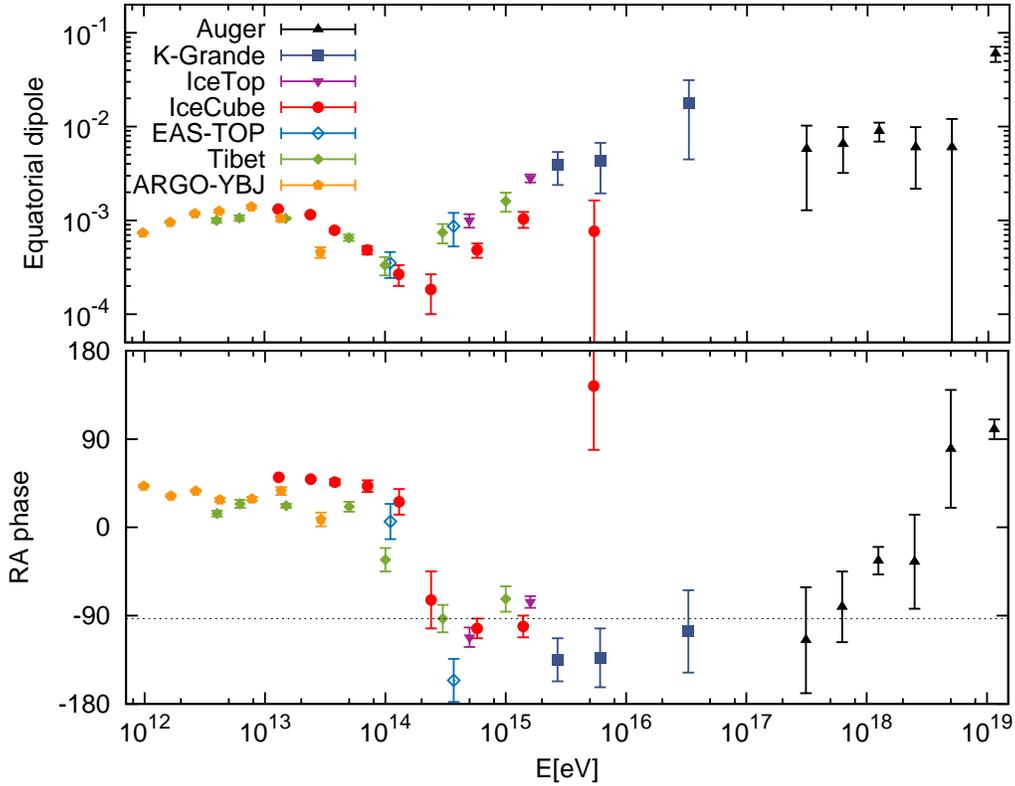,angle=-90,scale=0.7}}
\end{minipage}
\begin{minipage}[t]{16.5 cm}
\caption{Equatorial component of the dipole (top panel) and its phase in right ascension (bottom panel) for several cosmic ray experiments: Pierre Auger Observatory \cite{paoicrc15,
LS17},
KASCADE-Grande \cite{ch16}, IceTop \cite{ic12,ic16}, IceCube \cite{ic16}, EAS-TOP \cite{ag09}, Tibet-AS$\gamma$ \cite{am05,am15}, ARGO-YBJ \cite{argo15}. The dotted line in the bottom panel  indicates the RA of the Galactic center direction. \label{dipexp}}
\end{minipage}
\end{center}
\end{figure}

In the TeV to PeV energy range, for which cosmic rays  are of Galactic origin, several air shower experiments have detected large scale anisotropies  at the level of $10^{-4}$ to $10^{-3}$ \cite{am05,gu07,ag09,ic10,it13,am15,argo15,ic16}. Due to the lack of sensitivity to the variations in declination, they are not sensitive to the component of the dipolar anisotropy in the direction of the rotation axis of the Earth but can instead determine the equatorial component of a dipole $d_\perp$.\footnote{Due to the different latitude and zenith angle coverage of the sky by different experiments, the amplitude of the modulation of the flux as a function of the right ascension determined by each experiment is expected to be different. In the case that the dipolar component of the modulation is the dominant one, the equatorial component of the dipole (that should be the same for all the experiments) can be estimated as $d_{\perp} = r_1/\langle\cos\delta\rangle$, with $r_1$ the measured amplitude of the first harmonic modulation in right ascension and $\langle\cos\delta\rangle$ the mean value of the cosine of the declination of the events. This is the quantity plotted in Fig.~\ref{dipexp}.} The results for the amplitude and phase of the equatorial dipole component from several experiments are displayed in Fig.~\ref{dipexp}, covering seven orders of magnitude in energy.

The amplitude $d_\perp$ is found to increase with energy up to about 10 TeV, where it reaches values of $10^{-3}$, and decreases at higher energies up to about 100 TeV, where it reaches values of $10^{-4}$. In addition, the data indicate that the direction undergoes a phase flip at an energy around 100--300~TeV. At the lower energies the data show a trend for the dipole to be aligned with the local magnetic field \cite{ah16}. This is consistent with the predictions  of anisotropic diffusion which implies that the dipole should align with the projection of the CR gradient onto the local magnetic field. The local magnetic field direction is inferred from the IBEX measurement and is along Galactic coordinates $(l,b)\sim (210.5^\circ,-57.1^\circ)$, corresponding to  equatorial coordinates $(\alpha,\delta)=(48^\circ,-21^\circ)$ \cite{ibex}, in good agreement with the RA phases determined. This ordered magnetic field corresponds to the sum of the large scale regular magnetic field and the contribution of the turbulent component averaged over distance scales set by the CR gyroradius. At energies above $\sim 300$~TeV the equatorial dipole aligns with the right ascension of the Galactic center direction,  at $\alpha_{\rm GC} \simeq 266^\circ$. This is consistent with the expectations for the case of isotropic diffusion with a smooth distribution of sources in the Galaxy, as the diffusive dipole anisotropy  is expected to simply align with the direction of the larger concentration of sources, i.e. towards the Galactic center.

Another possible source for a dipolar anisotropy is due to the relative motion of the observer with respect to the frame in which the CR distribution is isotropic, what induces a weak dipole anisotropy pointing in the direction of the motion, known as the Compton-Getting effect \cite{CG}. The amplitude of the resulting dipole depends on the velocity of the observer $v$ and on the CR spectral index $\gamma$, being  $d=(\gamma+2)v/c$. Considering a value of $\gamma \simeq 2.7$ and as the observer velocity that of the solar system around the Galaxy, $v \simeq 220$~km\,s$^{-1}$, would lead to an energy independent  dipole amplitude of $few \times 10^{-3}$. The much smaller observed amplitude of the dipolar component indicates that the  cosmic rays actually co-rotate with the local stars \cite{ti06}. Moreover, the orbital motion of the Earth around the Sun leads also to a dipolar anisotropy in the cosmic ray flux  that is a function of the hour of the day, and can then be observed when analyzed in the solar frequency. Taking into account that the Earth moves at a velocity of 29.8~km\,s$^{-1}$ the expected amplitude is about $5 \times 10^{-4}$, with a maximum at 6~h. This anisotropy has been detected, and can be used to obtain an indirect independent determination of the spectral index \cite{ti08}.

Besides the dipolar pattern expected from the diffusion in the presence of a CR density gradient, also structures at smaller angular scales, at least down to 10$^\circ$, have been observed in the TeV to PeV energy range \cite{milagro,argo,icsmall,hawc}. These are not predicted by the simple diffusion picture and they have been the topic of different theoretical modeling efforts (see \cite{am17} for a review).  The suggested explanations consider local effects of the heliosphere, possible non-diffusive propagation or that they may be the result of a combination of the dipolar pattern and the local inhomogeneities of the interstellar magnetic fields within the scale of the diffusion length \cite{gi12,ah14}. The idea is that turbulent  magnetic fields close to the observer could produce  lensing effects that distort the image of the large scale dipolar anisotropy.

\subsection{Anisotropies for energies above the PeV}

\subsubsection{Large angular scale anisotropies}

At energies above few PeV, where the CR spectrum becomes steeper, the outward CR diffusive escape should lead to a dipolar component having its maximum not far from the Galactic center direction, what seems compatible with results obtained in the PeV to 1 EeV range \cite{it13,ch16,LS12,LS11}. If the steepening of the spectrum of the different nuclear components taking place from the knee up to the second knee were due to a more efficient escape from the Galaxy, as a result of drift effects dominating the CR transport, it is expected that the dipole direction should smoothly change from the Galactic center  direction to one closer to the northern Galactic pole, at $(\alpha, \delta) = (193^\circ, 27^\circ)$ \cite{ca03}.
This results from drift currents being, as indicated by the third term in eq.~(\ref{difflux}), orthogonal to both the regular magnetic field direction, which is mostly azimuthal along the Galactic disk, and the density gradient, which points towards the Galactic center. Anisotropy measurements in the energy range $10^{15}$--$10^{18}$~eV have phases in right ascension lying in between the Galactic center one at $\alpha=266^\circ$ and the North Galactic pole one at $\alpha=193^\circ$. Given the large uncertainties it is not possible at present to distinguish between the different scenarios, but  more precise measurements in this energy range, specially if they could be performed separately for the different mass groups,  would be important to better understand the origin of the knee and of the second knee.

\begin{figure}
\begin{center}
\begin{minipage}[t]{8 cm}
\centerline{\epsfig{file=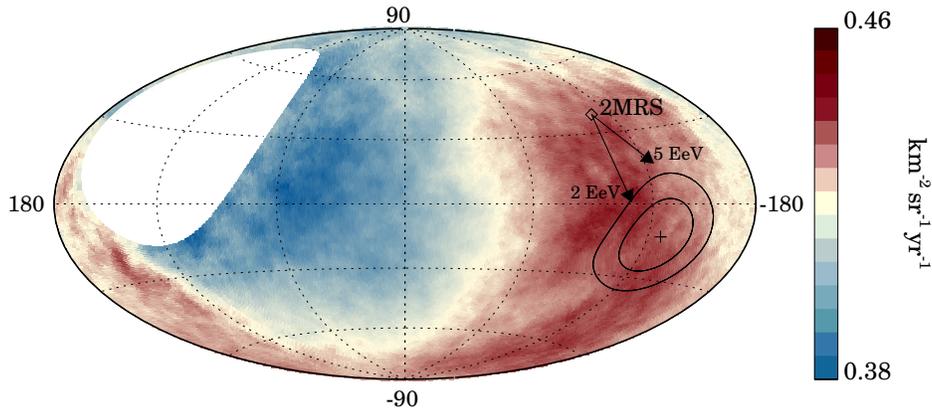,angle=0,scale=0.7}}
\end{minipage}
\begin{minipage}[t]{16.5 cm}
\caption{Map in Galactic coordinates of the flux of cosmic rays with $E > 8$~EeV, smoothed in  windows of $45^\circ$, as determined by the Auger Collaboration \cite{LS17}. The star indicates the direction of the reconstructed dipole and the ellipses the 68\% and 95\% CL regions. The diamond denotes the direction of the flux weighted dipole of IR selected galaxies from 2MRS \cite{erdogdu06}. The two arrows, labeled 5~EeV and 2~EeV, show the dipole direction of a flux of particles with those values of $E/Z$ arriving to the halo with a dipolar anisotropy in the 2MRS direction, which is at the origin of the arrows, would be modified after being deflected by the Galactic magnetic field modeled as in ref.~\cite{jf12}.\label{augerdip}}
\end{minipage}
\end{center}
\end{figure}

The change towards a predominantly extragalactic CR origin is expected to take place  somewhere above 0.1~EeV and  the study of the large scale  anisotropies should provide information about this transition. 
In fact, although at these energies the measured dipolar amplitudes are in general not significant, a change in the  phase of the anisotropies in right ascension  has been observed to take place at few EeV, going from a direction close to the Galactic center at lower energies to almost the opposite direction, with the maximum of the flux  pointing towards $\alpha \simeq 100^\circ$, at higher energies \cite{LS11,LS12,LS15}. Furthermore, by combining  the usual right ascension harmonic analysis with an harmonic analysis in the azimuthal coordinate of the CR arrival directions, which is sensitive to a north-south dipolar component, the Pierre Auger Observatory has reconstructed the three-dimensional CR dipole for energies above 4~EeV \cite{LS15}. In the highest energy bin analyzed, corresponding to $E > 8$~EeV, a significant dipolar amplitude of $6.5\%$ has been detected, pointing in Galactic coordinates towards $(l, b) = (233^\circ,−13^\circ)$ 
\cite{LS17}.
 This lies $\sim 125^\circ$ away from the Galactic center direction, indicating an extragalactic origin for this  flux. Fig.~\ref{augerdip} shows a map in 
Galactic 
coordinates of the flux for $E>8$~EeV smoothed in top-hat angular windows of $45^\circ$ radius, and in it the dipolar pattern is evident.

\begin{figure}
\begin{center}
\begin{minipage}[t]{8 cm}
\centerline{\epsfig{file=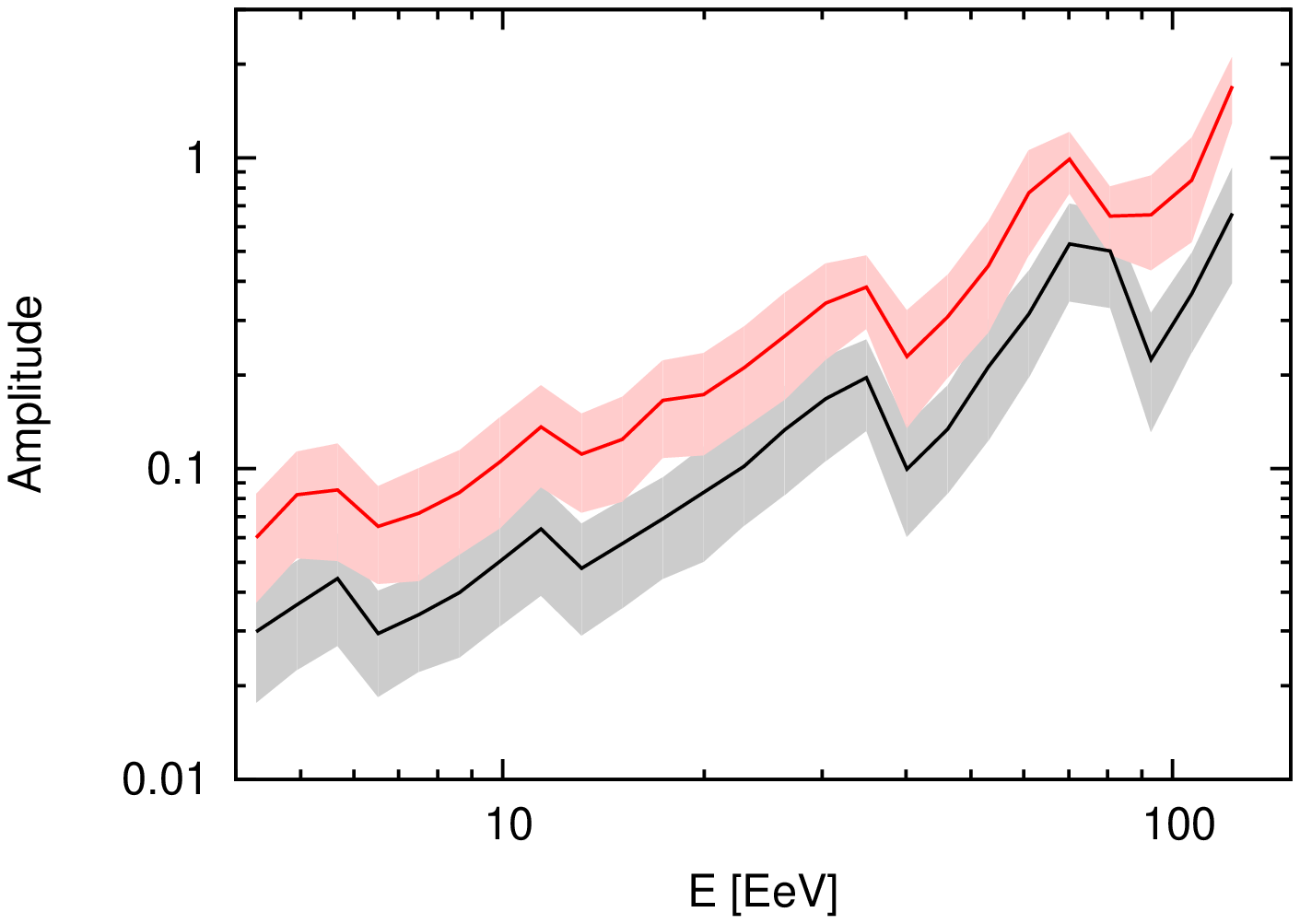,angle=-90,scale=0.6}\epsfig{file=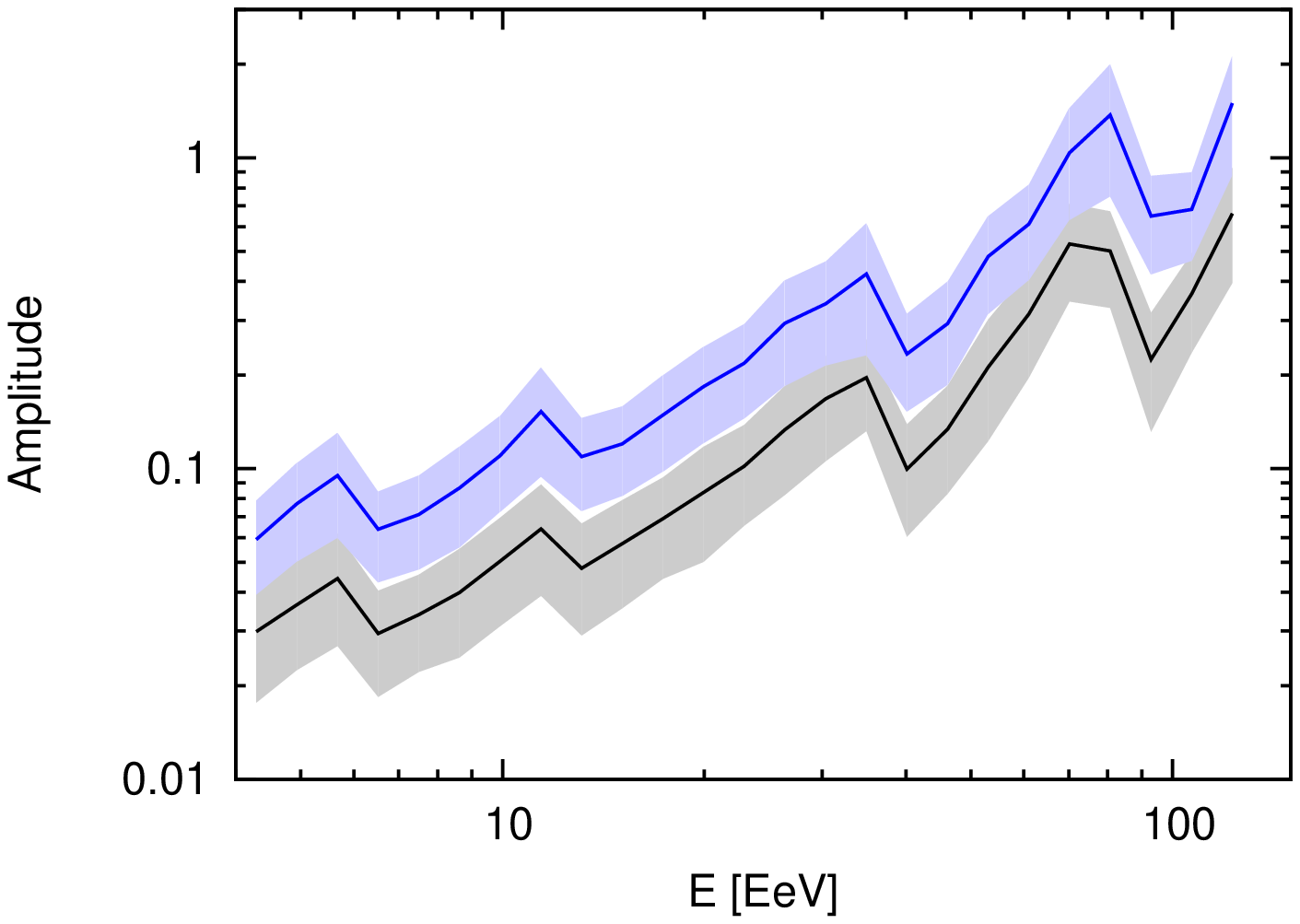,angle=-90,scale=0.6}}
\end{minipage}
\begin{minipage}[t]{16.5 cm}
\caption{Amplitude of the dipolar anisotropy in a mixed composition scenario for cosmic rays diffusing in the extragalactic magnetic field (from \cite{hmr15}). In the left panel the sources are distributed homogeneously with a density $10^{-4}$~Mpc$^{-3}$ (lower grey band) and $10^{-5}$~Mpc$^{-3}$ (upper red band). In the right panel the density is $10^{-4}$~Mpc$^{-3}$ and the sources are distributed homogeneously (lower grey band) or following the distribution of galaxies in the 2MRS catalog (upper blue band). The abrupt steps in the plots are just the consequence of the sharp rigidity cutoffs adopted. \label{difudip}}
\end{minipage}
\end{center}
\end{figure}

 There are different possible explanations for a dipolar anisotropy in the extragalactic CR flux. An anisotropy is expected from the Compton-Getting effect due to the motion of the observer. For instance, if the frame in which the extragalactic cosmic rays are isotropic were the same as that of the CMB radiation, then our peculiar motion would lead to a dipolar anisotropy of amplitude 0.6\%  \cite{ks06}, which is however an order of magnitude smaller than the observed one. A more promising explanation is that the dipolar flux could result from cosmic rays propagating diffusively in the extragalactic turbulent magnetic fields. This could happen  if the amplitude of the field is large and/or if the cosmic rays have a component with large electric charge \cite{ha14,hmr15}. The left panel of Fig.~\ref{difudip} shows the resulting dipole amplitude for a mixed composition scenario with maximum rigidity $R_{\rm c}=6$~EV and fractions $f_p = f_{He} =
f_{Si} = 0.19$, $f_{C} = 0.4$ and $f_{Fe} = 0.03$. A turbulent extragalactic magnetic field of amplitude $B_{\rm rms}=1$~nG and coherence length $l_{\rm c}=1$~Mpc was adopted. The dipolar amplitude obtained depends on the spatial density of sources, and results are shown for $n_s=10^{-4}$~Mpc$^{-3}$ and $10^{-5}$~Mpc$^{-3}$. The anisotropy is typically larger for lower densities, as the total flux gets shared between less sources in this case, and is expected to point towards the closest sources. 

 Due to the fact that in our local neighborhood galaxies are distributed inhomogeneously,  if the CR sources are distributed similarly to the matter in the Universe and enhancement in the large angular scale anisotropy in the CR arrival direction distribution is expected. Moreover, a dipolar component of the anisotropy could appear in this case even in the case in which magnetic deflections are small. The existence of inhomeogeneities in the distribution of nearby (within $\sim 100$~Mpc) galaxies is known to lead in particular to the non-vanishing acceleration of the Local Group which is responsible for the peculiar velocity that gives rise to the observed CMB dipole \cite{erdogdu06}. A similar non-isotropic distribution of the nearby extragalactic cosmic ray sources, following the anisotropies of the matter distribution,  would lead to an excess CR flux towards the direction with the highest concentration of nearby sources and this should contribute to the dipolar component of the large scale distribution of CR arrival directions.
The right panel of Fig.~\ref{difudip} illustrates this point, showing the expected amplitude of the dipole for the case of sources distributed homogeneously or following the galaxy distribution in the 2MRS catalog \cite{hu12}, adopting an average CR source density $n_s=10^{-4}$~Mpc$^{-3}$.

 As the energy threshold is increased, the maximum redshift from which extragalactic cosmic rays can arrive at Earth progressively decreases. This is a consequence of the energy losses due to interactions with the CMB and EBL backgrounds. Thus, the overall contribution to the flux coming from the nearby sources becomes more important as the energy increases, leading to a larger expected anisotropy at higher energies. It is clear from  Fig.~\ref{difudip} that dipolar anisotropies of 5--20\% amplitude at 10~EeV can naturally arise in these scenarios for reasonable values of the parameters \cite{hmr15}. 

The Galactic magnetic field is expected to modify the large angular scale anisotropies of extragalactic cosmic rays.
 Both the direction and amplitude of the dipolar component of the distribution of the CR flux arriving outside the Galaxy will be changed upon arrival at the Earth \cite{hmr10}. Moreover, even if only a dipolar component is present outside the Galaxy  also higher order multipoles, i.e. anisotropies on smaller angular scales,  will be generated by the action of the Galactic magnetic field. These effects are expected to be relevant for cosmic rays with $E/Z$ smaller than few EeV. 
 As an illustration of the effects of the Galactic magnetic field, the arrows in  Fig.~\ref{augerdip}  indicate how a dipolar CR distribution pointing towards the direction of the 2MRS flux weighted dipole, which lies at $(l,b)=(251^\circ, 38^\circ)$,  would be displaced when observed from the Earth, for CRs with $E/Z=2$~EeV or 5~EeV. It is worth noting that the agreement with the observed direction of the dipole above 8~EeV (plus sign and ellipses corresponding to 1 and 2$\sigma$ regions) improves after accounting for the Galactic magnetic field deflections.

\subsubsection{Small angular scales: intrinsic anisotropies}

At the highest energies, for which the propagation should become more rectilinear, anisotropies at smaller angular scales are expected to appear. Searches for localized excess fluxes have been performed both by the Pierre Auger and the Telescope Array observatories in the regions of the sky observable by each experiment. In these analyses the fluxes above different energy thresholds and around the  different sky directions are sampled using circular windows with varying angular radii. To detect an excess, for every window and energy threshold the number of observed events, $n_{\rm obs}$, is compared with that expected from an isotropic flux of cosmic rays, $n_{\rm exp}$. For each window, the binomial probability, $p$, of observing by chance in an isotropic flux an equal or larger number of events than that found in the data is computed. The minimum $p$-value corresponds then to  the most significant overdensity. This $p$-value does not represent the chance probability of the result, as it is still necessary to account for the `look-elsewhere' effect coming from the fact that many directions in the sky, with different angular radii and for different energy thresholds, have been sampled. A `penalized' probability, $P$, is obtained by performing identical searches  in simulated datasets where the direction of each event is chosen randomly following an isotropic distribution. The value of $P$ corresponds to the fraction of the simulations for which the minimum $p$-value is smaller than that found in the data.

\begin{figure}
\begin{center}
\begin{minipage}[t]{8 cm}
\centerline{\epsfig{file=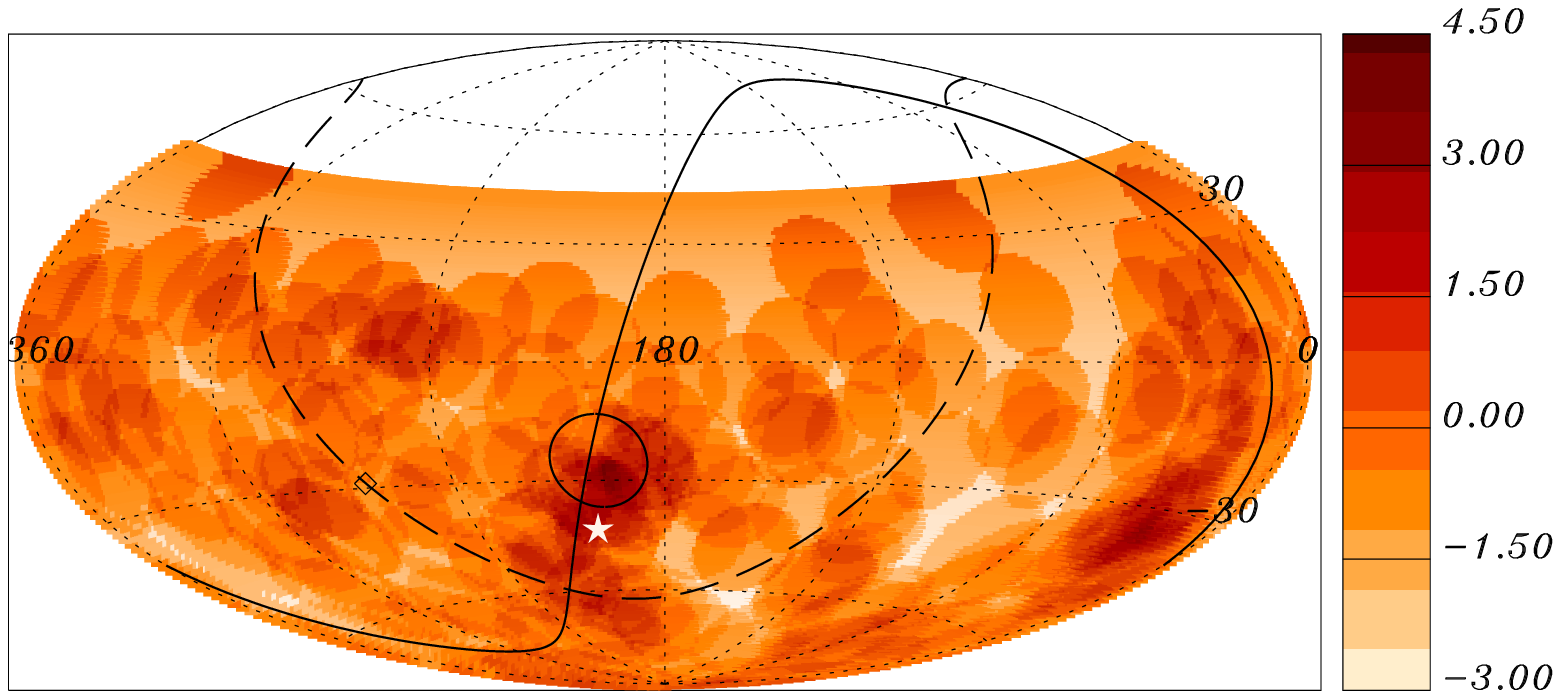,angle=0,scale=0.55}
\epsfig{file=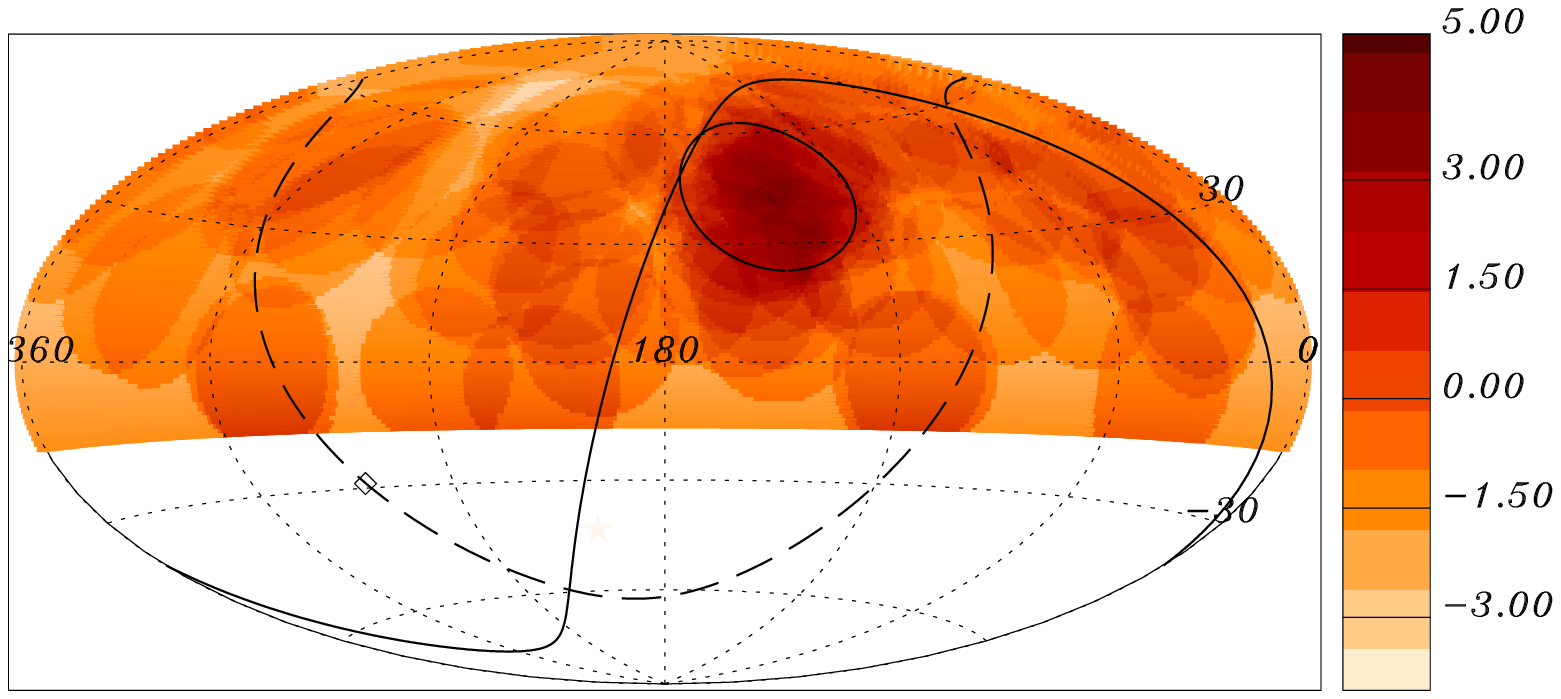,angle=0,scale=0.55}}
\end{minipage}
\begin{minipage}[t]{16.5 cm}
\caption{Significance map in equatorial coordinates for localized flux excesses in circular angular windows. Left panel: results of the Pierre Auger Observatory for an angular radius of $12^\circ$ and an energy threshold of 54 EeV, the black circle corresponds to the window with the maximum significance, the star shows the location of Cen~A, the dashed line denotes the Galactic plane and the solid line the Supergalactic plane (from \cite{SS15}). Right panel: results of the Telescope Array for an angular radius of $20^\circ$ and an energy threshold of 57~EeV. \label{od}}
\end{minipage}
\end{center}
\end{figure}
 
 The Pierre Auger Observatory performed the search using angular windows with radii varying from $1^\circ$ up to $30^\circ$, in $1^\circ$ steps. The centers of the windows were taken on a $1^\circ \times 1^\circ$ grid. The energy threshold of the events used to build the maps was varied from 40~EeV up to 80~EeV in steps of 1~EeV. The largest departure from isotropy was found for a minimum  $p = 5.9 \times 10^{-6}$, at an energy threshold of 54~EeV and in a $12^\circ$ radius window centered at right ascension and declination  $(\alpha,\delta)=(198^\circ ,-25^\circ  )$, i.e., for Galactic longitude and latitude $( l , b ) = ( - 51.1^\circ, 37.6^\circ)$, for which $n_{\rm obs}/n_{\rm exp} = 14/3.23$. The map of the Li-Ma \cite{lima83} significances of the excesses of events with $E \geq 54$ EeV in windows of $12^\circ$ radius is shown in the left panel of Fig.~\ref{od}. The highest excess region, having a Li-Ma significance of $4.3\sigma$, is indicated with a black circle. It is close to the Supergalactic plane, indicated with a solid line, and centered at about $18^\circ$ from the direction of Cen~A, indicated with a white star. The penalized probability, after accounting for the scan of all the parameters, is $P=0.69$, and is hence not significant \cite{SS15}.
 
 The Telescope Array performed the search at a fixed energy of 57~EeV, and considered 5 values for the angular window radius between $15^\circ$ and $35^\circ$ in steps of $5^\circ$. The maximum departure from isotropy for 7 years of observations is found for an angular radius of $20^\circ$ around the direction $(\alpha,\delta)= (148.5^\circ, 43.2^\circ)$ for which $n_{\rm obs}/n_{\rm exp} = 24/6.88$. The unpenalized statistical Li-Ma significance is $5.1\sigma$. The corresponding significance map is shown in the right panel of Fig.~\ref{od}. Taking into account the scan in the sky direction and the different angular windows considered the penalized chance probability of the observed hotspot in an isotropic cosmic-ray sky is $P = 3.7\times 10^{-4}$, equivalent to $3.4 \sigma$ \cite{taicrc15}. At the ICRC 2017 meeting an update of this search with two more years was reported, and the significance of the excess did not grow with additional data \cite{icrc17ta}.

A complementary method to look for clustering at small and intermediate angular scales is through the 2-pt correlation function, or autocorrelation of the events,  that measures the excess in the number of pairs with respect to that expected in the isotropic case, as a function of the angular separation between events. This statistics could be useful to detect clustering at some angular scale coming from  several regions of the sky rather than from just a single or a few localized excesses, which could go undetected with the search for overdensities in individual directions that was discussed before. The Pierre Auger Observatory performed this autocorrelation analysis counting the number of pairs of events that are within a certain angular distance $\psi$, $N_p (\psi, E_{\rm th})$, and above a given energy threshold $E_{\rm th}$. A scan like the one for the overdensity analysis was then performed but no significant excess was observed, indicating that the autocorrelation is compatible with the expectations from an isotropic distribution of arrival directions \cite{SS15}. The Telescope Array performed a similar analysis, considering three energy thresholds, 10, 40 and 57 EeV and scanning in all angular scales from $1^\circ$ up to $180^\circ$ \cite{taicrc15b}. While for the 10 and 40 EeV thresholds the results were compatible with the isotropic expectations, a moderate deviation from isotropy at angular scales between $20^\circ$ and $30^\circ$ was apparent in the highest-energy dataset.

The lack of significant clustering in the arrival directions of the highest energy events can be used to derive lower bounds on the density of sources of ultra-high energy cosmic rays, as the smaller the number of sources the larger is the flux expected from each of them. These bounds depend on the magnitude of the magnetic deflections and the spatial distribution of the sources. The Pierre Auger Collaboration found that, if the spread
due to magnetic deflections is smaller than $3^\circ$ the density of equal intrinsic intensity sources uniformly distributed in space should be larger than $5 \times 10^{-4}$ Mpc$^{-3}$ at $95\%$ CL. For larger angular spreads the bound is less restrictive, reaching $6 \times 10^{-6}$ Mpc$^{-3}$ for angular smearing of $30^\circ$. For a distribution of UHECR sources following the local matter in the Universe, as traced by the 2MRS catalog of galaxies, one could expect a slight enhancement in the expected number of pairs. Hence, slightly stronger lower bounds on the source density result,  ranging from $7 \times 10^{-4}$~Mpc$^{-3}$ if deflections are smaller than  $3^\circ$, up to $2 \times 10^{-5}$ Mpc$^{-3}$~for $30^\circ$
deflections \cite{augersd}. 

\subsubsection{Small angular scales: correlation with possible sources}

Besides the previous studies of intrinsic anisotropies, different analyses have been performed searching for correlations with known structures in the sky or with catalogs of possible CR sources.  Models of Galactic origin of UHECRs might give rise to an excess of events close to the Galactic plane or Galactic center, while the Supergalactic plane contains several nearby clusters and, thus, extragalactic cosmic rays could arrive preferentially near the Supergalactic plane. No significant correlation with any of these targets was found by the Pierre Auger Observatory at any energy threshold above 40~EeV \cite{SS15}. The Telescope Array found a (non-significant) deviation from isotropy around the Supergalactic plane for $E>57$~EeV, with an unpenalized $p$-value of 0.01 \cite{taicrc15}. 

\begin{figure}
\begin{center}
\begin{minipage}[t]{8 cm}
\centerline{\epsfig{file=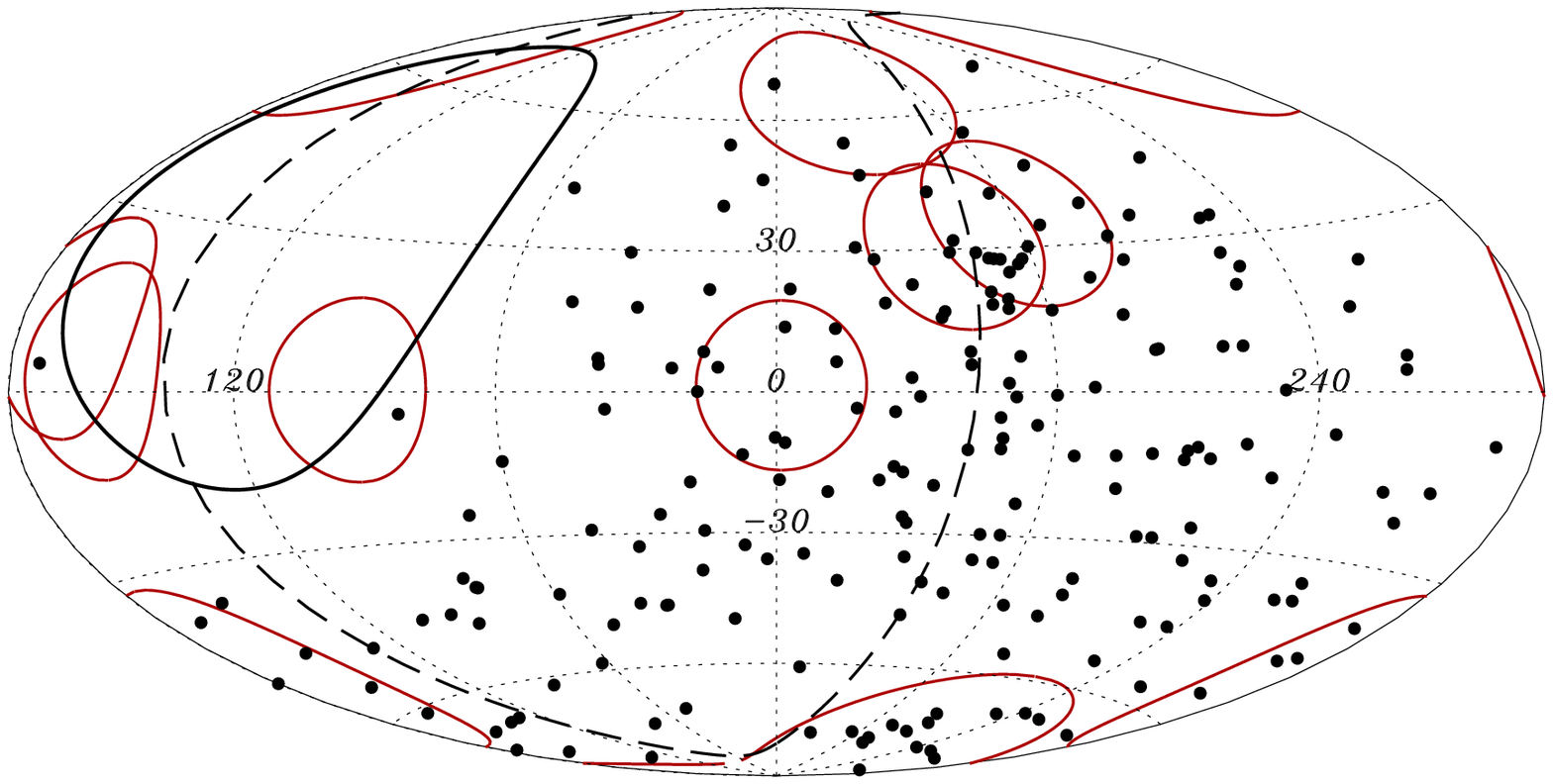,angle=0,scale=0.5}\epsfig{file=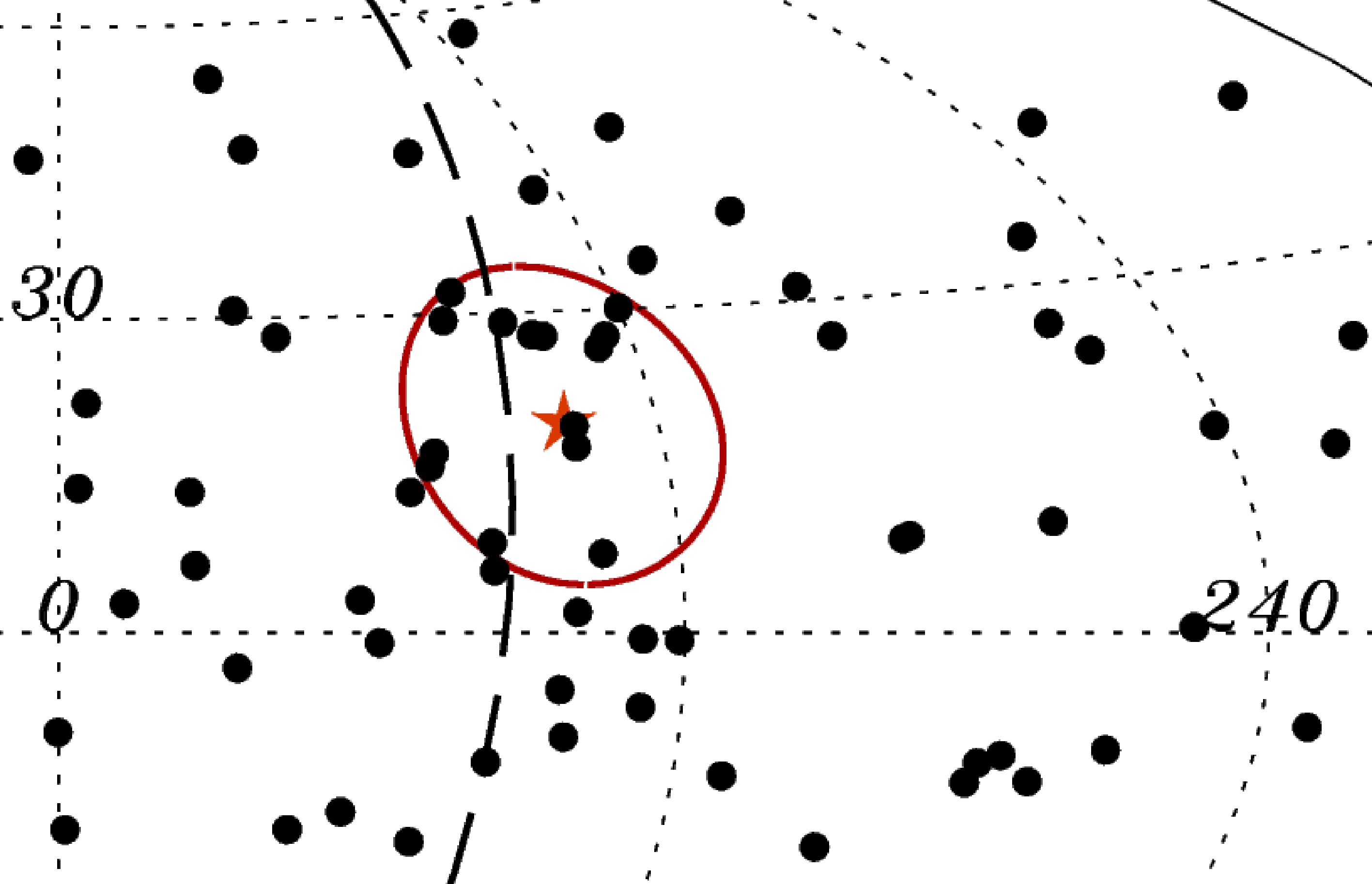,angle=0,scale=0.25}}
\end{minipage}
\begin{minipage}[t]{16.5 cm}
\caption{Left panel: Sky map in Galactic coordinates indicating with black dots the arrival directions of cosmic rays detected by the Pierre Auger observatory with energies above 58~EeV and red circles of $18^\circ$ radius around Swift AGNs with luminosity ${\cal L}_X  > 10^{44}$~erg\,s$^{-1}$ and closer than 130~Mpc (from \cite{SS15}). Right panel: Events with energies above 58~EeV in the direction near Cen~A (star), indicating a circle of $15^\circ$ radius around it.\label{swiftagn}}
\end{minipage}
\end{center}
\end{figure}

The Pierre Auger Observatory has also looked for correlations with galaxies in the 2MRS catalog \cite{hu12}, the Swift-BAT X-ray catalog of AGNs \cite{ba13} and a catalog of radiogalaxies with jets \cite{va12}. They test different possible scenarios for the sources: the normal galaxies in the 2MRS catalog could trace the locations of gamma-ray bursts and/or fast-spinning newborn pulsars, while Swift includes X-ray emitting AGNs, hosted mainly by spiral galaxies, and the radiogalaxy catalog selects AGNs with extended jets and radio lobes which are hosted mainly by elliptical galaxies. A cross-correlation analysis, searching for an excess in the number of pairs between the catalog objects and the high energy events, has shown no significant excess after scanning over the angular distance of the pairs, the maximum distance of the catalog objects considered and the minimum energy of the events. When considering a further scan in the minimum intrinsic luminosity of the objects to account for the possibility that only the more luminous sources could be able to accelerate cosmic rays up to the highest energies, it turned out that for the Swift AGNs the most significant excess is obtained for objects closer than $D = 130$ Mpc and more luminous than ${\cal L}_X  = 10^{44}$ erg\,s$^{-1}$, at angular distances smaller than $18^\circ$ and energies larger than $58$~EeV. For these parameters, there are 10 AGNs and 155 events, and 62 pairs are obtained between them, while the isotropic expectation is 32.8. The unpenalized $p$-value is $2 \times 10^{-6}$  and the penalized probability for the scan in the parameters performed is $1.3\%$. The left panel of Fig. \ref{swiftagn} shows the sky map in galactic coordinates indicating with black dots the arrival direction of the 155 events above 58~EeV and red circles of $18^\circ$ radius around the positions of the 10 AGNs are also indicated.
 At the ICRC 2017 meeting an update of this search was reported, with the penalized probability becoming a factor $\sim 20$ times smaller with three more years of additional data \cite{icrc17ugo}. In addition a search of correlations with starburst galaxies, using a likelihood method in which the sources are weighted by their radio flux, suggests a possible correlation with few nearby starburst galaxies.

The radio-loud active galaxy Cen~A, at a distance of about 4~Mpc, is the nearest radiogalaxy and is hence a prominent candidate source for
UHECRs in the southern sky \cite{ro96}. In addition, the nearby Centaurus cluster, at a distance of about 50~Mpc,  lies in a very similar direction. It is hence interesting to note that the most significant localized excess of UHECR
arrival directions reported by the Pierre Auger
Collaboration \cite{SS15} is very close to the direction of Cen~A (see left panel in Fig.~\ref{od}). A specific search of flux excesses around the Cen~A direction, scanning on angular scales and energy thresholds, led to the most significant departure from isotropy at an angle of $15^\circ$  for $E_{\rm th} = 58$~EeV, with a $p$-value of $2 \times 10^{-4}$ (with $n_{\rm obs}/n_{\rm exp} = 14/4.5$). The penalized probability after the scan in angle and energy threshold is $1.4\%$ \cite{SS15}. The right panel of Fig.~\ref{swiftagn} shows a region of the sky map in galactic coordinates indicating with black dots the arrival directions of events with energies above 58~EeV and a red circle of $15^\circ$ radius around Cen~A is indicated.  At the ICRC 2017 meeting an update of this search was reported, with the penalized probability becoming a factor $\sim 10$ times smaller with additional data \cite{icrc17ugo}.

In summary, at the highest energies the arrival distribution of cosmic rays has shown to be remarkably isotropic, at odds with the original expectations that assumed that only a few cosmic ray proton sources would be contributing at the highest energies. 
The most significant anisotropy detection is the dipolar modulation of $6.5\%$ amplitude for $E > 8$~EeV. The lack of small scale anisotropies, with just some hints of localized flux excesses at intermediate angular scales, implies that probably the Galactic and/or extragalactic magnetic fields have a significant effect on UHECRs trajectories. This is in fact expected in mixed composition scenarios, where the composition becomes heavier as the energy increases, in agreement with composition measurements \cite{augerxmax,sein}. Extragalactic magnetic fields with amplitudes in the nG range would significantly spread the arrival direction of heavy CR nuclei up to the highest energies, even for nearby extragalactic sources, washing out small scale anisotropies while still leading to significant large scale anisotropies and, hopefully, some intermediate angular scale anisotropies to be uncovered with more statistics.

\section{Outlook}

More than a century has passed since the discovery of cosmic rays and a lot of progress has been made towards understanding their nature and possible origin, but anyhow many fundamental issues still remain to be answered. This makes this field of research a very active one at the frontier both in physics and astrophysics, with new major experiments being constantly conceived and developed to find additional clues to solve the puzzle. The main question is of course that of the sources, i.e. where and how CRs get accelerated. To learn about this one needs to understand also how the CRs propagate from the sources up to the Earth in order to know what is the impact of magnetic field deflections and CR interactions, either at or near the sources or in the Galaxy or the inter-galactic space.

The vast majority of CRs are ionized atomic nuclei and in the last decade an impressive progress has been made in the determination of the composition as a function of the energy. These measurements provide crucial information to understand several issues, such as the effects of spallations in the low energy Galactic CRs, the rigidity dependent suppression responsible for the knee and second knee of the spectrum, the presence of a sizable light component at EeV energies of likely extragalactic origin, the mechanism responsible for the ankle feature and the transition to a heavier composition above it, as well as the observed spectral suppression at the highest energies.

Although the distribution of CR arrival directions has been found to be extremely isotropic at all the energies, the determination of the small existing anisotropies has shown a remarkable progress in the last decades, with significant large scale modulation measurements obtained for energies ranging from $10^{12}$~eV up to $10^{19}$~eV. These encode crucial information about the distribution of the sources and the propagation effects in the magnetic fields permeating the space between the sources and us. The possibility to perform these measurements separating the CRs by composition would be an extremely valuable improvement in the future. Anisotropies on smaller angular scales  have also been measured in recent years in the TeV to PeV energy range, leading to the possibility of detailed tests of models of CR propagation in the magnetic fields in the solar neighborhood. Small scale anisotropy studies at UHE are also of fundamental importance to understand the CR propagation and to eventually pin-down the sources. It is at the highest energies that the diffusion prevalent at lower energies may finally give way to a regime of moderate deflections in which some of the dominant sources may show up. Although some hints have appeared, no significant detection has yet been made. This is indicative that magnetic field effects on the CR trajectories are relevant up to the highest observed energies.

Among the issues that are still open and deserve continuous studies one could mention:
\begin{itemize}
\item Whether the knee is due to a rigidity dependent cutoff of the Galactic CR sources or is instead due to the onset of a more efficient escape of the CRs from the Galaxy. To answer this the study of the phase of the large scale anisotropy, and its determination for different CR compositions, could be useful.

\item The energy at which the galactic-extragalactic transition takes place and the origin of the light component present at EeV energies. In particular, to determine whether a significant fraction of this light component could be due to nucleons from  higher energy nuclei that get photodisintegrated at or near their sources. For these studies it would be important to have detailed composition and spectral information at EeV energies and also obtain it for each nuclear component up to its eventual cutoff.

\item The origin of the ankle and whether there are significant amounts of protons above it: is it related to the proton cutoff, to the shaping of the extragalactic proton spectrum by the pair production interactions with the CMB, is there a non-negligible surviving Galactic component still present, etc.

\item What is the origin of the spectral suppression observed at the highest energies? Is it due to a source cutoff or to the CR energy losses during their propagation from very faraway sources?

\item Another point to understand is if the Galactic CRs are just due to one dominant type of source (e.g. the acceleration in the SNR of some particular type of supernovae) or if different types of sources (including e.g. pulsars or the black hole at the Galactic center) could contribute at different energies.
Similarly the extragalactic component may get contributions from different sources with different cutoffs or compositions (e.g. GRBs, AGNs of different types, starburst galaxies or galaxy clusters).

\end{itemize}

The identification of  individual powerful UHECR sources could open a new window to do astronomy, with the deflections suffered by the charged particles making the search more challenging but also leading to interesting new features, such as the multiple imaging of the sources due to magnetic lensing or the spectrometer effect that could help to study the magnetic fields and the CR composition. On the other hand, an improved knowledge of Galactic and extragalactic magnetic fields will also be quite important. In addition, a multi-messenger approach, exploiting the strong links with high energy gamma rays and neutrinos, which are the natural byproducts of the CR acceleration process, can be fundamental to identify the CR sources. The joint analyses of gravitational-waves, electromagnetic radiation at different wavelengths and neutrinos are also becoming a new possibility to learn about very energetic processes that could be related to cosmic ray accelerators.

Many experimental efforts will help to push the limits, in particular with more precise and larger detectors in space pushing the direct detection up to few hundred TeV (e.g. with ISS-CREAM), high-altitude dense air-shower arrays pushing the indirect search down to a few TeV (such as ARGO-YBJ or LHAASO) and large sparse arrays improving and enlarging the capabilities at the highest energies (Auger and TA upgrades), so that further progress in this field is certainly expected to take place in the near future.

\section*{Acknowledgments}
We are very grateful to Diego Harari for all the work done together during the past years. Also to J. Candia, G. Golup and O. Taborda who did their PhD in our group. Thanks to M. Unger, R. Ulrich, T. Pierog, J. Beatty, J. Matthews, O. Taborda, D. Harari, J. Alvarez-Mu\~niz and J.C. Arteaga for help with the figures for this review. Thanks to Roger Clay for a careful reading of the manuscript and to our colleagues from the Pierre Auger Collaboration for many discussions.
This work was supported by CONICET (PIP 0447) and ANPCyT (PICT 2013-0213).

\end{document}